\documentclass[aps,prl,reprint,superscriptaddress,longbibliography,nofootinbib]{revtex4-2}

\usepackage{graphicx}
\usepackage{dcolumn}
\usepackage{bm}

\usepackage[figurename=Figure]{caption}
\usepackage{ragged2e}
\DeclareCaptionJustification{plain}{\justifying}
\usepackage[percent]{overpic}
\usepackage{float} 
\usepackage{verbatim} 
\usepackage{amsmath} 
\usepackage{bbold}
\usepackage{upgreek} 
\usepackage{booktabs}
\usepackage{amssymb} 
\usepackage{svg}
\usepackage{enumitem} 
\usepackage{latexsym,epsfig}
\usepackage{subcaption}
\usepackage{stackrel}
\usepackage{mathtools}

\usepackage{placeins} 

\usepackage[colorlinks=true,breaklinks=true,allcolors=blue]{hyperref}
\usepackage[capitalize]{cleveref}
\usepackage{lipsum}
\usepackage{dsfont}
\usepackage{coffeestains}

\usepackage[scr=boondox]{mathalfa}

\newtheorem{theorem}{Theorem}

\newcommand{\be}{\begin{equation}}
	\newcommand{\ee}{\end{equation}}
\newcommand{\bea}{\begin{equation}\begin{aligned}}
		\newcommand{\eea}{\end{aligned}\end{equation}}
\newcommand{\ben}{\begin{enumerate}}
	\newcommand{\een}{\end{enumerate}}

\DeclareDocumentCommand{\nint}{ O{} O{} m }{\ensuremath{ \int_{\mbox{\scriptsize $#1$}}^{\mbox{\scriptsize$#2$}}\!\!\! \mbox{\small $\,\mathrm{d}#3$\! }}}

\usepackage{tikz,bm} 
\usepackage{circuitikz}
\usetikzlibrary{decorations.markings}
\usetikzlibrary{decorations.pathreplacing,calligraphy}

\definecolor{mycolor}{rgb}{1,0.2,0.3}
\definecolor{brightgreen}{rgb}{0.4, 1.0, 0.0}
\definecolor{britishracinggreen}{rgb}{0.0, 0.26, 0.15}
\definecolor{cadmiumgreen}{rgb}{0.0, 0.42, 0.24}
\definecolor{ceruleanblue}{rgb}{0.16, 0.32, 0.75}
\definecolor{darkelectricblue}{rgb}{0.33, 0.41, 0.47}
\definecolor{darkpowderblue}{rgb}{0.0, 0.2, 0.6}
\definecolor{darktangerine}{rgb}{1.0, 0.66, 0.07}
\definecolor{emerald}{rgb}{0.31, 0.78, 0.47}
\definecolor{palatinatepurple}{rgb}{0.41, 0.16, 0.38}
\definecolor{pastelviolet}{rgb}{0.8, 0.6, 0.79}

\begin{document}

	\title{Magic Without Entanglement: Exact Revivals and Their Fisher Information Origin}
	\author{Rishabh Jha}
	\email{rishabh.jha@usc.edu}
	\affiliation{%
		Department of Physics and Astronomy, University of Southern California, Los Angeles, CA 90089-0484, USA
	}
	\author{Karun Gadge}
	\email{karun.gadge@uni-goettingen.de}
	\affiliation{%
		Institute for Theoretical Physics, Georg-August-Universit\"{a}t G\"{o}ttingen, Friedrich-Hund-Platz 1, 37077 G\"{o}ttingen, Germany
	}

	%


	\begin{abstract}
		Magic and entanglement are independent quantum resources, yet their exact relation in many-body dynamics has remained elusive. We uncover two structural principles. First, at any stabilizer state, the curvature of the second stabilizer R\'enyi entropy under an arbitrary Hermitian generator equals the quantum Fisher information up to a fixed normalization, creating a bidirectional bridge between computational and metrological resources. Second, for commuting Ising evolution on any forest graph, a Clifford pruning circuit yields the full stabilizer-R\'enyi family at arbitrary size and in any spatial embedding, thereby furnishing a graph-theoretic construction of families with finite magic density and vanishing entanglement density in the thermodynamic limit. We solve two paradigmatic one-dimensional realizations central to quantum simulation---an Ising quench and a kicked Floquet chain---exactly for arbitrary system size and directly in the thermodynamic limit, revealing finite magic density with vanishing entanglement density, distinct magic and entanglement revival periods, and Clifford points with zero magic but finite bipartite entanglement. The same tangent geometry fixes initial growth, perturbative revival lifting, and stability of thermodynamic magic minima. Large-scale Pauli-basis matrix-product-state calculations verify all predictions, and the tangent bridge yields a concrete protocol for detecting magic through established quantum-Fisher-information measurements.
	\end{abstract}

	\maketitle
	


	\textit{Introduction.---} Quantum many-body states support several inequivalent resources. Entanglement organizes nonlocal correlations and underlies the description of quenches, thermalization, and information spreading~\cite{calabrese2005evolution,calabrese2007quenches,bertini2019entanglement}. Magic quantifies departure from the stabilizer sector and supplies the non-Clifford resource required for universal fault-tolerant computation, linking quantum advantage, classical simulation complexity, and resource theory~\cite{gottesman1997stabilizer,bravyi2005universal,Bravyi2016Jun,veitch2014resource,howard2014contextuality,chitambar2019quantum,zheng2026gate,xia2026reservoir}. Stabilizer R\'enyi entropies provide an especially useful family of magic measures, defined from the distribution of Pauli expectation values and accessible without an optimization over stabilizer states~\cite{leone2022stabilizer}, with recent extensions across spin, bosonic, and fermionic systems~\cite{unifiedMagicCFT,seibert2026correlation,chu2026phaseless}.
	
	The two resources are logically independent already for few qubits: Bell states are maximally entangled stabilizer states and therefore have no magic, whereas a nonstabilizer product state has magic without entanglement. The many-body problem is harder: can local dynamics generate a finite magic density while entanglement stays nonextensive, do the two resources recur with distinct periods, and what geometric quantity controls the loss of a magic-free stabilizer point? These questions are timely given results on many-body magic generation, magic spreading, exact circuit dynamics, and entanglement-nonstabilizerness separations~\cite{liu2022manybody,rattacaso2023stabilizer,fux2024separation,frau2024nonstabilizerness,montanalopez2024exact,turkeshi2025magic,borda2026magicprotected}; an exact framework resolving resource density, dynamical revivals, and stability together has remained unavailable.

	Here we provide such a framework. First, we prove a tangent theorem: at every stabilizer state, the leading growth of the second stabilizer R\'enyi entropy under an arbitrary Hermitian generator equals the quantum Fisher information (QFI), which measures the speed of the trajectory in projective Hilbert space~\cite{braunstein1994statistical,Paris2011Nov,pezze2018quantum,toth2014quantum}. The theorem thus makes the QFI the exact local susceptibility of magic, giving a two-way bridge in which measuring either quantity fixes the other, and a direct experimental route to magic through established QFI measurement protocols. This relation also controls real-time evolution near a stabilizer state and detuning-induced lifting of Clifford revivals.

	Second, we establish a forest theorem for commuting Ising dynamics on arbitrary acyclic interaction graphs. A Clifford pruning circuit maps the full many-body problem to independent single-qubit rotations, yielding the complete stabilizer-R\'enyi family for every system size and every spatial embedding. The theorem identifies graph families with finite magic density and vanishing entanglement density, thereby realizing magic without entanglement density as an exact many-body phenomenon. For the open Ising chain, this structure gives exact quench and Floquet dynamics directly at arbitrary size and in the thermodynamic limit. Magic and entanglement exhibit distinct recurrence periods, including Clifford points with vanishing magic but finite bipartite entanglement. The tangent theorem then determines the initial magic curvature and the perturbative lifting of quench and Floquet revivals. Pauli-basis matrix-product-state calculations systematically verify the exact formulas, the quantum-Fisher-information benchmarks, and the perturbative lifting of the quench and Floquet revivals~\cite{haug2023mpsmagic,lami2023perfectsampling,tarabunga2024paulibasis}.

	\textit{Magic, stabilizers, and tangent geometry.---} Magic is used to mean nonstabilizerness: the part of a pure state that cannot be removed by Clifford gates and is absent in stabilizer states \cite{gottesman1997stabilizer,veitch2014resource}. Let \(I,X,Y,Z\) denote the single-qubit identity and Pauli matrices, and let \(\mathcal P_L=\{P_a=\sigma^{a_1}\otimes\cdots\otimes\sigma^{a_L}:a_j\in\{0,1,2,3\}\}\), with \(\sigma^0=I\), \(\sigma^1=X\), \(\sigma^2=Y\), and \(\sigma^3=Z\), be the \(4^L\) phase-free Pauli strings on \(L\) qubits. For a pure state \(|\psi\rangle\), define
	\begin{equation}
		\Xi_P(\psi)\coloneq 2^{-L}|\langle\psi|P|\psi\rangle|^2,\qquad P\in\mathcal P_L.
		\label{eq:pauli_probability}
	\end{equation}
	The Pauli expansion of \(|\psi\rangle\langle\psi|\) gives \(\sum_{P\in\mathcal P_L}\Xi_P=1\), so \(\Xi_P\) is a probability distribution over Pauli strings. The stabilizer R\'enyi entropy is
	\begin{equation}
		M_\alpha(\psi)=\frac{1}{1-\alpha}\log_2\sum_{P\in\mathcal P_L}\Xi_P(\psi)^\alpha-L,\qquad \alpha>0,
		\label{eq:sre}
	\end{equation}
	with the Shannon limit at \(\alpha=1\) understood by continuity \cite{leone2022stabilizer}. A stabilizer state is the simultaneous eigenstate of \(L\) independent commuting Pauli strings, whose phase-free group contains exactly \(2^L\) strings with \(|\langle P\rangle|=1\); every string outside it has zero expectation, so \(M_\alpha=0\). A Clifford unitary maps Pauli strings to Pauli strings under conjugation. It therefore only permutes the probabilities in Eq.~\eqref{eq:pauli_probability}, so \(M_\alpha\) is Clifford invariant. We focus on \(M_2\) whenever local geometry is needed. For pure states, \(M_2\) is a faithful nonstabilizerness quantifier, reduces to a simple fourth moment of Pauli expectations, and is the most directly computable and experimentally accessible member of the stabilizer R\'enyi family \cite{leone2022stabilizer}. We now show that \(M_2\) also obeys an exact tangent law at every stabilizer point.

	\begin{theorem}[Tangent geometry of magic]
		\label{thm:tangent}
		Let \(|s\rangle\) be any stabilizer state, \(K=K^\dagger\) and $F_Q$ denote the QFI. Along the path \(|\psi(\theta)\rangle=e^{-i\theta K}|s\rangle\), where \(\theta\) is any real control parameter,
		\begin{equation}
			\begin{aligned}
				M_2(\psi(\theta))=& \, \frac{\theta^2}{\ln2}F_Q(|s\rangle,K)+O(\theta^3), \\
				F_Q(|s\rangle,K)=& \, 4\left(\langle K^2\rangle_s-\langle K\rangle_s^2\right).
			\end{aligned}
			\label{eq:tangent_formula}
		\end{equation}
	\end{theorem}
	
	Here \(\langle A\rangle_s=\langle s|A|s\rangle\). For pure states, the QFI \(F_Q\) is four times the Fubini--Study metric along the trajectory and therefore measures its squared speed in projective Hilbert space~\cite{braunstein1994statistical,Paris2011Nov}. The parameter \(\theta\) is not tied to one protocol: it may be time, a Floquet detuning, or a variational angle. Eq.~\eqref{eq:tangent_formula} thus gives a bidirectional local relation at a stabilizer point,
	\begin{equation}
		F_Q(|s\rangle,K)=\frac{\ln 2}{2}\,
		\left.\frac{\partial^2 M_2(\psi(\theta))}{\partial\theta^2}\right|_{\theta=0}.
		\label{eq:qfi_magic_curvature}
	\end{equation}
	This is an explicit experimental proposal: an independently measured QFI, via established interferometric, response-based, or randomized-measurement protocols already demonstrated in many-body settings~\cite{hauke2016measuring,strobel2014fisher,Yu2021qfi}, directly fixes the leading tangential magic without reconstructing the full Pauli distribution, while conversely the measured curvature of \(M_2\) determines the QFI. This is a local, quadratic-order statement and does not imply that QFI determines magic at finite \(\theta\). Components of \(K\) that only change the global phase have zero variance and create no magic. The proof in the SM~\cite{SM} expands Eq.~\eqref{eq:sre} at \(\alpha=2\). Stabilizer Pauli strings give the universal order--\(\theta^2\) term, while all Pauli strings outside the stabilizer group enter first at order \(\theta^4\).

	\begin{figure}[t]
		\centering
		\includegraphics[width=0.49\textwidth]{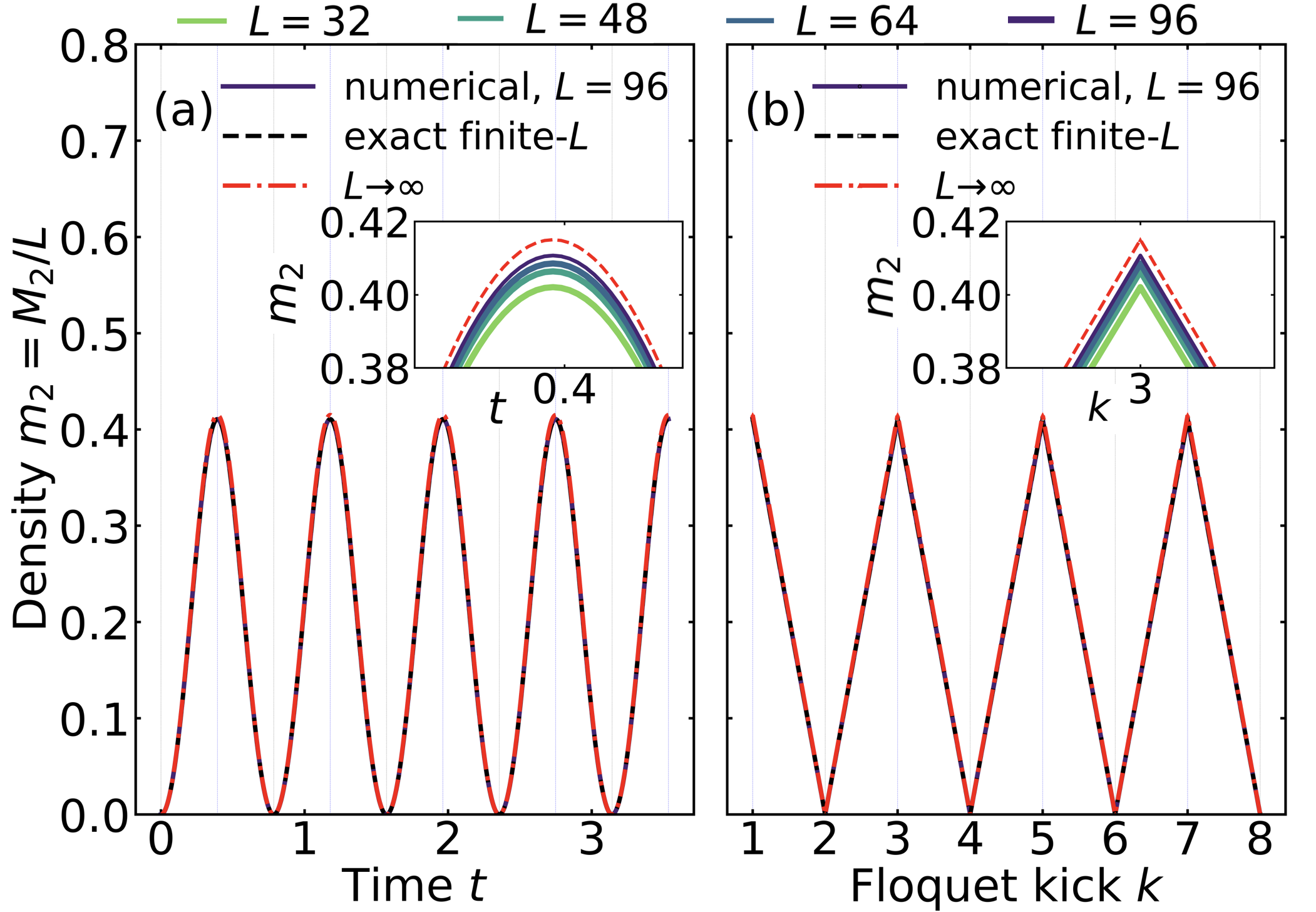}
		\caption{Exact magic-density revivals. Left: open-chain quench with \(J=1\) and \(h=0\), where \(M_2^{(L)}(t)/L=(1-1/L)m_2(Jt)\). Right: stroboscopic kicked-Ising dynamics at \(g=\pi/2\) and \(\theta=\pi/8\), where \(M_{2,{\rm F}}^{(L)}(k)/L=(1-1/L)m_2(k\theta)\); the magic revives at even kick number, with the first nontrivial revival at \(k=2\). System sizes are \(L=32,48,64,96\); the red curve is the thermodynamic prediction \(m_2(\vartheta)\), with \(\vartheta=Jt\) for the quench and \(\vartheta=k\theta\) for the kicked circuit.}
		\label{fig:magic_density}
	\end{figure}
	
	\textit{The forest theorem.---} The second result is an exact many-body normal form that does not rely on dimension, translation invariance, or a thermodynamic approximation. A graph \(G=(V,E)\) has vertices \(V\) and edges \(E\); here vertices are qubits and edges carry, e.g. two-qubit Ising gates. A tree is a connected graph without a closed loop, and a forest is a disjoint union of trees. Let \(c(G)\) be the number of connected components. For the representative stabilizer input \(|+\rangle^{\otimes L}\), with \(X|+\rangle=|+\rangle\), we define
	\begin{equation}
		|\psi_G(\theta)\rangle=\exp\left[-i\theta\sum_{(ij)\in E}Z_iZ_j\right]|+\rangle^{\otimes L}.
		\label{eq:graph_state}
	\end{equation}
	This is a representative choice, not a unique input. By Clifford invariance, if \(|s\rangle=C|+\rangle^{\otimes L}\) with generators conjugated as \(CZ_iZ_jC^\dagger\), Eq.~\eqref{eq:forest_formula} holds for the conjugated family too. Let \(m_\alpha(\theta)\coloneq M_\alpha(e^{-i\theta Z}|+\rangle)\) be the one-qubit SRE. Evaluating Eq.~\eqref{eq:sre} for this state gives (see SM~\cite{SM})
	\begin{equation}
		m_\alpha(\theta)=\frac{\log_2\left[1+\cos^{2\alpha}(2\theta)+\sin^{2\alpha}(2\theta)\right]-1}{1-\alpha},\qquad \alpha\neq1,
		\label{eq:one_qubit_magic}
	\end{equation}
	with \(\alpha=1\) obtained by continuity.
	
	\begin{theorem}[Forest normal form]
		\label{thm:forest}
		If \(G\) is a forest, then for every \(\alpha>0\),
		\begin{equation}
			M_\alpha(\psi_G(\theta))=\bigl[L-c(G)\bigr]m_\alpha(\theta).
			\label{eq:forest_formula}
		\end{equation}
		In particular,
		\begin{equation}
			M_2(\psi_G(\theta))=\bigl[L-c(G)\bigr]\left[-\log_2\left(1-\frac{1}{4}\sin^2(4\theta)\right)\right].
			\label{eq:forest_m2}
		\end{equation}
	\end{theorem}
	Eq.~\eqref{eq:forest_formula} thus has a well-defined thermodynamic limit whenever \(c(G)/L\) converges; for a connected tree, \(c(G)=1\) and \(M_\alpha/L\to m_\alpha(\theta)\). The same structure bounds entanglement: for any bipartition \(A|A^c\), only crossing edges raise the Schmidt rank, so \(S_{\rm EE}(A)\leq|\partial_G A|\), the number of edges between \(A\) and \(A^c\). Thus forests can carry extensive magic while the entanglement density vanishes whenever \(|\partial_G A|=o(L)\). This is the graph-theoretic form of magic without entanglement density. The proof is given in the SM~\cite{SM}; the mechanism is a Clifford elimination of all independent edge parities. Root each tree, and order the non-root vertices from leaves toward the root. For a non-root vertex \(v\), let \(p(v)\) be its parent. Applying \(\mathrm{CNOT}_{p(v)\to v}\) maps \(Z_{p(v)}Z_v\) to \(Z_v\). Iterating this leaf removal, or pruning, produces a Clifford circuit \(C_G\) such that
	\begin{equation}
		C_G\left(\prod_{(ij)\in E}e^{-i\theta Z_iZ_j}\right)C_G^\dagger
		=\prod_{v\notin R}e^{-i\theta Z_v},
		\label{eq:forest_normal_form}
	\end{equation}
	where \(R\) is the set of roots and \(|R|=c(G)\). Since each CNOT preserves \(|+\rangle^{\otimes L}\), Clifford invariance and additivity reduce the entire \(L\)-qubit Pauli distribution to \(L-c(G)\) identical one-qubit factors, proving Eq.~\eqref{eq:forest_formula}.

	The conjugated-stabilizer statement above is exact, but arbitrary stabilizer inputs with the unrotated \(Z_iZ_j\) gates need not obey Eq.~\eqref{eq:forest_formula}: the Clifford circuit may rearrange entanglement but cannot change magic, and the theorem evaluates the \(4^L\)-term Pauli moment by exposing the acyclic set of \(L-c(G)\) independent one-qubit rotations. Cycles are the obstruction: they preserve zero-magic Clifford revivals at \(\theta\in\pi\mathbb Z/4\), but away from these angles generally destroy the product formula.
	

	\begin{figure*}[t]
		\centering
		\includegraphics[width=\textwidth]{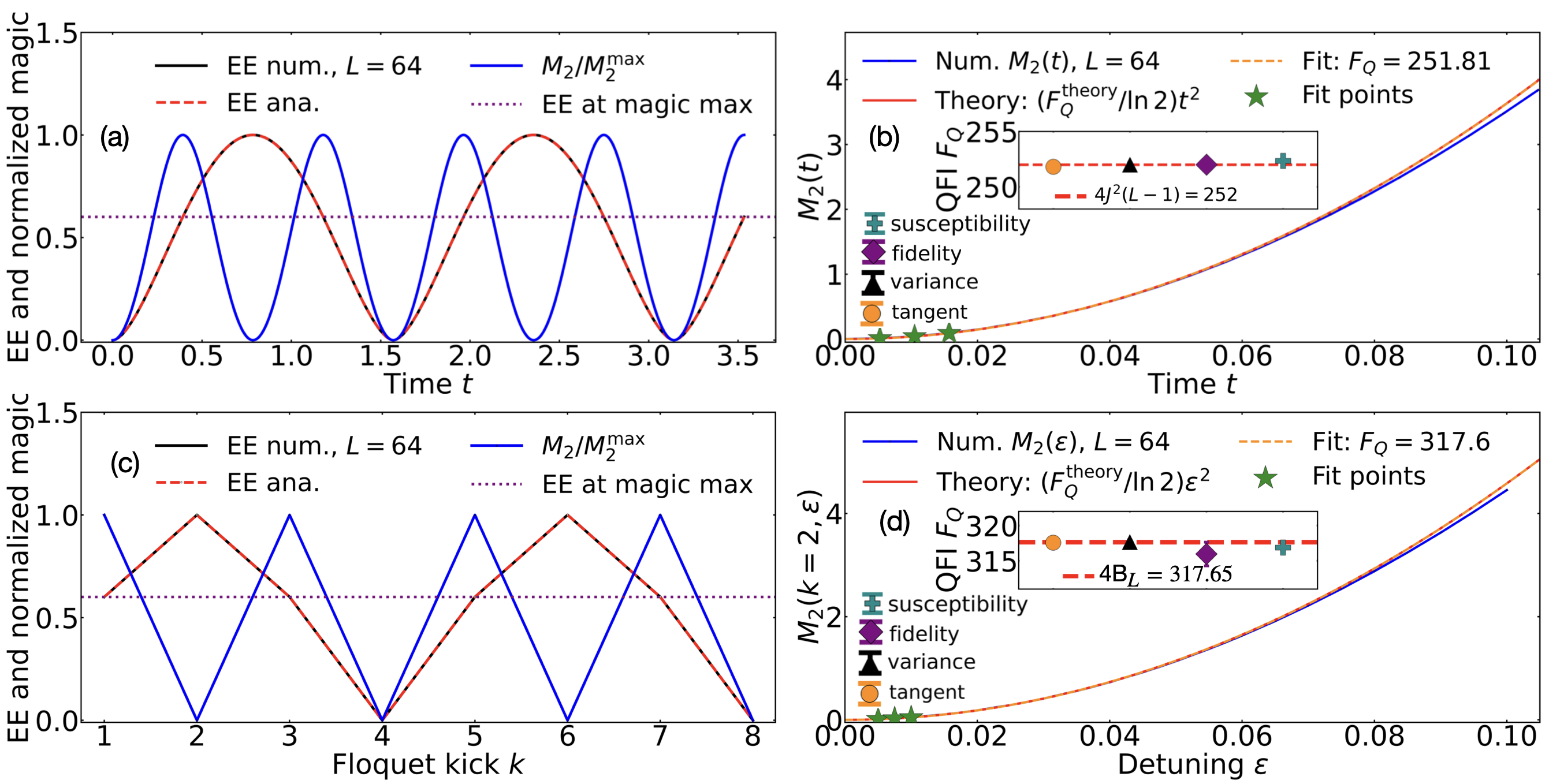}
		\caption{Magic, entanglement, and tangent curvature. Top row: open-chain quench with \(J=1\), \(h=0\), and \(L=64\). Panel (a) compares the normalized magic \(M_2/M_2^{\max}\) with the entanglement entropy, showing their different revival patterns. Panel (b) shows the initial quadratic growth of \(M_2(t)\) for small time \(t\) near the stabilizer point \(t=0\), together with the tangent-theorem prediction and independent quantum-Fisher-information benchmarks from the tangent coefficient, the variance, the fidelity curvature, and the dynamical susceptibility; here \(F_Q=4(L-1)=252\). Bottom row: stroboscopic kicked-Ising dynamics at \(g=\pi/2\), \(\theta=\pi/8\), and \(L=64\). Panel (c) compares the corresponding Floquet magic and entanglement revivals, while panel (d) shows the quadratic lifting of the two-kick revival as a function of the kick detuning \(\varepsilon=g-\pi/2\), together with the tangent-theorem prediction and the same independent benchmarks.}
		\label{fig:magic_ee_qfi}
	\end{figure*}

	\textit{One-dimensional realizations: quench and kicks.---} The open chain is the \textit{path graph} \(G_{\rm path}\), hence a connected tree with \(c(G_{\rm path})=1\). The forest theorem therefore applies directly to a standard one-dimensional spin chain. We first consider the commuting Ising quench
	\begin{equation}
		|\psi_L(t)\rangle=e^{-iJtH_{ZZ}}|+\rangle^{\otimes L},\qquad
		H_{ZZ}=\sum_{j=1}^{L-1}Z_jZ_{j+1}.
		\label{eq:ising_quench}
	\end{equation}
	The accumulated Ising angle is \(\vartheta=Jt\). To test how the exact revival is lifted by a noncommuting perturbation, we also use
	\begin{equation}
		H_{\rm TFIM}=JH_{ZZ}+hH_X,\qquad H_X=\sum_{j=1}^L X_j.
		\label{eq:tfim_hamiltonian}
	\end{equation}
	Finally, we study the kicked-Ising circuit at integer, stroboscopic time \(k\),
	\begin{equation}
		U_F(g,\theta)=e^{-i\theta H_{ZZ}}e^{-igH_X},\quad |\psi_L^{\rm F}(k)\rangle=U_F(g,\theta)^k|+\rangle^{\otimes L}.
		\label{eq:floquet_unitary}
	\end{equation}
	Here \(\theta\) is the Ising angle per kick. At \(g=\pi/2\), \(e^{-i(\pi/2)H_X}=(-i)^L\prod_jX_j\) is Clifford: this global product, not each \(X_j\), commutes with every \(Z_\ell Z_{\ell+1}\) and leaves \(|+\rangle^{\otimes L}\) invariant up to phase, so the stroboscopic state is Clifford-equivalent to the quench at angle \(k\theta\). We use \(\mu_2^{\rm Q}\) and \(\mu_2^{\rm F}\) for the quench and stroboscopic Floquet densities, respectively; \(|+\rangle^{\otimes L}\) is used throughout, with conjugated realizations given by the construction above. For \(h\neq0\) or \(g\neq\pi/2\), the dynamics is no longer reduced by the forest normal form. Although the TFIM dynamics is free-fermion solvable, \(M_2\) remains a fourth moment over the full spin-Pauli distribution, so we evaluate it via matrix-product-state Pauli-basis methods~\cite{haug2023mpsmagic,lami2023perfectsampling,tarabunga2024paulibasis}, with the Jordan-Wigner perspective, implementation, and convergence checks in the SM~\cite{SM}.
	
	\textit{Exact magic revivals and entanglement contrast.---} With \(\vartheta=Jt\) for the quench and \(\vartheta=k\theta\) for the kicked circuit at \(g=\pi/2\), the exact result is
	\begin{equation}
		M_2^{(L)}(\vartheta)=(L-1)m_2(\vartheta),\,\,
		m_2(\vartheta)=-\log_2\left[1-\frac{1}{4}\sin^2(4\vartheta)\right].
		\label{eq:one_dimensional_magic}
	\end{equation}
	Hence the thermodynamic magic densities are finite and exact,
	\begin{equation}
		\mu_2^{\rm Q}(t)=m_2(Jt),\qquad \mu_2^{\rm F}(k)=m_2(k\theta).
		\label{eq:magic_density}
	\end{equation}
	They vanish whenever \(\vartheta\in\pi\mathbb Z/4\), since every Ising gate is then Clifford, and peak at \(\log_2(4/3)\) for \(\vartheta=\pi/8\) mod \(\pi/4\) --- hence \(\theta=\pi/8\) is the natural stroboscopic choice: one kick reaches the magic maximum, two kicks give the first nontrivial Clifford revival at \(2\theta=\pi/4\). These are exact many-body revivals of the full Pauli distribution, not just of a few observables; Fig.~\ref{fig:magic_density} shows the quench curve together with the stroboscopic Floquet sequence at \(g=\pi/2\) and \(\theta=\pi/8\).

	The surprise is that the entanglement tells a different story. For any interior cut of the open chain, only the Ising gate crossing the cut can change the Schmidt rank. A two-term Schmidt decomposition, derived in the SM~\cite{SM}, gives
	\begin{equation}
		S_{\rm EE}^{(L)}(\vartheta)=h_2(\sin^2\vartheta),\,\,
		h_2(p)=-p\log_2p-(1-p)\log_2(1-p),
		\label{eq:entanglement_entropy}
	\end{equation}
	and the same expression holds stroboscopically for the kicked circuit after \(\vartheta\to k\theta\). The magic has period \(\pi/4\) in \(\vartheta\), while \(S_{\rm EE}\) has period \(\pi/2\). At odd Clifford angles, \(\vartheta=(2n+1)\pi/4\), the magic is exactly zero while \(S_{\rm EE}=1\) ebit across every cut. At \(\vartheta=\pi/8\), the magic density is maximal while \(S_{\rm EE}=h_2[(2-\sqrt2)/4]\simeq0.6009\). The thermodynamic-limit contrast is sharper: \(M_2^{(L)}(\vartheta)/L\to m_2(\vartheta)\) but \(S_{\rm EE}^{(L)}(\vartheta)/L\to0\); for generic \(\vartheta\), this is the one-dimensional realization of magic without entanglement density. For example, at \(\theta=\pi/8\), \(g=\pi/2\): two kicks return the magic to zero, four return the entanglement pattern, and eight return the full unitary up to phase, \(U_F^8\propto I\) --- so \(M_2\) repeats every two kicks while \(S_{\rm EE}\) repeats every four in one stroboscopic cycle. The detuning results below show that this period structure is exact at the Clifford point and quadratically lifted away from it.

	\begin{figure}[t]
		\centering
		\includegraphics[width=0.96\columnwidth]{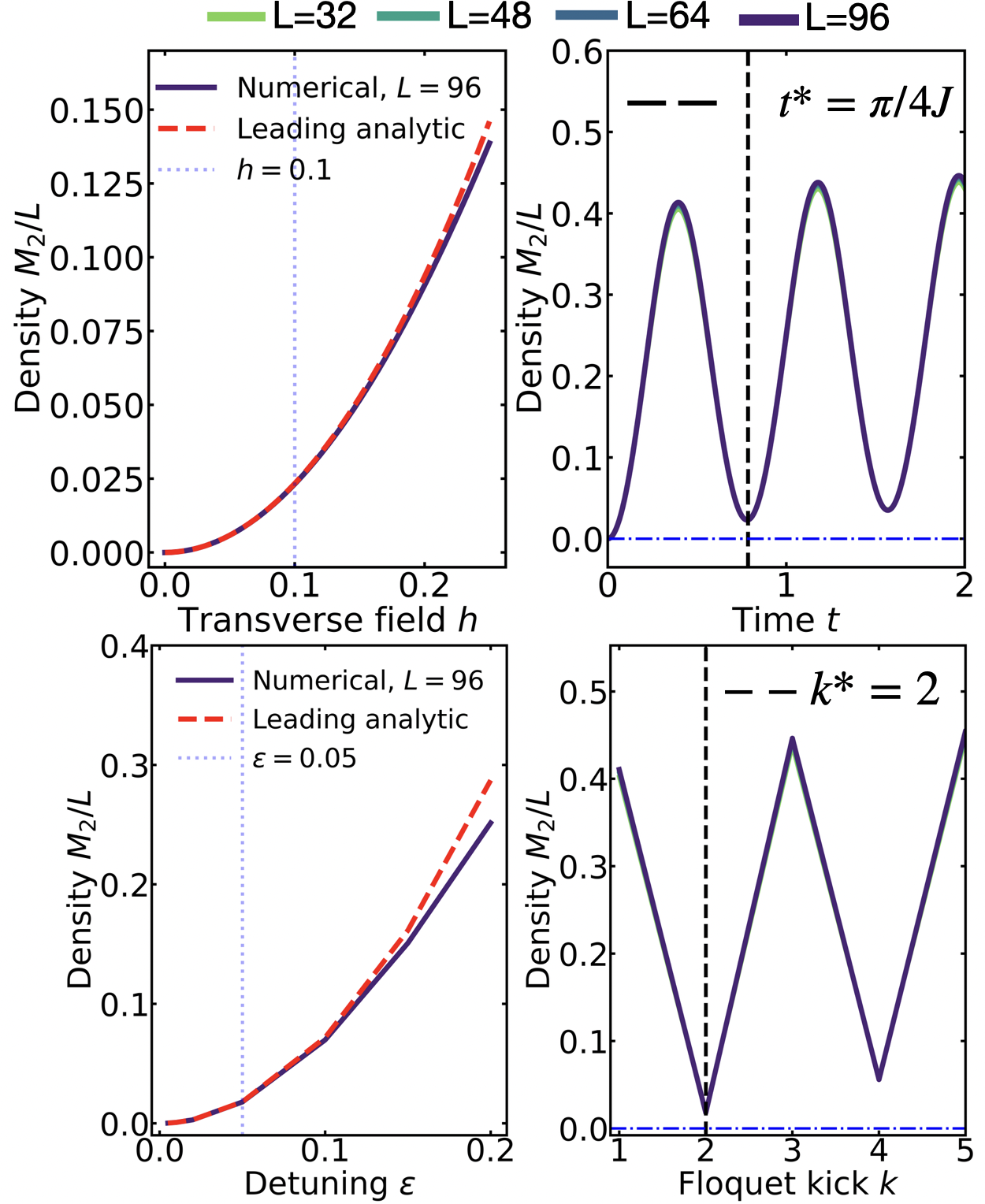}
		\caption{Quadratic lifting of exact magic revivals. Top row: quench revival at \(t_\star=\pi/(4J)\), detuned by the transverse field \(h/J\) where \(J=1\). Bottom row: two-kick Floquet revival at \(g=\pi/2\), \(\theta=\pi/8\), detuned by \(\varepsilon=g-\pi/2\). System sizes are \(L=32,48,64,96\); the leading quadratic predictions are Eqs.~\eqref{eq:static_lifting} and \eqref{eq:floquet_lifting}.}
		\label{fig:revival_lifting}
	\end{figure}
	
	\textit{Quantum Fisher curvature and revival lifting.---} The tangent theorem fixes the initial growth of magic without fitting any microscopic detail. Taylor expanding Eq.~\eqref{eq:one_dimensional_magic} near \(\vartheta=0\), with \(\vartheta=Jt\), gives the same coefficient as the tangent formula for \(h=0\). For the transverse-field Ising Hamiltonian in Eq.~\eqref{eq:tfim_hamiltonian}, the state \(|+\rangle^{\otimes L}\) is an eigenstate of \(H_X\). Thus the field has zero initial variance, and the same tangent formula gives
	\begin{equation}
		\begin{aligned}
			&M_2^{(L)}(t,h)=\frac{4J^2(L-1)}{\ln2}t^2+O(t^3),\\
			\Rightarrow &F_Q(|+\rangle^{\otimes L},H_{\rm TFIM})=4J^2(L-1).
		\end{aligned}
		\label{eq:initial_curvature}
	\end{equation}
	The same value is independently obtained from the variance, squared-fidelity curvature, and dynamical susceptibility using the generator \(K=JH_{ZZ}\) (all standard QFI diagnostics~\cite{pezze2018quantum,hauke2016measuring,Paris2011Nov}). At the initial state \(|+\rangle^{\otimes L}\), this equals \(F_Q(|+\rangle^{\otimes L},H_{\rm TFIM})\), because \(|+\rangle^{\otimes L}\) is an eigenstate of \(H_X\), so the transverse-field term contributes neither variance nor covariance. The \(4/\pi\) spectral convention is fixed in the SM~\cite{SM} and the comparison is shown in Fig.~\ref{fig:magic_ee_qfi}. The same geometry also controls the local stability and destruction of a revival. At \(h=0\), the first quench revival occurs at \(t_\star=\pi/(4J)\). The local stability analysis near this quench minimum treats \(h/J\) as the tangent detuning at the stabilizer point and gives, as derived in the SM~\cite{SM},
	\begin{equation}
		\begin{aligned}
			M_2^{(L)}(t_\star,h)
			&=\frac{4A_L}{\ln2}\left(\frac{h}{J}\right)^2+O((h/J)^3),\\
			A_L
			&=\begin{cases}
				1, & L=2,\\
				\dfrac{16L+24+(L-2)\pi^2}{64}, & L\geq3.
			\end{cases}
		\end{aligned}
		\label{eq:static_lifting}
	\end{equation}
	For the kicked circuit, choose \(\theta=\pi/8\), so a single kick reaches the maximum of \(m_2\), while two kicks accumulate \(2\theta=\pi/4\), the first nontrivial stroboscopic magic revival at \(g=\pi/2\). The local stability analysis near this Floquet minimum is obtained by detuning the kick as \(g=\pi/2+\varepsilon\), which lifts the two-kick revival as
	\begin{equation}
		\begin{aligned}
			M_{2,{\rm F}}^{(L)}(k=2)
			&=\frac{4B_L}{\ln2}\varepsilon^2+O(\varepsilon^3),\\
			B_L
			&=\begin{cases}
				2, & L=2,\\
				\dfrac{5L-8+4\sqrt2}{4}, & L\geq3.
			\end{cases}
		\end{aligned}
		\label{eq:floquet_lifting}
	\end{equation}
	Equivalently, the thermodynamic densities admit the following local two-parameter expansions around the exact revival points \((Jt,h/J)=(\pi/4,0)\) and \((\theta,g)=(\pi/8,\pi/2)\):
	\begin{equation}
		\begin{aligned}
			\mu_2^{\rm Q}(\delta,\eta)
			&=\frac{4}{\ln2}\left[\delta^2+\left(\frac{1}{4}+\frac{\pi^2}{64}\right)\eta^2\right]+o(\delta^2+\eta^2),\\
			\mu_2^{\rm F}(\delta_\theta,\varepsilon)
			&=\frac{1}{\ln2}\left(16\delta_\theta^2+5\varepsilon^2\right)+o(\delta_\theta^2+\varepsilon^2),
		\end{aligned}
		\label{eq:local_instability}
	\end{equation}
	where \(\delta=Jt-\pi/4\), \(\eta=h/J\), \(\delta_\theta=\theta-\pi/8\), and \(\varepsilon=g-\pi/2\). The mixed quadratic terms are absent because the corresponding covariances vanish, as derived in the SM~\cite{SM}. Thus the quench and Floquet revivals are isolated quadratic minima of the thermodynamic magic density; Fig.~\ref{fig:revival_lifting} compares the two liftings.

	\textit{Conclusion and outlook.---} We have established an exact framework for the generation, recurrence, and local stability of many-body magic. The tangent theorem identifies the QFI as the exact curvature of the second stabilizer R\'enyi entropy at any stabilizer state and along any Hermitian direction, thereby establishing a two-way local bridge between a quantum-computational resource and an experimentally mature metrological observable. The forest theorem evaluates the full stabilizer-R\'enyi family for commuting Ising dynamics on every acyclic interaction graph independent of spatial embedding, producing broad graph families with finite magic density and vanishing entanglement density. In the open Ising chain, these structural results yield exact quench and Floquet revivals, distinct periodicities of magic and entanglement, and a genuinely many-body magic density encoded in the complete Pauli distribution despite nonextensive bipartite entanglement.
	
	The same geometric result controls dynamics away from the exactly solvable points. It fixes the initial growth of magic, gives the finite-size coefficients governing the perturbative lifting of quench and Floquet revivals, and proves positive quadratic stability of the corresponding thermodynamic minima. Pauli-basis matrix-product-state simulations provide independent numerical verification of the analytic magic, entanglement, Fisher-information, and perturbative predictions at system sizes beyond direct Pauli enumeration~\cite{haug2023mpsmagic,lami2023perfectsampling,tarabunga2024paulibasis}. Together, these results show that magic, entanglement, and QFI furnish complementary rather than interchangeable diagnostics of quantum many-body dynamics.

	Several directions follow naturally. The forest theorem, being graph-theoretic, already applies to forests in any spatial dimension; extending it to graphs with cycles, hypergraph interactions, and qudit Pauli modules is a concrete route to higher-dimensional settings. A further extension is to subsystem-resolved, nonlocal magic, recently characterized via entanglement/spectral structure and Clifford-circuit generation~\cite{huang2026nonlocal,torre2026nonlocal,dallas2025nonlocal}. These analytic revival minima also provide a controlled reference point for when noncommuting or long-range perturbations generate nonanalytic reorganizations of magic. It will be particularly interesting to compare such behavior with temporal entanglement transitions and Page-curve dynamics, where an entanglement-Hamiltonian perspective has recently exposed sharp dynamical structures~\cite{kehrein2024page,jha2025page,gadge2026temporal,benedetti2026universality}. Experimentally, the tangent theorem is itself an explicit proposal for detecting magic: at any stabilizer reference point, QFI extracted from dynamical susceptibility measurements, spin-squeezing interferometry, or randomized measurements~\cite{hauke2016measuring,strobel2014fisher,Yu2021qfi} directly yields the leading magic curvature, without stabilizer tomography. Combined with existing protocols for stabilizer and entanglement R\'enyi entropies~\cite{oliviero2022measuring,brydges2019probing}, this makes a joint, experimentally well-defined test of magic, entanglement, and QFI across a tunable Clifford revival.

	\textit{Acknowledgment.---}
	R.J. acknowledges partial support by the U.S. Department of Energy, Office of Science, Office of Advanced Scientific Computing Research via the Exploratory Research for Extreme Scale Science (EXPRESS) program under Award Number DE-SC0026337. 
	K. G. acknowledges financial support by Deutsche Forschungsgemeinschaft (DFG, German Research Foundation) Grants No. 436382789, and No. 493420525, via large equipment grants (GOEGrid).
	

	\textit{Data Availability.---}
	The data that support the findings of this article are available upon reasonable request from the authors.
	
	\bibliography{refs}
	
\end{document}


\title{Supplemental Material --- Magic Without Entanglement: Exact Revivals and Their Fisher Information Origin}
	\author{Rishabh Jha}
	\email{rishabh.jha@usc.edu}
	\affiliation{%
		Department of Physics and Astronomy, University of Southern California, Los Angeles, CA 90089-0484, USA
	}
	\author{Karun Gadge}
	\email{karun.gadge@uni-goettingen.de}
	\affiliation{%
		Institute for Theoretical Physics, Georg-August-Universit\"{a}t G\"{o}ttingen, Friedrich-Hund-Platz 1, 37077 G\"{o}ttingen, Germany
	}
	
	\maketitle
	
	\onecolumngrid

	\tableofcontents
	
	
	\section{Conventions and stabilizer R\'enyi entropy}
	
	\subsection{Pauli strings and the Pauli probability distribution}
	
	We work with \(L\) qubits and Hilbert space \(\mathcal H_L=(\mathbb C^2)^{\otimes L}\). The single-qubit identity and Pauli matrices are
	\begin{equation}
		I=\begin{pmatrix}1&0\\0&1\end{pmatrix},\quad
		X=\begin{pmatrix}0&1\\1&0\end{pmatrix},\quad
		Y=\begin{pmatrix}0&-\mathrm{i}\\ \mathrm{i}&0\end{pmatrix},\quad
		Z=\begin{pmatrix}1&0\\0&-1\end{pmatrix}.
		\label{eq:SM_pauli_matrices}
	\end{equation}
	We write \(\sigma^0=I\), \(\sigma^1=X\), \(\sigma^2=Y\), and \(\sigma^3=Z\). The phase-free \(L\)-qubit Pauli-string set is
	\begin{equation}
		\mathcal P_L=\left\{P_a=\sigma^{a_1}\otimes\cdots\otimes\sigma^{a_L}:a_j\in\{0,1,2,3\}\right\}.
		\label{eq:SM_pauli_set}
	\end{equation}
	Thus \(|\mathcal P_L|=4^L\). Overall phases \(\pm1,\pm\mathrm{i}\) are not included in \(\mathcal P_L\), because the stabilizer R\'enyi entropies used below depend only on squared expectation values. The Pauli strings form an orthogonal operator basis:
	\begin{equation}
		\operatorname{Tr}(PQ)=2^L\delta_{P,Q},\qquad P,Q\in\mathcal P_L.
		\label{eq:SM_hilbert_schmidt}
	\end{equation}
	For a pure state \(|\psi\rangle\), define \(\rho_\psi=|\psi\rangle\langle\psi|\) and \(\langle P\rangle_\psi=\langle\psi|P|\psi\rangle\). Orthogonality gives the Pauli expansion
	\begin{equation}
		\rho_\psi=2^{-L}\sum_{P\in\mathcal P_L}\langle P\rangle_\psi P.
		\label{eq:SM_pauli_expansion}
	\end{equation}
	Since \(\operatorname{Tr}\rho_\psi^2=1\), Eq.~\eqref{eq:SM_hilbert_schmidt} implies
	\begin{equation}
		1=\operatorname{Tr}\rho_\psi^2=2^{-L}\sum_{P\in\mathcal P_L}|\langle P\rangle_\psi|^2.
		\label{eq:SM_purity_sum}
	\end{equation}
	This motivates the Pauli probability distribution
	\begin{equation}
		\Xi_P(\psi)\coloneq2^{-L}|\langle P\rangle_\psi|^2,\qquad P\in\mathcal P_L,
		\label{eq:SM_pauli_probability}
	\end{equation}
	which satisfies \(\sum_{P\in\mathcal P_L}\Xi_P(\psi)=1\).
	
	The stabilizer R\'enyi entropy is the R\'enyi entropy of this Pauli distribution, shifted so that stabilizer states have zero magic:
	\begin{equation}
		M_\alpha(\psi)=\frac{1}{1-\alpha}\log_2\sum_{P\in\mathcal P_L}\Xi_P(\psi)^\alpha-L,\qquad \alpha>0.
		\label{eq:SM_sre}
	\end{equation}
	The case \(\alpha=1\) is always understood by continuity. We restrict to \(\alpha>0\), because the Pauli distribution can have exact zeros.
	
	\subsection{Stabilizer states, Clifford invariance, and additivity}
	
A stabilizer state \(|s\rangle\) is defined as the simultaneous eigenstate of \(L\) independent commuting phase-free Pauli strings, called generators, \(G_1,\dots,G_L\). Taking all products of these generators and discarding any overall phase gives the phase-free stabilizer set
\begin{equation}
	\mathcal S(s)=\left\{S\in\mathcal P_L\;:\;S\text{ is equal, up to an overall phase, to }\prod_{j=1}^L G_j^{n_j}\text{ for some }n_j\in\{0,1\}\right\}.
\end{equation}
This set has exactly \(2^L\) elements, far fewer than the \(4^L\) strings in the full phase-free Pauli set \(\mathcal P_L\). Every \(S\in\mathcal S(s)\) satisfies
\begin{equation}
	S|s\rangle=s_S|s\rangle,\qquad s_S=\pm1,
	\label{eq:s_S}
\end{equation}
so \(|\langle s|S|s\rangle|=1\) for every stabilizer string. Equivalently, the signed operators \(s_S S\) form the usual stabilizer group of \(|s\rangle\). If instead \(P\notin\mathcal S(s)\), then \(P\) must anticommute with at least one generator. Indeed, a phase-free Pauli string that commutes with all \(L\) independent generators belongs to their maximal commuting Pauli set and therefore lies in \(\mathcal S(s)\). Let \(G\) be a generator with \(G|s\rangle=\pm|s\rangle\) and \(GP=-PG\). Then
\begin{equation}
	\langle P\rangle_s=\langle s|GPG|s\rangle=-\langle s|PGG|s\rangle=-\langle P\rangle_s\;\Longrightarrow\;\langle P\rangle_s=0,
\end{equation}
where \(G^2=I\) was used. Thus a stabilizer state has exactly \(2^L\) phase-free Pauli strings with unit squared expectation, namely those in \(\mathcal S(s)\), while every other phase-free Pauli string has zero expectation. Substituting this two-valued distribution into the definition of the stabilizer Renyi entropy \(M_\alpha\) gives \(M_\alpha(s)=0\) for all \(\alpha>0\), which is the defining property of a stabilizer state with zero magic.
	
	A Clifford unitary \(C\) maps Pauli strings to Pauli strings under conjugation:
	\begin{equation}
		C^\dagger P C=\pm Q,\qquad P,Q\in\mathcal P_L.
		\label{eq:SM_clifford_map}
	\end{equation}
	The sign is irrelevant in \(\Xi_P\). Thus
	\begin{equation}
		\Xi_P(C|\psi\rangle)=\Xi_Q(\psi),
		\label{eq:SM_clifford_probability}
	\end{equation}
	so Clifford conjugation only permutes the Pauli probability distribution. Hence
	\begin{equation}
		M_\alpha(C|\psi\rangle)=M_\alpha(\psi).
		\label{eq:SM_clifford_invariance}
	\end{equation}
	
	Additivity follows from factorization. If \(|\psi\rangle=|\psi_A\rangle\otimes|\psi_B\rangle\), \(L=L_A+L_B\), and \(P=P_A\otimes P_B\), then
	\begin{equation}
		\Xi_{P_A\otimes P_B}(\psi_A\otimes\psi_B)=\Xi_{P_A}(\psi_A)\Xi_{P_B}(\psi_B).
		\label{eq:SM_probability_factorization}
	\end{equation}
	The R\'enyi entropy of a product probability distribution is additive, and the subtraction by \(L=L_A+L_B\) is also additive. Therefore
	\begin{equation}
		M_\alpha(\psi_A\otimes\psi_B)=M_\alpha(\psi_A)+M_\alpha(\psi_B).
		\label{eq:SM_additivity}
	\end{equation}
	
	\subsection{The \(M_2\) fourth moment}
	
	For \(\alpha=2\), Eq.~\eqref{eq:SM_sre} becomes
	\begin{equation}
		M_2(\psi)=-\log_2\left[2^{-L}\sum_{P\in\mathcal P_L}\langle P\rangle_\psi^4\right].
		\label{eq:SM_m2_fourth_moment}
	\end{equation}
	Here the absolute value can be dropped because every phase-free Pauli string is Hermitian, so \(\langle P\rangle_\psi\) is real. It is useful to define
	\begin{equation}
		Q(\psi)\coloneq2^{-L}\sum_{P\in\mathcal P_L}\langle P\rangle_\psi^4,
		\label{eq:SM_Q_definition}
	\end{equation}
	so that \(M_2(\psi)=-\log_2 Q(\psi)\). The tangent formula below is obtained by expanding this Pauli fourth moment near a stabilizer state.
	
	
	\section{Tangent geometry of magic}

	\subsection{Projective tangent vectors and quantum Fisher information}
	
	Let
	\begin{equation}
		|\psi(\theta)\rangle=e^{-\mathrm{i}\theta K}|s\rangle,
		\label{eq:SM_tangent_path}
	\end{equation}
	where \(|s\rangle\) is a stabilizer state and \(K=K^\dagger\). We use
	\begin{equation}
		\langle A\rangle_s\coloneq\langle s|A|s\rangle.
		\label{eq:SM_expectation_s}
	\end{equation}
	Since \(K\) is Hermitian, \(e^{-\mathrm{i}\theta K}\) is unitary, so the family \(|\psi(\theta)\rangle\) is a normalized pure-state path through \(|s\rangle\).
	
	Differentiating Eq.~\eqref{eq:SM_tangent_path} gives
	\begin{equation}
		\frac{d}{d\theta}|\psi(\theta)\rangle
		=
		-\mathrm{i}K e^{-\mathrm{i}\theta K}|s\rangle.
	\end{equation}
	Evaluating at \(\theta=0\) yields the raw tangent vector
	\begin{equation}
		\left.\frac{d}{d\theta}|\psi(\theta)\rangle\right|_{\theta=0}
		=
		-\mathrm{i}K|s\rangle.
		\label{eq:SM_raw_tangent}
	\end{equation}
	
	The vector \(K|s\rangle\) naturally decomposes into a component parallel to \(|s\rangle\) and a component orthogonal to \(|s\rangle\):
	\begin{equation}
		K|s\rangle
		=
		\langle K\rangle_s |s\rangle
		+
		\bigl(K-\langle K\rangle_s\bigr)|s\rangle.
	\end{equation}
	The first term is parallel to \(|s\rangle\). After multiplication by \(-\mathrm{i}\), it generates only an infinitesimal change of overall phase and therefore does not change the physical ray. The physically relevant first-order change is the orthogonal component,
	\begin{equation}
		|\dot\psi_\perp\rangle
		=
		-\mathrm{i}\bigl(K-\langle K\rangle_s\bigr)|s\rangle.
		\label{eq:SM_projective_tangent}
	\end{equation}
	Indeed,
	\begin{equation}
		\langle s|\dot\psi_\perp\rangle
		=
		-\mathrm{i}\bigl(\langle K\rangle_s-\langle K\rangle_s\bigr)
		=
		0,
	\end{equation}
	so \( |\dot\psi_\perp\rangle \) is orthogonal to \(|s\rangle\), as required for a projective tangent vector.
	
	Its squared norm is
	\begin{align}
		\langle\dot\psi_\perp|\dot\psi_\perp\rangle
		&=
		\langle s|
		\bigl(K-\langle K\rangle_s\bigr)^2
		|s\rangle \\
		&=
		\langle s|
		\left(
		K^2-2\langle K\rangle_s K+\langle K\rangle_s^2
		\right)
		|s\rangle \\
		&=
		\langle K^2\rangle_s
		-
		2\langle K\rangle_s^2
		+
		\langle K\rangle_s^2 \\
		&=
		\langle K^2\rangle_s-\langle K\rangle_s^2.
	\end{align}
	Therefore
	\begin{equation}
		\langle\dot\psi_\perp|\dot\psi_\perp\rangle
		=
		\langle K^2\rangle_s-\langle K\rangle_s^2
		\equiv
		\operatorname{Var}_s(K).
		\label{eq:SM_variance_metric}
	\end{equation}
	Thus the variance of \(K\) is precisely the squared norm of the projective tangent vector.
	
	For a general pure-state family \(|\psi(\theta)\rangle\), the quantum Fisher information may be written as
	\begin{equation}
		F_Q
		=
		4\left(
		\langle \partial_\theta \psi|\partial_\theta \psi\rangle
		-
		|\langle \psi|\partial_\theta \psi\rangle|^2
		\right),
	\end{equation}
	which is four times the Fubini--Study metric along the path~\cite{Paris2011Nov}. For the unitary family in Eq.~\eqref{eq:SM_tangent_path}, evaluated at
	\(\theta=0\), one has
	\begin{equation}
		|\partial_\theta \psi\rangle\big|_{\theta=0}
		=
		-\mathrm{i}K|s\rangle,
	\end{equation}
	and hence
	\begin{align}
		\langle \partial_\theta \psi|\partial_\theta \psi\rangle\big|_{\theta=0}
		&=
		\langle K^2\rangle_s, \\
		|\langle \psi|\partial_\theta \psi\rangle|^2\big|_{\theta=0}
		&=
		|-\mathrm{i}\langle K\rangle_s|^2
		=
		\langle K\rangle_s^2.
	\end{align}
	Substituting these expressions gives
	\begin{equation}
		F_Q(|s\rangle,K)
		=
		4\operatorname{Var}_s(K)
		=
		4\left(\langle K^2\rangle_s-\langle K\rangle_s^2\right).
		\label{eq:SM_QFI_definition}
	\end{equation}
	Equivalently, \(F_Q\) is four times the squared norm of the projective tangent vector in Eq.~\eqref{eq:SM_projective_tangent}. This is the standard pure-state convention for quantum Fisher information~\cite{Paris2011Nov}.

	\subsection{Proof of the tangent formula}
	
	\begin{theorem}[Tangent formula at a stabilizer point]
		Let \(|s\rangle\) be a stabilizer state and let \(K=K^\dagger\). Along the path \(|\psi(\theta)\rangle=e^{-\mathrm{i}\theta K}|s\rangle\),
		\begin{equation}
			\boxed{ M_2(\psi(\theta))=\frac{\theta^2}{\ln2}F_Q(|s\rangle,K)+O(\theta^3)}.
			\label{eq:SM_tangent_formula}
		\end{equation}
	\end{theorem}
	
\begin{proof}
Let \(|\psi(\theta)\rangle=e^{-\mathrm{i}\theta K}|s\rangle\), in agreement with the path defined in Eq.~\eqref{eq:SM_tangent_path}. For any Pauli string \(P\in\mathcal P_L\), define its time-dependent expectation value by
\begin{equation}
	r_P(\theta)\coloneq\langle\psi(\theta)|P|\psi(\theta)\rangle=\langle s|e^{\mathrm{i}\theta K}Pe^{-\mathrm{i}\theta K}|s\rangle.
	\label{eq:SM_rP_definition}
\end{equation}
This is the Heisenberg-picture expectation value of \(P\) along the unitary path generated by \(K\). The quantity \(Q\) defined in Eq.~\eqref{eq:SM_Q_definition} is the Pauli fourth moment that determines the second stabilizer Renyi entropy through Eq.~\eqref{eq:SM_m2_fourth_moment}. Evaluating that definition on \(|\psi(\theta)\rangle\) gives
\begin{equation}
	Q(\theta)\coloneq Q(\psi(\theta))=2^{-L}\sum_{P\in\mathcal P_L}r_P(\theta)^4,
	\label{eq:SM_Qtheta}
\end{equation}
and therefore
\begin{equation}
	M_2(\psi(\theta))=-\log_2 Q(\theta).
\end{equation}
We now expand this fourth moment to second order in \(\theta\).
	
	\textbf{Step 1:} isolating the surviving terms. At \(\theta=0\), \(r_P(0)=\langle s|P|s\rangle\). As shown above in Eq.~\eqref{eq:s_S}, this equals \(s_S=\pm1\) if \(P=S\in\mathcal S(s)\), and equals \(0\) otherwise. So for any \(P\notin\mathcal S(s)\), the function \(r_P(\theta)\) starts at zero, meaning \(r_P(\theta)=O(\theta)\) as \(\theta\to0\), hence \(r_P(\theta)^4=O(\theta^4)\). These terms cannot affect \(Q(\theta)\) below order \(\theta^4\), so up to and including order \(\theta^2\), the sum in Eq.~\eqref{eq:SM_Qtheta} only needs the \(2^L\) stabilizer strings \(S\in\mathcal S(s)\), each of which starts at \(r_S(0)=s_S=\pm1\) rather than zero.
	
	\textbf{Step 2:} Taylor expanding \(r_S(\theta)\) for \(S\in\mathcal S(s)\). Differentiating Eq.~\eqref{eq:SM_rP_definition} once with respect to \(\theta\) gives the standard Heisenberg equation of motion,
	\begin{equation}
		r_S'(\theta)=\mathrm{i}\langle s|e^{\mathrm{i}\theta K}[K,S]e^{-\mathrm{i}\theta K}|s\rangle,
	\end{equation}
	so that at \(\theta=0\),
	\begin{equation}
		r_S'(0)=\mathrm{i}\langle s|[K,S]|s\rangle=\mathrm{i}\left(\langle s|KS|s\rangle-\langle s|SK|s\rangle\right).
		\label{eq:SM_stabilizer_first_derivative_raw}
	\end{equation}
	Since \(S|s\rangle=s_S|s\rangle\) and \(S\) is Hermitian so \(\langle s|S=s_S\langle s|\), both terms in the parenthesis equal \(s_S\langle s|K|s\rangle=s_S\langle K\rangle_s\), so they cancel exactly and
	\begin{equation}
		r_S'(0)=0.
		\label{eq:SM_stabilizer_first_derivative}
	\end{equation}
	Differentiating once more and setting \(\theta=0\), using the same Heisenberg equation of motion applied to \([K,S]\) in place of \(S\), gives the second derivative as a nested, or double, commutator,
	\begin{equation}
		r_S''(0)=-\langle s|[K,[K,S]]|s\rangle.
		\label{eq:SM_second_derivative}
	\end{equation}
	Since \(r_S(0)=s_S\) and \(r_S'(0)=0\), the Taylor expansion of \(r_S(\theta)\) to second order is
	\begin{equation}
		r_S(\theta)=s_S+\tfrac{1}{2}r_S''(0)\,\theta^2+O(\theta^3).
	\end{equation}
	Raising this to the fourth power and using \(s_S^2=1\) and \(s_S^4=1\), only the cross term between the zeroth and second order pieces survives to order \(\theta^2\),
	\begin{equation}
		r_S(\theta)^4=\left(s_S+\tfrac{1}{2}r_S''(0)\theta^2\right)^4+O(\theta^3)=1+2\,s_S\,r_S''(0)\,\theta^2+O(\theta^3).
		\label{eq:SM_stabilizer_fourth_expansion}
	\end{equation}
	
\textbf{Step 3:} summing over the stabilizer group using the projector identity. We now need \(\sum_{S\in\mathcal S(s)}s_S\,r_S''(0)\). The key algebraic tool is the stabilizer projector identity,\footnote{For completeness, let \(G_1,\ldots,G_L\) be independent commuting Pauli-string generators of \(\lvert s\rangle\), with \(G_j\lvert s\rangle=s_j\lvert s\rangle\) and \(s_j=\pm1\). Since \(G_j^2=I\), the operator \(\Pi_j=(I+s_jG_j)/2\) projects onto the eigenspace of \(G_j\) with eigenvalue \(s_j\): it gives \(\Pi_j\lvert s\rangle=\lvert s\rangle\), while it annihilates any eigenvector with eigenvalue \(-s_j\). Because all generators commute, the product of these projectors selects their common eigenspace, which is one-dimensional because the generators are independent. Hence \(\lvert s\rangle\langle s\rvert=\prod_{j=1}^{L}(I+s_jG_j)/2\). Expanding the product gives \(2^{-L}\sum_{n_1,\ldots,n_L\in\{0,1\}}\prod_{j=1}^{L}(s_jG_j)^{n_j}\). For each binary choice \((n_1,\ldots,n_L)\), let \(S\) be the phase-free Pauli string proportional to \(\prod_jG_j^{n_j}\), and let \(s_S\) be its eigenvalue on \(\lvert s\rangle\). Then \(\prod_j(s_jG_j)^{n_j}=s_SS\). The \(2^L\) possible products are precisely the elements of \(\mathcal S(s)\), so the expansion becomes \(\lvert s\rangle\langle s\rvert=2^{-L}\sum_{S\in\mathcal S(s)}s_SS\).}
	\begin{equation}
		|s\rangle\langle s|=2^{-L}\sum_{S\in\mathcal S(s)}s_S\,S.
		\label{eq:SM_stabilizer_projector}
	\end{equation}
	This identity holds because each generator \(G_j\) has a projector \((I+s_j G_j)/2\) onto its \(s_j\) eigenspace, where \(s_j=\pm1\) is the eigenvalue of \(G_j\) on \(|s\rangle\); multiplying all \(L\) such projectors together gives the projector onto the unique simultaneous eigenspace, which is the one-dimensional space spanned by \(|s\rangle\) itself, and expanding the product reproduces exactly the sum on the right-hand side of Eq.~\eqref{eq:SM_stabilizer_projector}. Substituting the double commutator from Eq.~\eqref{eq:SM_second_derivative} and expanding it as \([K,[K,S]]=K^2S-2KSK+SK^2\) gives
	\begin{equation}
		\sum_{S\in\mathcal S(s)}s_S\,r_S''(0)=-\sum_{S\in\mathcal S(s)}s_S\,\langle s|\left(K^2S-2KSK+SK^2\right)|s\rangle.
		\label{eq:SM_second_derivative_sum_start}
	\end{equation}
	For the first term, pull \(K^2\) outside the sum and use Eq.~\eqref{eq:SM_stabilizer_projector} on what remains,
	\begin{equation}
		\sum_{S\in\mathcal S(s)}s_S\,\langle s|K^2S|s\rangle=\langle s|K^2\left(\sum_{S}s_S S\right)|s\rangle=2^L\langle s|K^2|s\rangle\langle s|s\rangle=2^L\langle K^2\rangle_s,
	\end{equation}
	where we used \(\sum_S s_S S=2^L|s\rangle\langle s|\) so that acting on \(|s\rangle\) simply returns \(2^L|s\rangle\), and the third term \(SK^2\) gives the identical result \(2^L\langle K^2\rangle_s\) by the mirror argument acting on the bra \(\langle s|\). For the middle term, the sum sits between two factors of \(K\), so it converts directly into a projector sandwiched by \(K\) on both sides,
	\begin{equation}
		\sum_{S\in\mathcal S(s)}s_S\,\langle s|KSK|s\rangle=\langle s|K\left(\sum_{S\in\mathcal S(s)}s_S S\right)K|s\rangle=2^L\langle s|K|s\rangle\langle s|K|s\rangle=2^L\langle K\rangle_s^2,
		\label{eq:SM_middle_term}
	\end{equation}
	where the projector \(|s\rangle\langle s|\) sitting between the two \(K\) factors splits the matrix element into two separate expectation values of \(K\), each equal to \(\langle K\rangle_s\). Combining the first, middle, and third terms with their respective signs and prefactors from Eq.~\eqref{eq:SM_second_derivative_sum_start} gives
	\begin{equation}
		\sum_{S\in\mathcal S(s)}s_S\,r_S''(0)=-\left(2^L\langle K^2\rangle_s-2\cdot2^L\langle K\rangle_s^2+2^L\langle K^2\rangle_s\right)=-2^{L+1}\left(\langle K^2\rangle_s-\langle K\rangle_s^2\right).
		\label{eq:SM_second_derivative_sum}
	\end{equation}
	The combination \(\langle K^2\rangle_s-\langle K\rangle_s^2\) is by definition the variance \(\operatorname{Var}_s(K)\).
	
	\textbf{Step 4:} assembling \(Q(\theta)\). Insert Eq.~\eqref{eq:SM_stabilizer_fourth_expansion} into Eq.~\eqref{eq:SM_Qtheta}, restricted to the \(2^L\) stabilizer strings as justified in Step 1,
	\begin{equation}
		Q(\theta)=2^{-L}\sum_{S\in\mathcal S(s)}\left(1+2\,s_S\,r_S''(0)\,\theta^2\right)+O(\theta^3)
		=2^{-L}\left(2^L+2\theta^2\sum_{S\in\mathcal S(s)}s_S\,r_S''(0)\right)+O(\theta^3).
	\end{equation}
	Substituting Eq.~\eqref{eq:SM_second_derivative_sum} for the sum gives
	\begin{equation}
		Q(\theta)=1+2^{-L}\cdot2\theta^2\cdot\left(-2^{L+1}\operatorname{Var}_s(K)\right)+O(\theta^3)=1-4\theta^2\operatorname{Var}_s(K)+O(\theta^3).
		\label{eq:SM_Q_expansion}
	\end{equation}
	
	\textbf{Step 5:} converting to \(M_2\) and to the quantum Fisher information. Using the small-\(x\) expansion \(-\log_2(1-x)=x/\ln2+O(x^2)\) with \(x=4\theta^2\operatorname{Var}_s(K)\),
	\begin{equation}
		M_2(\psi(\theta))=-\log_2 Q(\theta)=\frac{4\theta^2}{\ln2}\operatorname{Var}_s(K)+O(\theta^3).
		\label{eq:SM_tangent_final_step}
	\end{equation}
	Recall that the quantum Fisher information of the path \(|\psi(\theta)\rangle=e^{-\mathrm{i}\theta K}|s\rangle\) at \(\theta=0\) is \(F_Q(|s\rangle,K)=4\operatorname{Var}_s(K)\), so Eq.~\eqref{eq:SM_tangent_final_step} is equivalently
	\begin{equation}
		M_2(\psi(\theta))=\frac{\theta^2}{\ln2}F_Q(|s\rangle,K)+O(\theta^3),
	\end{equation}
	which is the tangent formula. 
\end{proof}
	
	The proof also explains why the coefficient is universal at stabilizer points as can be seen directly from Step 1. The quadratic term comes entirely from the \(2^L\) stabilizer Pauli strings whose expectation values are initially \(\pm1\), while every Pauli string outside the stabilizer group has zero initial expectation and can only enter the fourth moment at order \(\theta^4\) or higher.
	
	\subsection{Independent QFI benchmarks}
	\label{subsec:SM_independent_QFI_benchmarks}
	
	The QFI benchmark in the main text is evaluated independently of the Pauli fourth moment. We use the same stabilizer state and Ising generator,
	\begin{equation}
		|s\rangle=|+\rangle^{\otimes L},\qquad
		K=JH_{ZZ}=J\sum_{j=1}^{L-1}Z_jZ_{j+1}.
		\label{eq:SM_QFI_benchmark_generator}
	\end{equation}
	For the transverse-field Ising generator
	\begin{equation}
		H_{\rm TFIM}=JH_{ZZ}+hH_X,\qquad H_X=\sum_{j=1}^{L}X_j,
		\label{eq:SM_TFIM_generator}
	\end{equation}
	the same initial QFI is obtained, because \(|+\rangle^{\otimes L}\) is an eigenstate of \(H_X\). Thus the transverse field contributes no initial variance and no initial covariance.
	
	The first check is the direct variance. Since \(|s\rangle=|+\rangle^{\otimes L}\) is a product state and \(\langle +|Z|+\rangle=0\), one has
	\begin{equation}
		\langle Z_jZ_{j+1}\rangle_s
		=
		\langle +|Z|+\rangle\,\langle +|Z|+\rangle
		=
		0.
	\end{equation}
	Moreover,
	\begin{equation}
		(Z_jZ_{j+1})^2=Z_j^2Z_{j+1}^2=I,
	\end{equation}
	and therefore
	\begin{equation}
		\langle Z_jZ_{j+1}\rangle_s=0,\qquad
		\langle (Z_jZ_{j+1})^2\rangle_s=1.
		\label{eq:SM_QFI_bond_rules}
	\end{equation}
	Now write
	\begin{equation}
		K=J\sum_{j=1}^{L-1} B_j,\qquad B_j\coloneq Z_jZ_{j+1}.
	\end{equation}
	Then
	\begin{equation}
		\langle K\rangle_s
		=
		J\sum_{j=1}^{L-1}\langle B_j\rangle_s
		=
		0.
	\end{equation}
	For \(\langle K^2\rangle_s\), we have
	\begin{equation}
		K^2
		=
		J^2\sum_{j=1}^{L-1} B_j^2
		+
		J^2\sum_{j\neq k} B_jB_k.
	\end{equation}
	The diagonal terms give \(B_j^2=I\), hence \(\langle B_j^2\rangle_s=1\). For \(j\neq k\), the cross terms vanish. If the two bonds are disjoint, the expectation value factorizes and is zero because each bond has zero expectation. If they are adjacent, then
	\begin{equation}
		B_jB_{j+1}
		=
		(Z_jZ_{j+1})(Z_{j+1}Z_{j+2})
		=
		Z_jZ_{j+2},
	\end{equation}
	whose expectation also vanishes in \(|+\rangle^{\otimes L}\), because \(\langle +|Z|+\rangle=0\). Thus \(\langle B_jB_k\rangle_s=0\) for all \(j\neq k\), and
	\begin{equation}
		\langle K^2\rangle_s
		=
		J^2\sum_{j=1}^{L-1}\langle B_j^2\rangle_s
		=
		J^2(L-1).
	\end{equation}
	Hence
	\begin{equation}
		\boxed{	F_Q^{\rm var}=4\left(\langle K^2\rangle_s-\langle K\rangle_s^2\right)=4J^2(L-1) }.
		\label{eq:SM_QFI_variance_check}
	\end{equation}
	
	The second check uses the wave-function fidelity. Along the path \(|\psi(t)\rangle=e^{-\mathrm{i}tK}|s\rangle\),
	\begin{equation}
		F(t,t+\epsilon)=|\langle\psi(t)|\psi(t+\epsilon)\rangle|^2
		=|\langle s|e^{-\mathrm{i}\epsilon K}|s\rangle|^2.
		\label{eq:SM_fidelity_definition}
	\end{equation}
	Expanding the exponential gives
	\begin{equation}
		\langle s|e^{-\mathrm{i}\epsilon K}|s\rangle
		=1-\mathrm{i}\epsilon\langle K\rangle_s-\frac{\epsilon^2}{2}\langle K^2\rangle_s+O(\epsilon^3),
		\label{eq:SM_fidelity_amplitude_expansion}
	\end{equation}
	and therefore
	\begin{equation}
		F(t,t+\epsilon)
		=1-\epsilon^2\left(\langle K^2\rangle_s-\langle K\rangle_s^2\right)+O(\epsilon^3).
		\label{eq:SM_squared_fidelity_expansion}
	\end{equation}
	Thus
	\begin{equation}
		\boxed{
			F_Q^{\rm fid}=\lim_{\epsilon\to0}\frac{4\left[1-F(t,t+\epsilon)\right]}{\epsilon^2}.
		}
		\label{eq:SM_QFI_fidelity_check}
	\end{equation}
	This formula uses the squared fidelity.

	The third check uses the zero-temperature dynamical susceptibility. The reference Hamiltonian is
	\begin{equation}
		H_{\rm ref}=-\sum_j X_j,
		\label{eq:SM_reference_hamiltonian}
	\end{equation}
	whose ground state is \(|0\rangle=|+\rangle^{\otimes L}\). We probe this state with \(K=JH_{ZZ}\). Let \(\{|n\rangle\}\) be a complete set of eigenstates of \(H_{\rm ref}\),
	\begin{equation}
		H_{\rm ref}|n\rangle=E_n|n\rangle.
	\end{equation}
	Inserting completeness into \(\langle 0|K^2|0\rangle\) gives
	\begin{equation}
		\langle 0|K^2|0\rangle
		=
		\sum_n \langle 0|K|n\rangle\langle n|K|0\rangle
		=
		\sum_n |\langle 0|K|n\rangle|^2.
	\end{equation}
	Subtracting the \(n=0\) contribution, one obtains
	\begin{equation}
		\operatorname{Var}_0(K)
		=
		\langle K^2\rangle_0-\langle K\rangle_0^2
		=
		\sum_{n\neq0} |\langle 0|K|n\rangle|^2.
	\end{equation}
	With the spectral convention
	\begin{equation}
		\chi_{KK}''(\omega)=\pi\sum_{n\neq0}|\langle0|K|n\rangle|^2\delta\!\left[\omega-(E_n-E_0)\right],
		\label{eq:SM_susceptibility_convention}
	\end{equation}
	the positive-frequency spectral weight is manifestly nonnegative term by term, since each contribution is a modulus squared times a delta function. Integrating over \(\omega>0\) yields
	\begin{equation}
		\int_0^\infty d\omega\,\chi_{KK}''(\omega)
		=
		\pi\sum_{n\neq0}|\langle0|K|n\rangle|^2
		=
		\pi\,\operatorname{Var}_0(K).
	\end{equation}
	For the pure ground state \(|0\rangle\), the quantum Fisher information is \(F_Q=4\,\operatorname{Var}_0(K)\). Therefore
	\begin{equation}
		\boxed{	F_Q^{\rm susc}=\frac{4}{\pi}\int_0^\infty d\omega\,\chi_{KK}''(\omega) } .
		\label{eq:SM_QFI_susceptibility_check}
	\end{equation}
	Equation~\eqref{eq:SM_QFI_susceptibility_check} is the zero-temperature specialization, in the susceptibility convention of Eq.~\eqref{eq:SM_susceptibility_convention}, of the standard QFI--response connection discussed in Refs.~\cite{hauke2016measuring,Paris2011Nov}. Some implementations define the retarded response with the opposite overall sign, which makes the imaginary part negative at positive frequency. In that convention one should use \(-\chi_{KK}''\) in Eq.~\eqref{eq:SM_QFI_susceptibility_check}, equivalently integrate the positive spectral weight.
	
	The three estimates in Eqs.~\eqref{eq:SM_QFI_variance_check}, \eqref{eq:SM_QFI_fidelity_check}, and \eqref{eq:SM_QFI_susceptibility_check} should match the tangent-theorem extraction. Writing \(M_2^{(L)}(t,h)\) for the second stabilizer Renyi entropy of the \(L\)-site state evolved under \(H_{\rm TFIM}\), Eq.~\eqref{eq:SM_tangent_formula} gives
	\begin{equation}
		\boxed{
			F_Q^{\rm magic}=(\ln2)\cdot\lim_{t\to0}\frac{M_2^{(L)}(t,h)}{t^2}.
		}
		\label{eq:SM_QFI_magic_extraction}
	\end{equation}
	For the initial TFIM curvature, all four routes give \(F_Q=4J^2(L-1)\).
	
	\section{The one-qubit building block}
	\label{sec:one_qubit}
	
	\subsection{Pauli expectations of \(e^{-i\theta Z}|+\rangle\)}
	
	The elementary state that appears after the forest reduction is
	\begin{equation}
		|\phi_\theta\rangle=e^{-i\theta Z}|+\rangle.
		\label{eq:SM_one_qubit_state}
	\end{equation}
	Since
	\begin{equation}
		e^{-i\theta Z}=
		\begin{pmatrix}
			e^{-i\theta} & 0\\
			0 & e^{i\theta}
		\end{pmatrix},
		\qquad
		|+\rangle=\frac{|0\rangle+|1\rangle}{\sqrt2},
		\label{eq:SM_single_qubit_rotation}
	\end{equation}
	we have
	\begin{equation}
		|\phi_\theta\rangle=\frac{e^{-i\theta}|0\rangle+e^{i\theta}|1\rangle}{\sqrt2}.
		\label{eq:SM_phi_explicit}
	\end{equation}
	The Pauli expectations are
	\begin{equation}
		\langle I\rangle_{\phi_\theta}=1,\qquad
		\langle X\rangle_{\phi_\theta}=\cos(2\theta),\qquad
		\langle Y\rangle_{\phi_\theta}=\sin(2\theta),\qquad
		\langle Z\rangle_{\phi_\theta}=0.
		\label{eq:SM_one_qubit_expectations}
	\end{equation}
	For example,
	\begin{equation}
		\langle X\rangle_{\phi_\theta}
		=\frac{1}{2}\left(e^{2i\theta}+e^{-2i\theta}\right)
		=\cos(2\theta),
		\label{eq:SM_X_expectation}
	\end{equation}
	and
	\begin{equation}
		\langle Y\rangle_{\phi_\theta}
		=\frac{1}{2}\left(-ie^{2i\theta}+ie^{-2i\theta}\right)
		=\sin(2\theta).
		\label{eq:SM_Y_expectation}
	\end{equation}
	Thus the one-qubit Pauli probability distribution is
	\begin{equation}
		\Xi_I=\frac{1}{2},\qquad
		\Xi_X=\frac{1}{2}\cos^2(2\theta),\qquad
		\Xi_Y=\frac{1}{2}\sin^2(2\theta),\qquad
		\Xi_Z=0.
		\label{eq:SM_one_qubit_distribution}
	\end{equation}
	
	\subsection{Derivation of \(m_\alpha(\theta)\)}
	
	By definition,
	\begin{equation}
		m_\alpha(\theta)\coloneq M_\alpha(e^{-i\theta Z}|+\rangle).
		\label{eq:SM_malpha_definition}
	\end{equation}
	For one qubit, the stabilizer R\'enyi entropy is the R\'enyi entropy of the four probabilities in Eq.~\eqref{eq:SM_one_qubit_distribution}, minus one. For \(\alpha>0\) and \(\alpha\neq1\),
	\begin{equation}
		\begin{aligned}
			H_\alpha(\Xi_{\phi_\theta})
			&=\frac{1}{1-\alpha}\log_2
			\left[
			\left(\frac{1}{2}\right)^\alpha
			+\left(\frac{\cos^2(2\theta)}{2}\right)^\alpha
			+\left(\frac{\sin^2(2\theta)}{2}\right)^\alpha
			\right]\\
			&=\frac{1}{1-\alpha}\log_2
			\left[
			2^{-\alpha}\left(1+\cos^{2\alpha}(2\theta)+\sin^{2\alpha}(2\theta)\right)
			\right].
		\end{aligned}
		\label{eq:SM_one_qubit_renyi}
	\end{equation}
	Here the \(Z\)-probability gives no contribution because \(0^\alpha=0\) for \(\alpha>0\). Subtracting the one-qubit normalization gives
	\begin{equation}
		\boxed{	m_\alpha(\theta)
			=H_\alpha(\Xi_{\phi_\theta})-1=\frac{\log_2\left[1+\cos^{2\alpha}(2\theta)+\sin^{2\alpha}(2\theta)\right]-1}{1-\alpha},
			\qquad \alpha>0,\quad \alpha \neq1 }.
		\label{eq:SM_one_qubit_magic}
	\end{equation}
	This is the one-qubit function used in the forest theorem.
	
	\subsection{The limits \(\alpha=1\) and \(\alpha=2\)}
	
	The Shannon limit is obtained by taking \(\alpha\to1\). Since the probability distribution is Eq.~\eqref{eq:SM_one_qubit_distribution},
	\begin{equation}
		\boxed{	m_1(\theta)
			=-\frac{1}{2}\left[
			\cos^2(2\theta)\log_2\cos^2(2\theta)
			+\sin^2(2\theta)\log_2\sin^2(2\theta)
			\right]},
		\label{eq:SM_m1}
	\end{equation}
	with the convention \(0\log_2 0=0\). For \(\alpha=2\),
	\begin{equation}
		m_2(\theta)=1-\log_2\left[1+\cos^4(2\theta)+\sin^4(2\theta)\right].
		\label{eq:SM_m2_intermediate}
	\end{equation}
	Using
	\begin{equation}
		\cos^4 x+\sin^4 x=1-\frac{1}{2}\sin^2(2x),
		\label{eq:SM_trig_identity}
	\end{equation}
	with \(x=2\theta\), we obtain
	\begin{equation}
		\boxed{	m_2(\theta)
			=-\log_2\left[1-\frac{1}{4}\sin^2(4\theta)\right] }.
		\label{eq:SM_m2}
	\end{equation}

	
	\section{Forest normal form}
	\label{sec:forest_normal_form}
	
	\subsection{Graph terminology and pruning setup}
	
	A finite simple graph is a pair \(G=(V,E)\), where \(V\) is a finite set of vertices and \(E\) is a set of unordered pairs of distinct vertices. In the quantum circuit, vertices are qubits and each edge \((ij)\in E\) carries an Ising gate \(e^{-i\theta Z_iZ_j}\). A path is a chain of vertices. A connected graph is one in which any two vertices are joined by a path. A connected component is a maximal connected piece of a graph. A cycle is a closed path with no repeated vertices except the starting and ending vertex. A tree is a connected graph with no cycles. A forest is a disjoint union of trees. We write \(c(G)\) for the number of connected components. If \(G\) is a forest with \(|V|=L\), then
	\begin{equation}
		|E|=L-c(G).
		\label{eq:SM_forest_edge_count}
	\end{equation}
	Indeed, each connected component is a tree, and a tree with \(L_a\) vertices has \(L_a-1\) edges. Summing over all components gives \(|E|=\sum_a(L_a-1)=L-c(G)\). This is also the number of non-root vertices after choosing one root in each component.
	
\begin{figure}[t]
		\centering
		\begin{tikzpicture}[
			scale=0.92,
			q/.style={circle,draw,minimum size=5.5mm,inner sep=0pt,font=\scriptsize},
			lbl/.style={draw=none,rectangle,font=\scriptsize,align=center},
			elim/.style={dotted,thick}
			]
			\node[lbl] at (1.5,1.2) {(a) Basic graph types};
			\node[q] (p1) at (0,0) {1};
			\node[q] (p2) at (0.8,0) {2};
			\node[q] (p3) at (1.6,0) {3};
			\node[q] (p4) at (2.4,0) {4};
			\draw (p1)--(p2)--(p3)--(p4);
			\node[lbl] at (1.2,-0.55) {path};
			\node[q] (t1) at (4.0,0) {1};
			\node[q] (t2) at (4.8,0.55) {2};
			\node[q] (t3) at (4.8,-0.55) {3};
			\node[q] (t4) at (5.6,0) {4};
			\draw (t1)--(t2);
			\draw (t1)--(t3);
			\draw (t3)--(t4);
			\node[lbl] at (4.8,-1.05) {tree};
			\node[q] (f1) at (0,-2.1) {1};
			\node[q] (f2) at (0.8,-1.65) {2};
			\node[q] (f3) at (0.8,-2.55) {3};
			\node[q] (f4) at (2.2,-2.1) {4};
			\node[q] (f5) at (3.0,-2.1) {5};
			\draw (f1)--(f2);
			\draw (f1)--(f3);
			\draw (f4)--(f5);
			\node[lbl] at (1.5,-3.05) {forest};
			\node[q] (c1) at (4.0,-1.55) {1};
			\node[q] (c2) at (5.0,-1.55) {2};
			\node[q] (c3) at (5.4,-2.45) {3};
			\node[q] (c4) at (4.4,-2.85) {4};
			\node[q] (c5) at (3.6,-2.35) {5};
			\draw (c1)--(c2)--(c3)--(c4)--(c5)--(c1);
			\node[lbl] at (4.5,-3.35) {cycle};
			\node[lbl] at (8.6,1.5) {(b) Leaf-to-root pruning};
			\node[q] (a1) at (7.0,0) {1};
			\node[q] (a2) at (7.9,0.55) {2};
			\node[q] (a3) at (7.9,-0.55) {3};
			\node[q] (a4) at (8.8,0.95) {4};
			\node[q] (a5) at (8.8,0.15) {5};
			\node[q] (a6) at (8.8,-0.55) {6};
			\draw (a1)--(a2);
			\draw (a1)--(a3);
			\draw (a2)--(a4);
			\draw (a2)--(a5);
			\draw (a3)--(a6);
			\node[lbl] at (7.6,-1.1) {choose root \(1\)};
			\node[q] (b1) at (10.2,0) {1};
			\node[q] (b2) at (11.1,0.55) {2};
			\node[q] (b3) at (11.1,-0.55) {3};
			\node[q] (b4) at (12.0,0.95) {4};
			\node[q] (b5) at (12.0,0.15) {5};
			\node[q] (b6) at (12.0,-0.55) {6};
			\draw (b1)--(b2);
			\draw (b1)--(b3);
			\draw[elim] (b2)--(b4);
			\draw[elim] (b2)--(b5);
			\draw[elim] (b3)--(b6);
			\node[lbl] at (11.1,-1.1) {eliminate leaves};
			\node[q] (d1) at (8.6,-2.45) {1};
			\node[q] (d2) at (9.5,-1.9) {2};
			\node[q] (d3) at (9.5,-3.0) {3};
			\node[q] (d4) at (10.4,-1.5) {4};
			\node[q] (d5) at (10.4,-2.3) {5};
			\node[q] (d6) at (10.4,-3.0) {6};
			\draw[elim] (d1)--(d2);
			\draw[elim] (d1)--(d3);
			\draw[elim] (d2)--(d4);
			\draw[elim] (d2)--(d5);
			\draw[elim] (d3)--(d6);
			\node[lbl] at (9.5,-3.55) {all non-root edges become one-qubit rotations};
		\end{tikzpicture}
		\caption{Graph language and CNOT pruning. Solid edges are unevaluated \(Z_iZ_j\) parities. Dotted edges have been converted by \(\mathrm{CNOT}_{p(v)\to v}\) into one-qubit rotations \(Z_v\). Each non-root vertex is eliminated once, giving \(L-c(G)\) one-qubit factors.}
		\label{fig:SM_graph_pruning}
	\end{figure}
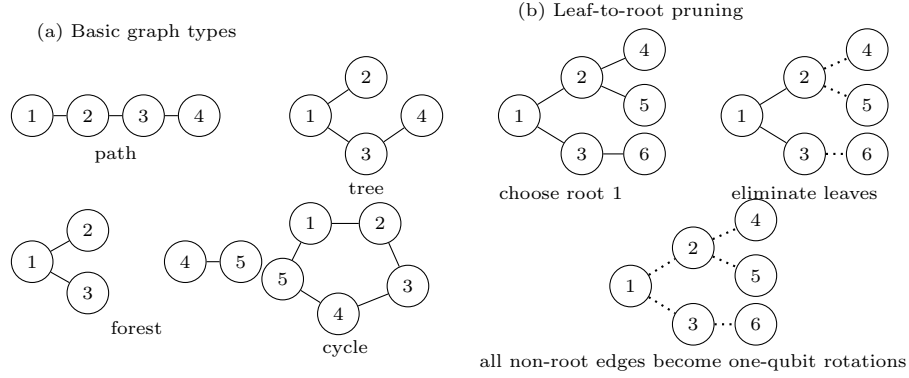

	A root is a bookkeeping choice, not a physical choice. Once one root is chosen in each connected component, every non-root vertex \(v\) has a unique parent \(p(v)\), defined as the previous vertex on the unique path from \(v\) to the root of its component. A leaf is a vertex with degree one in the remaining tree. Leaf-to-root order means that one starts with leaves farthest from the root, eliminates them, then eliminates the newly created leaves, and continues until only the root remains in each component. Thus every non-root vertex is eliminated exactly once, and the number of eliminated vertices is \(L-c(G)\). Figure~\ref{fig:SM_graph_pruning} illustrates the graph language and the pruning order.
	
	\subsection{The CNOT Clifford mechanism}
	
	The pruning step uses only the two-qubit CNOT gate. For two qubits, the CNOT gate with the first qubit as control and the second qubit as target acts as
	\begin{equation}
		|00\rangle\mapsto |00\rangle,\qquad
		|01\rangle\mapsto |01\rangle,\qquad
		|10\rangle\mapsto |11\rangle,\qquad
		|11\rangle\mapsto |10\rangle.
		\label{eq:SM_CNOT_basis_map}
	\end{equation}
	Equivalently, for a control qubit \(p\) and target qubit \(v\),
	\begin{equation}
		\mathrm{CNOT}_{p\to v}|x,y\rangle=|x,y\oplus x\rangle,\qquad x,y\in\{0,1\},
		\label{eq:SM_CNOT_action}
	\end{equation}
	where \(\oplus\) denotes addition modulo 2. In the ordered basis \(\{|00\rangle,|01\rangle,|10\rangle,|11\rangle\}\), the matrix is
	\begin{equation}
		\mathrm{CNOT}_{p\to v}=
		\begin{pmatrix}
			1&0&0&0\\
			0&1&0&0\\
			0&0&0&1\\
			0&0&1&0
		\end{pmatrix}.
		\label{eq:SM_CNOT_matrix}
	\end{equation}
	It is real, unitary, Hermitian, and self-inverse:
	\begin{equation}
		\mathrm{CNOT}_{p\to v}^\dagger=\mathrm{CNOT}_{p\to v},\qquad
		\mathrm{CNOT}_{p\to v}^2=I.
		\label{eq:SM_CNOT_self_inverse}
	\end{equation}
	Since CNOT is its own inverse, conjugating an operator \(O\) by CNOT means applying CNOT, then \(O\), then CNOT again:
	\begin{equation}
		O\mapsto \mathrm{CNOT}_{p\to v}\, O \, \mathrm{CNOT}_{p\to v}.
		\label{eq:SM_CNOT_conjugation_map}
	\end{equation}
	
	The important point is how CNOT conjugates \(Z\) operators. Since \(Z_p|x,y\rangle=(-1)^x|x,y\rangle\), Eq.~\eqref{eq:SM_CNOT_action} gives
	\begin{equation}
		\begin{aligned}
			\mathrm{CNOT}_{p\to v}Z_p\mathrm{CNOT}_{p\to v}|x,y\rangle
			&=\mathrm{CNOT}_{p\to v}Z_p|x,y\oplus x\rangle\\
			&=(-1)^x\mathrm{CNOT}_{p\to v}|x,y\oplus x\rangle\\
			&=(-1)^x|x,y\rangle\\
			&=Z_p|x,y\rangle.
		\end{aligned}
		\label{eq:SM_CNOT_Zp_derivation}
	\end{equation}
	Hence
	\begin{equation}
		\boxed{	\mathrm{CNOT}_{p\to v}Z_p\mathrm{CNOT}_{p\to v}=Z_p }.
		\label{eq:SM_CNOT_Zp_rule}
	\end{equation}
	Similarly, since \(Z_v|x,y\rangle=(-1)^y|x,y\rangle\),
	\begin{equation}
		\begin{aligned}
			\mathrm{CNOT}_{p\to v}Z_v\mathrm{CNOT}_{p\to v}|x,y\rangle
			&=\mathrm{CNOT}_{p\to v}Z_v|x,y\oplus x\rangle\\
			&=(-1)^{y\oplus x}\mathrm{CNOT}_{p\to v}|x,y\oplus x\rangle\\
			&=(-1)^{x+y}|x,y\rangle\\
			&=Z_pZ_v|x,y\rangle.
		\end{aligned}
		\label{eq:SM_CNOT_Zv_derivation}
	\end{equation}
	Therefore
	\begin{equation}
		\mathrm{CNOT}_{p\to v}Z_v\mathrm{CNOT}_{p\to v}=Z_pZ_v.
		\label{eq:SM_CNOT_Zv_rule}
	\end{equation}
	Multiplying Eqs.~\eqref{eq:SM_CNOT_Zp_rule} and \eqref{eq:SM_CNOT_Zv_rule} gives the crucial edge-parity reduction:
	\begin{equation}
		\begin{aligned}
			\mathrm{CNOT}_{p\to v}(Z_pZ_v)\mathrm{CNOT}_{p\to v}
			&=\left(\mathrm{CNOT}_{p\to v}Z_p\mathrm{CNOT}_{p\to v}\right)
			\left(\mathrm{CNOT}_{p\to v}Z_v\mathrm{CNOT}_{p\to v}\right)\\
			&=Z_p(Z_pZ_v)\\
			&=Z_p^2Z_v\\
			&=Z_v.
		\end{aligned}
		\label{eq:SM_CNOT_edge_derivation}
	\end{equation}
	Thus
	\begin{equation}
		\boxed{		\mathrm{CNOT}_{p\to v}(Z_pZ_v)\mathrm{CNOT}_{p\to v}=Z_v }.
		\label{eq:SM_CNOT_edge_rule}
	\end{equation}
	Consequently,
	\begin{equation}
		\mathrm{CNOT}_{p\to v}e^{-i\theta Z_pZ_v}\mathrm{CNOT}_{p\to v}=e^{-i\theta Z_v}.
		\label{eq:SM_CNOT_rotation_rule}
	\end{equation}
	This identity is the local mechanism behind the forest normal form: a CNOT from parent to child turns the two-qubit edge rotation on that parent-child edge into a one-qubit \(Z\) rotation on the child.
	
	Two further facts are needed. First, CNOT preserves the two-qubit plus state:
	\begin{equation}
		|+\rangle_p|+\rangle_v=\frac{1}{2}\left(|00\rangle+|01\rangle+|10\rangle+|11\rangle\right),
		\label{eq:SM_two_plus_expansion}
	\end{equation}
	and CNOT only permutes the four equal-amplitude basis states. Hence
	\begin{equation}
		\mathrm{CNOT}_{p\to v}|+\rangle_p|+\rangle_v=|+\rangle_p|+\rangle_v.
		\label{eq:SM_CNOT_plus}
	\end{equation}
	Any circuit built only from CNOT gates therefore leaves \(|+\rangle^{\otimes L}\) invariant. Second, CNOT is a Clifford gate: it maps Pauli strings to Pauli strings under conjugation. In the phase-free convention used for the SRE, this means \(C^\dagger PC=\pm Q\) for Pauli strings \(P,Q\in\mathcal P_L\), with phases irrelevant after taking squared expectation values. Products of CNOT gates are Clifford circuits. Since the SRE is Clifford invariant, a CNOT pruning circuit can change the Pauli coordinates of the state but cannot change \(M_\alpha\).
	
	\subsection{CNOT normal form for forests}
	
	For a graph \(G=(V,E)\), define
	\begin{equation}
		|\psi_G(\theta)\rangle
		=\exp\left[-i\theta\sum_{(ij)\in E}Z_iZ_j\right]|+\rangle^{\otimes L}.
		\label{eq:SM_graph_state}
	\end{equation}
	Since all \(Z_iZ_j\) terms commute,
	\begin{equation}
		|\psi_G(\theta)\rangle
		=\left(\prod_{(ij)\in E}e^{-i\theta Z_iZ_j}\right)|+\rangle^{\otimes L}.
		\label{eq:SM_graph_state_product}
	\end{equation}
	
	\begin{theorem}[CNOT normal form for forests]
		\label{thm:SM_CNOT_forest_normal_form}
		Let \(G=(V,E)\) be a forest with \(|V|=L\) and \(c(G)\) connected components. Choose one root in each component, and let \(R\) be the set of roots. Then there exists a CNOT Clifford circuit \(C_G\) such that
		\begin{equation}
			\boxed{	C_G\left(\prod_{(ij)\in E}e^{-i\theta Z_iZ_j}\right)C_G^\dagger
				=\prod_{v\notin R}e^{-i\theta Z_v}},
			\label{eq:SM_CNOT_normal_form}
		\end{equation}
		and
		\begin{equation}
			\boxed{	C_G|+\rangle^{\otimes L}=|+\rangle^{\otimes L}}.
			\label{eq:SM_CNOT_normal_form_plus}
		\end{equation}
	\end{theorem}
	
	\begin{proof}
		Root each connected component. For every non-root vertex \(v\), let \(p(v)\) be its parent. Apply \(\mathrm{CNOT}_{p(v)\to v}\) in leaf-to-root order. When \(v\) is a leaf of the remaining tree, the edge parity on its unique remaining incident edge is \(Z_{p(v)}Z_v\). By Eq.~\eqref{eq:SM_CNOT_edge_rule}, this edge parity is mapped to \(Z_v\), and therefore the corresponding edge gate is mapped to \(e^{-i\theta Z_v}\).
		
		The leaf-to-root order ensures that eliminated leaves stay eliminated. Once the factor \(Z_v\) has been produced, no later pruning step uses \(v\) as a target, because the algorithm never returns to removed leaves. Later CNOTs act closer to the root, on \(p(v)\) or its ancestors, and therefore cannot turn \(Z_v\) back into an edge parity. Iterating over all non-root vertices eliminates every edge in the forest. Since one root was chosen in each connected component, the number of non-root vertices is \(L-c(G)=|E|\), so every forest edge is eliminated exactly once. This proves Eq.~\eqref{eq:SM_CNOT_normal_form}. Equation~\eqref{eq:SM_CNOT_normal_form_plus} follows from Eq.~\eqref{eq:SM_CNOT_plus}, because \(C_G\) is a product of CNOT gates.
	\end{proof}

	\subsection{Proof of the forest SRE formula}
	
	We first record the notation used throughout this proof. For \(L\) qubits, a phase-free Pauli string is written \(P_{\mathbf a}=\sigma_{a_1}\otimes\sigma_{a_2}\otimes\cdots\otimes\sigma_{a_L}\), where each label \(a_j\in\{0,1,2,3\}\) selects \(\sigma_0=I\), \(\sigma_1=X\), \(\sigma_2=Y\), or \(\sigma_3=Z\) on site \(j\). For a pure state \(|\psi\rangle\), the associated Pauli probability distribution is \(\Xi_\psi(\mathbf a)=2^{-L}\langle\psi|P_{\mathbf a}|\psi\rangle^2\), and the stabilizer Renyi entropy is \(M_\alpha(\psi)=\frac{1}{1-\alpha}\log_2\!\Big(\sum_{\mathbf a}\Xi_\psi(\mathbf a)^\alpha\Big)-L\). We now restate and prove the theorem.
	
	\begin{theorem}[Forest normal form]
		\label{thm:SM_forest}
		If \(G=(V,E)\) is a forest with \(|V|=L\) and \(c(G)\) connected components, then for every \(\alpha>0\),
		\begin{equation}
			\boxed{	M_\alpha(\psi_G(\theta))=\left[L-c(G)\right]m_\alpha(\theta)},
			\label{eq:SM_forest_formula}
		\end{equation}
		with \(\alpha=1\) understood by continuity. Here $m_\alpha(\theta)$ is the one-qubit magic given in Eq.~\eqref{eq:SM_one_qubit_magic}. In particular,
		\begin{equation}
			\boxed{	M_2(\psi_G(\theta))=\left[L-c(G)\right]\left[-\log_2\left(1-\frac{1}{4}\sin^2(4\theta)\right)\right] }.
			\label{eq:SM_forest_m2}
		\end{equation}
	\end{theorem}
	
	\begin{proof}
		The proof proceeds in five explicit stages: (i) apply the CNOT normal form to trade the entangled forest state for an explicit product state, (ii) invoke Clifford invariance of \(M_\alpha\) to transfer the computation to this product state, (iii) prove additivity of \(M_\alpha\) across a tensor product from first principles, (iv) evaluate the two types of one-qubit factors, and (v) sum the contributions.
		
		{\bf Stage 1: the pruned product state.} By Theorem~\ref{thm:SM_CNOT_forest_normal_form}, there is a Clifford circuit \(C_G\), built only from CNOT gates along a chosen rooting of the forest \(G\), such that
		\begin{equation}
			C_G|\psi_G(\theta)\rangle
			=\left(\prod_{v\notin R}e^{-i\theta Z_v}\right)|+\rangle^{\otimes L}.
			\label{eq:SM_pruned_state}
		\end{equation}
		Here \(R\subset V\) is the set of chosen roots, one per connected component, so that \(|R|=c(G)\) and the number of non-root vertices is \(|V\setminus R|=L-c(G)\). Since all the operators \(e^{-i\theta Z_v}\) for different \(v\) act on different qubits, they commute trivially, and the right-hand side of Eq.~\eqref{eq:SM_pruned_state} can be written explicitly as a tensor product over individual qubits,
		\begin{equation}
			C_G|\psi_G(\theta)\rangle
			=
			\bigotimes_{r\in R}|+\rangle_r
			\otimes
			\bigotimes_{v\notin R}e^{-i\theta Z_v}|+\rangle_v.
			\label{eq:SM_pruned_product_state}
		\end{equation}
		Each root qubit is left untouched in the state \(|+\rangle\), while each non-root qubit has been rotated by the same angle \(\theta\) about the \(Z\) axis starting from \(|+\rangle\). Call this right-hand side \(|\Phi(\theta)\rangle\), so that \(C_G|\psi_G(\theta)\rangle=|\Phi(\theta)\rangle\).
		
		{\bf Stage 2: Clifford invariance.} Clifford unitaries permute the set of phase-free Pauli strings under conjugation, \(C P_{\mathbf a} C^\dagger=\pm P_{\mathbf b}\) for some relabeling \(\mathbf a\mapsto\mathbf b\), and the overall sign is irrelevant once expectation values are squared. Consequently, for any Clifford unitary \(C\) and any state \(|\chi\rangle\), the Pauli distribution \(\Xi_{C\chi}\) is just \(\Xi_\chi\) with its labels permuted, so the sum \(\sum_{\mathbf a}\Xi_{C\chi}(\mathbf a)^\alpha\) is unchanged, term for term, from \(\sum_{\mathbf a}\Xi_\chi(\mathbf a)^\alpha\). This proves the Clifford invariance identity \(M_\alpha(C\chi)=M_\alpha(\chi)\), which was already established earlier above and is simply being invoked here. Applying it to \(C=C_G\) and \(|\chi\rangle=|\psi_G(\theta)\rangle\) gives
		\begin{equation}
			M_\alpha(\psi_G(\theta))=M_\alpha(C_G\psi_G(\theta))=M_\alpha(\Phi(\theta)).
			\label{eq:SM_forest_clifford_step}
		\end{equation}
		This is the key step that replaces the original, entangled, hard-to-analyze state by the simple product state \(|\Phi(\theta)\rangle\) without changing the quantity we want to compute.
		
		{\bf Stage 3: additivity across a tensor product, derived explicitly.} Let \(|\Phi\rangle=\bigotimes_{k=1}^L|\phi_k\rangle\) be any product state of \(L\) qubits. For a Pauli string \(P_{\mathbf a}=\bigotimes_k \sigma_{a_k}\), the expectation value factorizes site by site because the state and the operator are both tensor products,
		\begin{equation}
			\langle\Phi|P_{\mathbf a}|\Phi\rangle=\prod_{k=1}^L\langle\phi_k|\sigma_{a_k}|\phi_k\rangle.
		\end{equation}
		Squaring both sides and multiplying by the overall normalization \(2^{-L}=\prod_{k=1}^L 2^{-1}\) gives
		\begin{equation}
			\Xi_\Phi(\mathbf a)=\prod_{k=1}^L \xi_k(a_k),\qquad
			\xi_k(a)\coloneq \tfrac12\langle\phi_k|\sigma_a|\phi_k\rangle^2.
			\label{eq:SM_product_distribution_factorization}
		\end{equation}
		Thus the full \(4^L\)-entry Pauli distribution of a product state is itself a product, one four-number distribution \(\xi_k\) per qubit. Raising to the power \(\alpha\) and summing over all \(4^L\) labels, the sum factorizes into a product of \(L\) independent single-qubit sums, because summing a product of independent factors over all label combinations is the same as multiplying the individual sums,
		\begin{equation}
			\sum_{\mathbf a}\Xi_\Phi(\mathbf a)^\alpha
			=
			\prod_{k=1}^L\left(\sum_{a=0}^3\xi_k(a)^\alpha\right).
		\end{equation}
		Taking \(\log_2\) turns this product into a sum, so
		\begin{equation}
			\frac{1}{1-\alpha}\log_2\!\left(\sum_{\mathbf a}\Xi_\Phi(\mathbf a)^\alpha\right)
			=
			\sum_{k=1}^L\frac{1}{1-\alpha}\log_2\!\left(\sum_{a=0}^3\xi_k(a)^\alpha\right).
		\end{equation}
		Subtracting \(L=\sum_{k=1}^L 1\) from both sides and recognizing that each term on the right, minus one, is precisely the one-qubit stabilizer Renyi entropy \(M_\alpha(\phi_k)\), gives the additivity identity used in the theorem,
		\begin{equation}
			M_\alpha(\Phi)=\sum_{k=1}^L M_\alpha(\phi_k).
			\label{eq:SM_additivity_derived}
		\end{equation}
		Applying Eq.~\eqref{eq:SM_additivity_derived} to \(|\Phi(\theta)\rangle\) from Eq.~\eqref{eq:SM_pruned_product_state}, whose factors are indexed by root and non-root qubits, gives
		\begin{equation}
			M_\alpha(\Phi(\theta))
			=\sum_{r\in R}M_\alpha(|+\rangle_r)+\sum_{v\notin R}M_\alpha(e^{-i\theta Z_v}|+\rangle_v).
			\label{eq:SM_forest_additivity_step}
		\end{equation}
		
		{\bf Stage 4: evaluating the two types of one-qubit factors.} The root factors are the state \(|+\rangle\), which is a stabilizer state: it is the simultaneous \(+1\) eigenstate of the single Pauli operator \(X\), and stabilizer states have vanishing magic \(M_\alpha=0\) for every \(\alpha>0\), because their Pauli distribution places all its weight on exactly \(2\) Pauli strings (namely \(I\) and \(X\), for one qubit) with squared expectation equal to one, and zero weight elsewhere. Hence
		\begin{equation}
			M_\alpha(|+\rangle_r)=0\qquad\text{for every }r\in R.
			\label{eq:SM_root_zero_magic}
		\end{equation}
		The non-root factors are all of the identical form \(e^{-i\theta Z}|+\rangle\), with the same angle \(\theta\) at every non-root vertex \(v\), since Eq.~\eqref{eq:SM_pruned_state} applies the same one-qubit rotation to every non-root qubit. By the one-qubit derivation given earlier in Sec.~\ref{sec:one_qubit}, this single-qubit state has stabilizer Renyi entropy equal to the function \(m_\alpha(\theta)\) derived in Eq.~\eqref{eq:SM_one_qubit_magic}. Because the angle and the initial state are identical at every non-root vertex, the value of \(M_\alpha\) is the same number at every non-root vertex,
		\begin{equation}
			M_\alpha(e^{-i\theta Z_v}|+\rangle_v)=m_\alpha(\theta)\qquad\text{for every }v\notin R.
			\label{eq:SM_nonroot_equal_magic}
		\end{equation}
		
		{\bf Stage 5: summing the contributions.} Substituting Eqs.~\eqref{eq:SM_root_zero_magic} and \eqref{eq:SM_nonroot_equal_magic} into Eq.~\eqref{eq:SM_forest_additivity_step}, and using \(|R|=c(G)\) so that the number of non-root vertices is \(L-c(G)\), gives
		\begin{equation}
			M_\alpha(\Phi(\theta))
			=
			\underbrace{c(G)\cdot 0}_{\text{root contribution}}
			+
			\underbrace{[L-c(G)]\cdot m_\alpha(\theta)}_{\text{non-root contribution}}
			=
			[L-c(G)]\,m_\alpha(\theta).
		\end{equation}
		Combining this with the Clifford invariance identity of Eq.~\eqref{eq:SM_forest_clifford_step} gives
		\begin{equation}
			M_\alpha(\psi_G(\theta))=\left[L-c(G)\right]m_\alpha(\theta).
			\label{eq:SM_forest_formula_proved}
		\end{equation}
		This proves Eq.~\eqref{eq:SM_forest_formula}. Substituting the explicit one-qubit function \(m_2(\theta)=-\log_2\!\left(1-\tfrac14\sin^2(4\theta)\right)\), derived earlier in Eq.~\eqref{eq:SM_m2}, into Eq.~\eqref{eq:SM_forest_formula_proved} gives Eq.~\eqref{eq:SM_forest_m2}, completing the proof.
	\end{proof}
	
	The theorem above establishes an identity for the single number \(M_\alpha\), but the argument in fact proves something stronger about the entire underlying Pauli distribution, and it is worth making this explicit since it clarifies exactly why the final count of contributions is \(L-c(G)\) rather than some other combination. Recall that \(M_\alpha\) is only a summary statistic, a single real number extracted from the full table of \(4^L\) probabilities \(\Xi_\psi(\mathbf a)\), one probability for every possible assignment of Pauli labels to the \(L\) qubits. Equation~\eqref{eq:SM_product_distribution_factorization} in Stage 3 of the proof already showed that for any product state, this entire \(4^L\)-entry table factorizes into a tensor product of \(L\) individual four-number tables, one per qubit. Applying this fact specifically to the pruned state \(|\Phi(\theta)\rangle\) of Eq.~\eqref{eq:SM_pruned_product_state}, whose factors are either \(|+\rangle\) at root vertices or \(e^{-i\theta Z}|+\rangle\) at non-root vertices, means the full Pauli distribution of \(|\Phi(\theta)\rangle\) is built from only two distinct four-number building blocks, repeated \(c(G)\) and \(L-c(G)\) times respectively. Explicitly, writing the single-qubit distribution in the fixed order \((I,X,Y,Z)\) and using \(\xi(a)=\tfrac12\langle\phi|\sigma_a|\phi\rangle^2\) from Eq.~\eqref{eq:SM_product_distribution_factorization}, the root building block follows from \(\langle+|I|+\rangle=1\), \(\langle+|X|+\rangle=1\), \(\langle+|Y|+\rangle=0\), \(\langle+|Z|+\rangle=0\), which gives \(\tfrac12(1)^2=\tfrac12\) for \(I\), \(\tfrac12(1)^2=\tfrac12\) for \(X\), and zero for \(Y\) and \(Z\), so that
	\begin{equation}
		\xi^{(+)}=\left(\frac{1}{2},\frac{1}{2},0,0\right).
		\label{eq:SM_root_block_derived}
	\end{equation}
	The non-root building block is obtained the same way, but now using the rotated expectation values \(\langle+|e^{i\theta Z}\sigma_a e^{-i\theta Z}|+\rangle\) computed earlier for the elementary one-qubit rotation, namely expectation \(1\) for \(I\), \(\cos(2\theta)\) for \(X\), \(\sin(2\theta)\) for \(Y\), and \(0\) for \(Z\); squaring each of these and multiplying by \(\tfrac12\) gives
	\begin{equation}
		\xi^{(\theta)}=\left(\frac{1}{2},\frac{1}{2}\cos^2(2\theta),\frac{1}{2}\sin^2(2\theta),0\right).
		\label{eq:SM_nonroot_block_derived}
	\end{equation}
	Since the full distribution of a product state is the tensor product of its single-qubit factors, as shown in Eq.~\eqref{eq:SM_product_distribution_factorization}, and since Eq.~\eqref{eq:SM_pruned_product_state} places the block \(\xi^{(+)}\) at every one of the \(c(G)\) root qubits and the block \(\xi^{(\theta)}\) at every one of the \(L-c(G)\) non-root qubits, the full \(4^L\)-entry distribution of \(|\Phi(\theta)\rangle\) is
	\begin{equation}
		\Xi_{\Phi(\theta)}=\left(\xi^{(+)}\right)^{\otimes c(G)}\otimes\left(\xi^{(\theta)}\right)^{\otimes[L-c(G)]}.
		\label{eq:SM_forest_distribution_factorization}
	\end{equation}
	It remains only to note that this factorized distribution belongs to the CNOT-rotated state \(|\Phi(\theta)\rangle=C_G|\psi_G(\theta)\rangle\), not literally to the original physical state \(|\psi_G(\theta)\rangle\) itself. Since \(C_G\) is a Clifford circuit, conjugation by \(C_G\) acts on Pauli strings by permuting their labels, up to an overall sign that is irrelevant after squaring, so the true distribution \(\Xi_{\psi_G(\theta)}\) of the original state consists of exactly the same multiset of numerical values as \(\Xi_{\Phi(\theta)}\) in Eq.~\eqref{eq:SM_forest_distribution_factorization}, only redistributed among the \(4^L\) Pauli labels according to the specific permutation induced by \(C_G\). This is the precise meaning of the qualifying phrase ``up to the Clifford permutation of Pauli labels'' and it is why Eq.~\eqref{eq:SM_forest_distribution_factorization} is stated as an equivalence up to relabeling rather than as a literal equality of the original distribution. As a consistency check, applying the additivity argument of Stage 3 directly to Eq.~\eqref{eq:SM_forest_distribution_factorization}, namely raising every entry to the power \(\alpha\), summing, and taking \(\frac{1}{1-\alpha}\log_2(\cdot)-L\), reproduces exactly \(c(G)\) copies of the one-qubit stabilizer Renyi entropy of \(\xi^{(+)}\), each equal to zero since \(\xi^{(+)}\) is a stabilizer distribution, plus \(L-c(G)\) copies of the one-qubit stabilizer Renyi entropy of \(\xi^{(\theta)}\), each equal to \(m_\alpha(\theta)\), giving back precisely \(M_\alpha(\psi_G(\theta))=[L-c(G)]m_\alpha(\theta)\) of Eq.~\eqref{eq:SM_forest_formula_proved}. This distribution-level statement is therefore not an independent new fact but a strictly finer version of the same product structure already used in the proof, and it is the underlying reason the final answer contains exactly \(L-c(G)\) identical one-qubit magic contributions rather than a more intricate combination depending on the detailed shape of the forest.

	\subsection{Thermodynamic limit and conjugated stabilizer inputs}
	
	Let \(G_L\) be a sequence of forests with \(L\) vertices and suppose
	\begin{equation}
		\lim_{L\to\infty}\frac{c(G_L)}{L}=\rho.
		\label{eq:SM_component_density}
	\end{equation}
	Dividing Eq.~\eqref{eq:SM_forest_formula} by \(L\) gives
	\begin{equation}
		\boxed{	\lim_{L\to\infty}\frac{M_\alpha(\psi_{G_L}(\theta))}{L}
			=(1-\rho)m_\alpha(\theta) }.
		\label{eq:SM_forest_density}
	\end{equation}
	For connected trees, \(c(G_L)=1\), so \(\rho=0\) and the thermodynamic-limit magic density is simply
	\begin{equation}
		\mu_\alpha(\theta)=m_\alpha(\theta).
		\label{eq:SM_connected_tree_density}
	\end{equation}
	
	The representative input in Eq.~\eqref{eq:SM_graph_state} is not the only stabilizer realization of the same formula. Let \(C\) be any Clifford unitary and let
	\begin{equation}
		|s\rangle=C|+\rangle^{\otimes L},\qquad
		\widetilde Z_i\widetilde Z_j=CZ_iZ_jC^\dagger.
		\label{eq:SM_conjugated_generators}
	\end{equation}
	To see how the conjugated family follows from this definition, start from \(|\psi_G(\theta)\rangle=\exp\big[-i\theta\sum_{(ij)\in E}Z_iZ_j\big]|+\rangle^{\otimes L}\) and apply \(C\) to both sides, inserting \(C^\dagger C=I\) immediately after the exponential, which is allowed since \(C^\dagger C=I\) for any unitary \(C\); this gives \(C|\psi_G(\theta)\rangle=C\exp\big[-i\theta\sum_{(ij)\in E}Z_iZ_j\big]C^\dagger\,\big(C|+\rangle^{\otimes L}\big)\). Next use the standard conjugation identity, valid for any Hermitian operator \(A\) and any unitary \(C\), \(Ce^{-i\theta A}C^\dagger=e^{-i\theta CAC^\dagger}\); this identity follows by expanding the exponential as a power series and using \(CA^nC^\dagger=(CAC^\dagger)^n\), since every pair of adjacent factors \(C^\dagger C=I\) inside \(CA^nC^\dagger=CAA\cdots AC^\dagger\) cancels once \(C^\dagger C=I\) is inserted between consecutive powers of \(A\), turning \(CA^nC^\dagger\) into \(n\) copies of \(CAC^\dagger\) multiplied together. Applying this identity with \(A=\sum_{(ij)\in E}Z_iZ_j\) gives \(C\exp\big[-i\theta\sum_{(ij)\in E}Z_iZ_j\big]C^\dagger=\exp\big[-i\theta\sum_{(ij)\in E}CZ_iZ_jC^\dagger\big]=\exp\big[-i\theta\sum_{(ij)\in E}\widetilde Z_i\widetilde Z_j\big]\), where the last step uses the definition of \(\widetilde Z_i\widetilde Z_j\) in Eq.~\eqref{eq:SM_conjugated_generators}. Substituting this back, and using \(C|+\rangle^{\otimes L}=|s\rangle\) from Eq.~\eqref{eq:SM_conjugated_generators}, gives exactly
	\begin{equation}
		\exp\left[-i\theta\sum_{(ij)\in E}\widetilde Z_i\widetilde Z_j\right]|s\rangle
		=
		C|\psi_G(\theta)\rangle.
		\label{eq:SM_conjugated_family}
	\end{equation}
	Thus Eq.~\eqref{eq:SM_conjugated_family} is nothing more than conjugating the entire forest circuit, generators and state together, by the same unitary \(C\); it is not a new physical example, only the original state and generators relabeled by \(C\). By Clifford invariance, this conjugated family has exactly the same SRE as \(|\psi_G(\theta)\rangle\), and Eq.~\eqref{eq:SM_forest_formula} applies unchanged. This statement is exact, but it blocks a tempting and false shortcut: one might think that since every stabilizer state \(|s\rangle\) is Clifford-equivalent to \(|+\rangle^{\otimes L}\), the forest formula should automatically hold for any stabilizer input \(|s\rangle\) driven by the plain, unrotated generators \(Z_iZ_j\), with no conjugation applied to the generators. This is false. Theorem~\ref{thm:SM_forest} requires the generators to be conjugated by the same \(C\) that produced \(|s\rangle\) from \(|+\rangle^{\otimes L}\), as in Eq.~\eqref{eq:SM_conjugated_generators}. If instead one drives an arbitrary stabilizer state \(|s\rangle\) with the bare, unrotated operators \(Z_iZ_j\), the resulting family \(\exp\big[-i\theta\sum_{(ij)\in E}Z_iZ_j\big]|s\rangle\) need not be related to \(|\psi_G(\theta)\rangle\) by any Clifford circuit at all, because \(|s\rangle\) need not lie in the Clifford orbit of \(|+\rangle^{\otimes L}\) with respect to these specific unrotated generators. Without that structural relationship, Clifford invariance gives no guarantee, and the forest formula need not hold.
	
	\subsection{Cycles and the obstruction to the product formula}
	\label{subsec:SM_cycles_obstruction}
	
	The forest proof works because every edge parity can be eliminated independently. Cycles create relations among edge parities. The simplest example is a triangle:
	\begin{equation}
		(Z_1Z_2)(Z_2Z_3)(Z_3Z_1)=I.
		\label{eq:SM_triangle_relation}
	\end{equation}
	Thus the three edge parities are not independent. Equivalently, represent an edge \((ij)\) by a binary vector with ones at sites \(i\) and \(j\). Multiplication of \(Z\)-strings corresponds to addition of these vectors over the field \(\mathbb F_2=\{0,1\}\), where addition is performed modulo \(2\). This correspondence follows from \(Z_j^2=I\), so that a site appearing twice cancels, exactly as \(1+1=0\) over \(\mathbb F_2\). For the triangle,
	\begin{equation}
		(1,1,0)+(0,1,1)+(1,0,1)=(0,0,0)\quad\text{over }\mathbb F_2.
		\label{eq:SM_F2_triangle}
	\end{equation}
	This linear dependence is the cycle obstruction. For a graph with cycles, \(|E|>L-c(G)\), while the rank of the edge-parity span is \(L-c(G)\). Hence the edge gates cannot all be converted into independent one-qubit rotations, and the all-time product formula generally fails.
	
	What remains true is the Clifford revival. If
	\begin{equation}
		\theta\in\frac{\pi}{4}\mathbb Z,
		\label{eq:SM_clifford_angles}
	\end{equation}
	then each two-qubit gate \(e^{-i\theta Z_iZ_j}\) is Clifford. Therefore the full graph circuit is Clifford for any graph, not only for forests. Since \(|+\rangle^{\otimes L}\) is a stabilizer state and Clifford unitaries map stabilizer states to stabilizer states,
	\begin{equation}
		\boxed{	M_\alpha(\psi_G(\theta))=0,\qquad
			\theta\in\frac{\pi}{4}\mathbb Z,\qquad \alpha>0 },
		\label{eq:SM_universal_graph_revivals}
	\end{equation}
	with \(\alpha=1\) understood by continuity. Thus forests give the stronger all-time product formula, while arbitrary graphs retain the exact zero-magic revivals at Clifford angles.

	\section{Entanglement bounds and magic without entanglement density}
	
	\subsection{Schmidt-rank bound for forest cuts}
	
	Let \(A|A^c\) be a bipartition of the vertices of a graph \(G=(V,E)\). The set of edges crossing the cut is
	\begin{equation}
		\partial_G A=\{(ij)\in E:i\in A,\ j\in A^c\}.
		\label{eq:SM_cut_boundary}
	\end{equation}
	All gates supported entirely inside \(A\) or entirely inside \(A^c\) are local with respect to this bipartition and cannot change the entanglement between the two sides. Only the gates on \(\partial_GA\) can increase the Schmidt rank.
	
	A single crossing gate has the operator decomposition
	\begin{equation}
		e^{-i\theta Z_iZ_j}=\cos\theta\,I_iI_j-i\sin\theta\,Z_iZ_j.
		\label{eq:SM_crossing_gate_schmidt}
	\end{equation}
	Across the cut, this operator has operator Schmidt rank at most two. Acting with such an operator can increase the Schmidt rank of a state by at most a factor of two. Therefore, after all crossing gates are applied, the Schmidt rank \(\chi_A\) across \(A|A^c\) obeys
	\begin{equation}
		\chi_A\leq 2^{|\partial_GA|}.
		\label{eq:SM_schmidt_rank_bound}
	\end{equation}
	For the von Neumann entanglement entropy,
	\begin{equation}
		S_{\rm EE}(A)\leq\log_2\chi_A\leq|\partial_GA|.
		\label{eq:SM_EE_boundary_bound}
	\end{equation}
	This bound applies to graph Ising states on any graph. For forests, it combines directly with the exact magic formula.
	
	\subsection{Area-law cuts with extensive magic}
	
	Consider a sequence of forests \(G_L\) with \(c(G_L)/L\to\rho\). From Eq.~\eqref{eq:SM_forest_density},
	\begin{equation}
		\lim_{L\to\infty}\frac{M_\alpha(\psi_{G_L}(\theta))}{L}
		=(1-\rho)m_\alpha(\theta).
		\label{eq:SM_magic_density_forest_again}
	\end{equation}
	Now choose a sequence of cuts \(A_L|A_L^c\) with subextensive graph boundary,
	\begin{equation}
		|\partial_{G_L}A_L|=o(L).
		\label{eq:SM_subextensive_boundary}
	\end{equation}
	Equation~\eqref{eq:SM_EE_boundary_bound} gives
	\begin{equation}
		\lim_{L\to\infty}\frac{S_{\rm EE}(A_L)}{L}=0.
		\label{eq:SM_zero_EE_density}
	\end{equation}
	Thus, whenever \((1-\rho)m_\alpha(\theta)\neq0\), the same sequence of states has finite magic density and zero entanglement entropy density:
	\begin{equation}
		\boxed{	\lim_{L\to\infty}\frac{M_\alpha(\psi_{G_L}(\theta))}{L}\neq0,
			\qquad
			\lim_{L\to\infty}\frac{S_{\rm EE}(A_L)}{L}=0 }.
		\label{eq:SM_magic_without_EE_density}
	\end{equation}
	This is the graph-theoretic form of magic without entanglement density. The open chain is the simplest connected-tree example: an interval cut has one crossing edge, so \(S_{\rm EE}=O(1)\), while \(M_\alpha\) is proportional to \(L-1\).


	\section{Open-chain Ising quench}
	
	\subsection{Exact magic formula and thermodynamic density}
	
	The open-chain Ising quench studied in the main text is
	\begin{equation}
		|\psi_L(t)\rangle=e^{-iJtH_{ZZ}}|+\rangle^{\otimes L},\qquad
		H_{ZZ}=\sum_{j=1}^{L-1}Z_jZ_{j+1}.
		\label{eq:SM_open_chain_quench}
	\end{equation}
	We write the accumulated Ising angle as
	\begin{equation}
		\vartheta=Jt.
		\label{eq:SM_accumulated_angle}
	\end{equation}
	The open chain is the \textit{path graph} \(G_{\rm path}\). It is a connected tree, so \(c(G_{\rm path})=1\). The forest formula in Eq.~\eqref{eq:SM_forest_formula} gives, for every \(\alpha>0\),
	\begin{equation}
		\boxed{	M_\alpha^{(L)}(\vartheta)=(L-1)m_\alpha(\vartheta)},
		\label{eq:SM_open_chain_all_alpha}
	\end{equation}
	with \(\alpha=1\) understood by continuity. For \(\alpha=2\), using Eq.~\eqref{eq:SM_m2},
	\begin{equation}
		\boxed{	M_2^{(L)}(\vartheta)
			=(L-1)\left[-\log_2\left(1-\frac{1}{4}\sin^2(4\vartheta)\right)\right]}.
		\label{eq:SM_open_chain_m2}
	\end{equation}
	The site density is therefore
	\begin{equation}
		\mu_\alpha^{(L)}(\vartheta)\coloneq\frac{M_\alpha^{(L)}(\vartheta)}{L}
		=\left(1-\frac{1}{L}\right)m_\alpha(\vartheta),
		\label{eq:SM_open_chain_finite_density}
	\end{equation}
	and the thermodynamic-limit density is
	\begin{equation}
		\boxed{	\mu_\alpha(\vartheta)=\lim_{L\to\infty}\frac{M_\alpha^{(L)}(\vartheta)}{L}=m_\alpha(\vartheta)}.
		\label{eq:SM_open_chain_density}
	\end{equation}
	In particular,
	\begin{equation}
		\boxed{	\mu_2(\vartheta)=-\log_2\left(1-\frac{1}{4}\sin^2(4\vartheta)\right)}.
		\label{eq:SM_open_chain_mu2}
	\end{equation}
	
	\subsection{Exact revivals and maximum \(M_2\)}
	
	At the Clifford angles
	\begin{equation}
		\vartheta\in\frac{\pi}{4}\mathbb Z,
		\label{eq:SM_open_chain_clifford_angles}
	\end{equation}
	one has either \(\cos^2(2\vartheta)=1\), \(\sin^2(2\vartheta)=0\), or \(\cos^2(2\vartheta)=0\), \(\sin^2(2\vartheta)=1\). Hence
	\begin{equation}
		1+\cos^{2\alpha}(2\vartheta)+\sin^{2\alpha}(2\vartheta)=2,
		\label{eq:SM_revival_identity}
	\end{equation}
	so \(m_\alpha(\vartheta)=0\). Therefore
	\begin{equation}
		\boxed{	M_\alpha^{(L)}(\vartheta)=0,\qquad
			\vartheta\in\frac{\pi}{4}\mathbb Z,\qquad \alpha>0}.
		\label{eq:SM_open_chain_revivals}
	\end{equation}
	These are exact revivals to the stabilizer manifold. For \(M_2\), the maximum occurs when \(\sin^2(4\vartheta)=1\), namely
	\begin{equation}
		\vartheta=\frac{\pi}{8}\mod \frac{\pi}{4}.
		\label{eq:SM_open_chain_max_angle}
	\end{equation}
	At these angles,
	\begin{equation}
		\boxed{	m_2^{\rm max}=\log_2\frac{4}{3},\qquad
			M_2^{(L),{\rm max}}=(L-1)\log_2\frac{4}{3}}.
		\label{eq:SM_open_chain_max_m2}
	\end{equation}
	
	\subsection{Exact entanglement entropy across an arbitrary cut}
	
	Now consider an arbitrary interior cut of the open chain,
	\begin{equation}
		A=\{1,\ldots,m\},\qquad A^c=\{m+1,\ldots,L\},\qquad 1\leq m\leq L-1.
		\label{eq:SM_open_chain_cut}
	\end{equation}
	All Ising gates commute. Gates entirely within \(A\) or entirely within \(A^c\) are local with respect to this cut and therefore cannot change the entanglement across it. The only gate that crosses the cut is \(e^{-i\vartheta Z_mZ_{m+1}}\). Hence the state can be written as
	\begin{equation}
		|\psi_L(\vartheta)\rangle
		=e^{-i\vartheta Z_mZ_{m+1}}
		\left(|\psi_A(\vartheta)\rangle\otimes|\psi_{A^c}(\vartheta)\rangle\right),
		\label{eq:SM_cut_factorized_form}
	\end{equation}
	where \(|\psi_A(\vartheta)\rangle\) contains only the gates internal to \(A\), \(|\psi_{A^c}(\vartheta)\rangle\) contains only the gates internal to \(A^c\), and both states are normalized because they are obtained from \(|+\rangle^{\otimes L}\) by unitary gates. Expanding the single crossing gate gives
	\begin{equation}
		e^{-i\vartheta Z_mZ_{m+1}}
		=\cos\vartheta\,I-i\sin\vartheta\,Z_mZ_{m+1}.
		\label{eq:SM_crossing_gate_expansion}
	\end{equation}
	Substituting this into Eq.~\eqref{eq:SM_cut_factorized_form} yields
	\begin{equation}
		|\psi_L(\vartheta)\rangle
		=\cos\vartheta\,|\psi_A(\vartheta)\rangle\otimes|\psi_{A^c}(\vartheta)\rangle
		-i\sin\vartheta\,\left(Z_m|\psi_A(\vartheta)\rangle\right)\otimes
		\left(Z_{m+1}|\psi_{A^c}(\vartheta)\rangle\right).
		\label{eq:SM_two_term_schmidt}
	\end{equation}
	To recognize this as a Schmidt decomposition, define
	\begin{equation}
		|a_1\rangle=|\psi_A(\vartheta)\rangle,\qquad
		|a_2\rangle=Z_m|\psi_A(\vartheta)\rangle,
		\label{eq:SM_a_vectors}
	\end{equation}
	and
	\begin{equation}
		|b_1\rangle=|\psi_{A^c}(\vartheta)\rangle,\qquad
		|b_2\rangle=Z_{m+1}|\psi_{A^c}(\vartheta)\rangle.
		\label{eq:SM_b_vectors}
	\end{equation}
	Then Eq.~\eqref{eq:SM_two_term_schmidt} becomes
	\begin{equation}
		|\psi_L(\vartheta)\rangle
		=\cos\vartheta\,|a_1\rangle|b_1\rangle-i\sin\vartheta\,|a_2\rangle|b_2\rangle.
		\label{eq:SM_two_term_compact}
	\end{equation}
	A Schmidt decomposition is an expansion \(|\psi\rangle=\sum_k\lambda_k|a_k\rangle|b_k\rangle\) in which the states on each side are orthonormal. We therefore check orthogonality and normalization separately. For orthogonality on side \(A\),
	\begin{equation}
		\langle a_1|a_2\rangle
		=\langle\psi_A(\vartheta)|Z_m|\psi_A(\vartheta)\rangle.
		\label{eq:SM_overlap_A_start}
	\end{equation}
	Every gate inside \(|\psi_A(\vartheta)\rangle\) is diagonal in the \(Z\) basis and therefore commutes with \(Z_m\). Hence this expectation value is unchanged from its value in the initial state \(|+\rangle^{\otimes L}\). Since \(|+\rangle\) is an eigenstate of \(X\), not of \(Z\), one has \(\langle+|Z|+\rangle=0\), so
	\begin{equation}
		\langle a_1|a_2\rangle=0.
		\label{eq:SM_overlap_A_zero}
	\end{equation}
	The same argument on side \(A^c\) gives
	\begin{equation}
		\langle b_1|b_2\rangle
		=\langle\psi_{A^c}(\vartheta)|Z_{m+1}|\psi_{A^c}(\vartheta)\rangle
		=0.
		\label{eq:SM_overlap_B_zero}
	\end{equation}
	Next we check normalization. Since \(Z_m\) is Hermitian and \(Z_m^2=I\),
	\begin{equation}
		\langle a_2|a_2\rangle
		=\langle\psi_A(\vartheta)|Z_m^\dagger Z_m|\psi_A(\vartheta)\rangle
		=\langle\psi_A(\vartheta)|Z_m^2|\psi_A(\vartheta)\rangle
		=\langle\psi_A(\vartheta)|\psi_A(\vartheta)\rangle
		=1.
		\label{eq:SM_norm_A2}
	\end{equation}
	Also \(\langle a_1|a_1\rangle=1\) by normalization of \(|\psi_A(\vartheta)\rangle\). Thus \(\{|a_1\rangle,|a_2\rangle\}\) is an orthonormal set. Exactly the same computation gives
	\begin{equation}
		\langle b_1|b_1\rangle=1,\qquad
		\langle b_2|b_2\rangle
		=\langle\psi_{A^c}(\vartheta)|Z_{m+1}^2|\psi_{A^c}(\vartheta)\rangle
		=1,
		\label{eq:SM_norm_B}
	\end{equation}
	so \(\{|b_1\rangle,|b_2\rangle\}\) is also orthonormal. Therefore Eq.~\eqref{eq:SM_two_term_compact} is a genuine Schmidt decomposition with Schmidt coefficients
	\begin{equation}
		\lambda_1=\cos\vartheta,\qquad \lambda_2=\sin\vartheta,
		\label{eq:SM_schmidt_coefficients}
	\end{equation}
	where the phase \(-i\) in the second term is irrelevant for the squared Schmidt coefficients. The reduced density matrix on side \(A\) is now immediate:
	\begin{equation}
		\rho_A
		=\mathrm{Tr}_{A^c}\left(|\psi_L(\vartheta)\rangle\langle\psi_L(\vartheta)|\right)
		=\lambda_1^2|a_1\rangle\langle a_1|+\lambda_2^2|a_2\rangle\langle a_2|,
		\label{eq:SM_rhoA_schmidt}
	\end{equation}
	because the cross terms vanish after tracing over \(A^c\), being proportional to \(\langle b_2|b_1\rangle=0\). Substituting Eq.~\eqref{eq:SM_schmidt_coefficients} gives
	\begin{equation}
		\rho_A
		=\cos^2\vartheta\,|a_1\rangle\langle a_1|
		+\sin^2\vartheta\,|a_2\rangle\langle a_2|.
		\label{eq:SM_rhoA_diagonal}
	\end{equation}
	Since this is already diagonal in an orthonormal basis, its eigenvalues are
	\begin{equation}
		\cos^2\vartheta,\qquad \sin^2\vartheta.
		\label{eq:SM_rhoA_eigenvalues}
	\end{equation}
	The von Neumann entanglement entropy is therefore
	\begin{equation}
		S_{\rm EE}^{(L)}(\vartheta)
		=-\mathrm{Tr}(\rho_A\log_2\rho_A) =-\cos^2\vartheta\log_2\cos^2\vartheta-\sin^2\vartheta\log_2\sin^2\vartheta.
		\label{eq:SM_open_chain_EE_expanded}
	\end{equation}
	Writing \(p=\sin^2\vartheta\), so that \(1-p=\cos^2\vartheta\), this becomes
	\begin{equation}
		\boxed{ S_{\rm EE}^{(L)}(\vartheta)=h_2(\sin^2\vartheta)},
		\label{eq:SM_open_chain_EE}
	\end{equation}
	where
	\begin{equation}
		h_2(p)=-p\log_2p-(1-p)\log_2(1-p).
		\label{eq:SM_binary_entropy}
	\end{equation}
	This result is independent of \(L\) and of the cut position, because neither the cut location \(m\) nor the system size \(L\) entered anywhere in the derivation above, only the single crossing gate. More generally, the same closed form holds for any initial state for which the two overlaps in Eqs.~\eqref{eq:SM_overlap_A_zero} and \eqref{eq:SM_overlap_B_zero} vanish; if they do not, one must diagonalize the resulting two-by-two reduced density matrix instead of reading off \(\cos^2\vartheta\) and \(\sin^2\vartheta\) directly.
	
	\subsection{Magic-entanglement contrast at Clifford angles}
	
	The magic and entanglement have different periods. From Eq.~\eqref{eq:SM_open_chain_m2}, \(M_2\) has period \(\pi/4\) in \(\vartheta\). From Eq.~\eqref{eq:SM_open_chain_EE}, \(S_{\rm EE}\) has period \(\pi/2\) in \(\vartheta\). At Clifford angles \(\vartheta=k\pi/4\),
	\begin{equation}
		\boxed{	S_{\rm EE}^{(L)}\left(\frac{k\pi}{4}\right)
			=h_2\left(\sin^2\frac{k\pi}{4}\right)
			=
			\begin{cases}
				0, & k\ \mathrm{even},\\
				1, & k\ \mathrm{odd}.
		\end{cases} }
		\label{eq:SM_EE_clifford_angles}
	\end{equation}
	Thus at odd Clifford angles, the magic is exactly zero while the entanglement across every interior cut is exactly one ebit. At the magic maximum,
	\begin{equation}
		\vartheta=\frac{\pi}{8},
		\label{eq:SM_magic_max_angle_EE}
	\end{equation}
	the entanglement entropy is
	\begin{equation}
		S_{\rm EE}^{(L)}\left(\frac{\pi}{8}\right)
		=h_2\left(\frac{2-\sqrt2}{4}\right)\simeq0.6009.
		\label{eq:SM_EE_at_magic_max}
	\end{equation}
	The thermodynamic-limit contrast is sharper:
	\begin{equation}
		\boxed{	\lim_{L\to\infty}\frac{M_2^{(L)}(\vartheta)}{L}=m_2(\vartheta),
			\qquad
			\lim_{L\to\infty}\frac{S_{\rm EE}^{(L)}(\vartheta)}{L}=0 }.
		\label{eq:SM_magic_EE_density_contrast}
	\end{equation}
	For generic \(\vartheta\), the first quantity is nonzero while the second is zero. This is the one-dimensional realization of magic without entanglement density.
	

	\section{Kicked-Ising Floquet dynamics}
	
	\subsection{The Clifford kick at \(g=\pi/2\)}
	
	The kicked-Ising unitary is
	\begin{equation}
		U_F(g,\theta)=e^{-i\theta H_{ZZ}}e^{-igH_X},\qquad
		H_X=\sum_{j=1}^{L}X_j.
		\label{eq:SM_floquet_unitary}
	\end{equation}
	At \(g=\pi/2\),
	\begin{equation}
		e^{-i(\pi/2)H_X}
		=\prod_{j=1}^{L}e^{-i(\pi/2)X_j}
		=(-i)^L\prod_{j=1}^{L}X_j.
		\label{eq:SM_pi_over_two_kick}
	\end{equation}
	This kick is Clifford. The product \(\prod_{j=1}^{L}X_j\), not each individual \(X_j\), commutes with every Ising bond \(Z_\ell Z_{\ell+1}\). Indeed, \(Z_\ell Z_{\ell+1}\) anticommutes with \(X_\ell\) and with \(X_{\ell+1}\), producing two minus signs:
	\begin{equation}
		\left(\prod_{j=1}^{L}X_j\right)Z_\ell Z_{\ell+1}
		=Z_\ell Z_{\ell+1}\left(\prod_{j=1}^{L}X_j\right).
		\label{eq:SM_global_X_commutes}
	\end{equation}
	Moreover,
	\begin{equation}
		\left(\prod_{j=1}^{L}X_j\right)|+\rangle^{\otimes L}=|+\rangle^{\otimes L}.
		\label{eq:SM_global_X_plus}
	\end{equation}
	Hence one kick at \(g=\pi/2\) differs from \(e^{-i\theta H_{ZZ}}\) only by a commuting Clifford factor and an overall phase. Because this factor also acts trivially on \(|+\rangle^{\otimes L}\), repeated kicks give, up to an overall phase,
	\begin{equation}
		U_F(\pi/2,\theta)^k|+\rangle^{\otimes L}
		=e^{-ik\theta H_{ZZ}}|+\rangle^{\otimes L}.
		\label{eq:SM_floquet_to_quench}
	\end{equation}
	Thus the stroboscopic state at \(g=\pi/2\) is equivalent to the open-chain quench at accumulated angle
	\begin{equation}
		\vartheta=k\theta.
		\label{eq:SM_floquet_accumulated_angle}
	\end{equation}
	
	\subsection{Exact stroboscopic magic and entanglement}
	
	Using Eq.~\eqref{eq:SM_open_chain_all_alpha} with \(\vartheta=k\theta\), the exact stroboscopic SRE is
	\begin{equation}
		\boxed{	M_\alpha^{(L),{\rm F}}(k)=(L-1)m_\alpha(k\theta) },
		\label{eq:SM_floquet_all_alpha}
	\end{equation}
	for every \(\alpha>0\), with \(\alpha=1\) understood by continuity. In particular,
	\begin{equation}
		\boxed{	M_{2,{\rm F}}^{(L)}(k)
			=(L-1)\left[-\log_2\left(1-\frac{1}{4}\sin^2(4k\theta)\right)\right] }.
		\label{eq:SM_floquet_m2}
	\end{equation}
	The thermodynamic-limit stroboscopic magic density is
	\begin{equation}
		\boxed{	\mu_2^{\rm F}(k)=m_2(k\theta)}.
		\label{eq:SM_floquet_density}
	\end{equation}
	The same substitution gives the entanglement entropy across any interior cut,
	\begin{equation}
		\boxed{	S_{\rm EE}^{(L),{\rm F}}(k)=h_2(\sin^2(k\theta))}.
		\label{eq:SM_floquet_EE}
	\end{equation}
	Thus the Floquet problem at \(g=\pi/2\) inherits the exact quench formulas, with the continuous angle \(\vartheta\) replaced by the discrete accumulated angle \(k\theta\).
	
	\subsection{Periodicity at \(\theta=\pi/8\)}
	
	We now choose
	\begin{equation}
		\theta=\frac{\pi}{8}.
		\label{eq:SM_theta_pi_over_8}
	\end{equation}
	This choice is not arbitrary. It is the smallest positive kick angle with two useful properties at once. After one kick, the accumulated angle is \(\vartheta=\pi/8\), where \(m_2(\vartheta)\) is maximal. After two kicks, the accumulated angle is \(\vartheta=\pi/4\), the first nontrivial Clifford revival angle at which the magic vanishes exactly. More generally, any \(\theta=n\pi/8\) makes the two-kick angle \(2\theta=n\pi/4\) a Clifford angle, while the odd multiples \(\theta=(2n+1)\pi/8\) also place the one-kick state at the magic maximum.
	
	At \(k=0\), the initial stabilizer state has zero magic. After one kick the magic is nonzero, while after two kicks the accumulated angle reaches the Clifford value \(\pi/4\), so the magic returns exactly to zero:
	\begin{equation}
		M_{2,{\rm F}}^{(L)}(k=2)=0.
		\label{eq:SM_two_kick_magic_revival}
	\end{equation}
	The reason the magic sequence then has period two is simply that increasing \(k\) by \(2\) shifts the accumulated angle by
	\begin{equation}
		(k+2)\theta-k\theta=2\theta=\frac{\pi}{4},
		\label{eq:SM_magic_period_shift}
	\end{equation}
	which is exactly one period of \(m_2(\vartheta)\). Hence
	\begin{equation}
		\boxed{	m_2\left((k+2)\frac{\pi}{8}\right)=m_2\left(k\frac{\pi}{8}\right) }.
		\label{eq:SM_magic_period_two}
	\end{equation}
	For the entanglement entropy, increasing \(k\) by \(4\) shifts the accumulated angle by
	\begin{equation}
		(k+4)\theta-k\theta=4\theta=\frac{\pi}{2},
		\label{eq:SM_EE_period_shift}
	\end{equation}
	which is the period of \(h_2(\sin^2\vartheta)\). Thus
	\begin{equation}
		\boxed{	S_{\rm EE}^{(L),{\rm F}}(k)=h_2\left(\sin^2\frac{k\pi}{8}\right) = h_2\left(\sin^2\frac{(k+4)\pi}{8}\right)},
		\label{eq:SM_floquet_EE_pi_over_8}
	\end{equation}
	\textit{has period four in kick number. Within the same stroboscopic cycle, the magic therefore revives twice as fast as the entanglement entropy.}
	
	At \(g=\pi/2\) and \(\theta=\pi/8\), the eighth power of the Floquet unitary is a global phase. Starting from Eq.~\eqref{eq:SM_pi_over_two_kick},
	\begin{equation}
		\begin{aligned}
			U_F(\pi/2,\pi/8)^8
			&=\left(e^{-i(\pi/8)H_{ZZ}}e^{-i(\pi/2)H_X}\right)^8\\
			&\propto\left(e^{-i(\pi/8)H_{ZZ}}\prod_{j=1}^{L}X_j\right)^8\\
			&\propto e^{-i\pi H_{ZZ}}\left(\prod_{j=1}^{L}X_j\right)^8\\
			&=e^{-i\pi H_{ZZ}},
		\end{aligned}
		\label{eq:SM_eight_kick_step}
	\end{equation}
	where we used that \(\prod_jX_j\) commutes with \(H_{ZZ}\) and that \((\prod_jX_j)^8=I\). Since all Ising bonds commute,
	\begin{equation}
		\boxed{ e^{-i\pi H_{ZZ}}
			=\prod_{j=1}^{L-1}e^{-i\pi Z_jZ_{j+1}}
			=\prod_{j=1}^{L-1}(-I)
			=(-1)^{L-1}I
			\propto I }. 
		\label{eq:SM_eight_kick_identity}
	\end{equation}
	\textit{Thus eight kicks return the system to itself up to a global phase.} The exact period structure is therefore a Clifford-point property. Away from this point, the later perturbative sections show that the magic revival is lifted quadratically.
	
	
	\section{Initial curvature and QFI benchmarks}
	
	\subsection{Initial variance of \(H_{ZZ}\) and \(H_{\rm TFIM}\)}
	
	The transverse-field Ising Hamiltonian is
	\begin{equation}
		H_{\rm TFIM}=JH_{ZZ}+hH_X,\qquad
		H_X=\sum_{j=1}^{L}X_j.
		\label{eq:SM_TFIM_initial_curvature}
	\end{equation}
	We evaluate its initial quantum Fisher information in the state \(|+\rangle^{\otimes L}\). For a pure state, the quantum Fisher information of a generator \(K\) is
	\begin{equation}
		F_Q\bigl(|\psi\rangle,K\bigr)=4\,\operatorname{Var}_{\psi}(K),
	\end{equation}
	so the task is simply to compute the variance of \(H_{\rm TFIM}\) in \(|+\rangle^{\otimes L}\).
	
	We begin with the transverse-field term. Since each site satisfies \(X_j|+\rangle=|+\rangle\), the product state \(|+\rangle^{\otimes L}\) is an eigenstate of \(H_X\):
	\begin{equation}
		H_X|+\rangle^{\otimes L}=L|+\rangle^{\otimes L}.
		\label{eq:SM_HX_eigenstate}
	\end{equation}
	It follows immediately that
	\begin{equation}
		\langle H_X\rangle_+=L,\qquad
		\langle H_X^2\rangle_+=L^2,
	\end{equation}
	and hence
	\begin{equation}
		\operatorname{Var}_+(H_X)
		=
		\langle H_X^2\rangle_+-\langle H_X\rangle_+^2
		=
		0.
	\end{equation}
	Thus \(H_X\) contributes no variance in the initial state.
	
	We now turn to the Ising part. Write
	\begin{equation}
		H_{ZZ}=\sum_{j=1}^{L-1}B_j,\qquad B_j\coloneq Z_jZ_{j+1}.
	\end{equation}
	Since \(|+\rangle^{\otimes L}\) is a product state and \(\langle +|Z|+\rangle=0\), each bond has zero expectation value:
	\begin{equation}
		\langle B_j\rangle_+
		=
		\langle Z_jZ_{j+1}\rangle_+
		=
		\langle +|Z|+\rangle\,\langle +|Z|+\rangle
		=
		0.
	\end{equation}
	Moreover, each bond squares to the identity,
	\begin{equation}
		B_j^2=(Z_jZ_{j+1})^2=Z_j^2Z_{j+1}^2=I,
	\end{equation}
	so
	\begin{equation}
		\langle Z_jZ_{j+1}\rangle_+=0,\qquad
		\langle (Z_jZ_{j+1})^2\rangle_+=1.
		\label{eq:SM_bond_expectations}
	\end{equation}
	
	To compute the full variance, expand
	\begin{equation}
		H_{ZZ}^2
		=
		\sum_{j=1}^{L-1}B_j^2
		+
		\sum_{j\neq \ell}B_jB_\ell.
	\end{equation}
	The diagonal terms are simple: \(B_j^2=I\), so each contributes \(1\) in expectation value. The only question is whether any cross term survives. For \(j\neq \ell\), the product \(B_jB_\ell\) always has zero expectation value in \(|+\rangle^{\otimes L}\). If the two bonds are disjoint, the expectation factorizes and vanishes because each bond has zero expectation. If they are adjacent, for example \(\ell=j+1\), then
	\begin{equation}
		B_jB_{j+1}
		=
		(Z_jZ_{j+1})(Z_{j+1}Z_{j+2})
		=
		Z_jZ_{j+2},
	\end{equation}
	whose expectation also vanishes because it still contains a single \(Z\) on site \(j\) and a single \(Z\) on site \(j+2\), each with \(\langle+|Z|+\rangle=0\). Equivalently, every product \(B_jB_\ell\) with \(j\neq \ell\) contains at least one site on which an uncancelled \(Z\) acts once, and that is enough to make its expectation vanish in the \(|+\rangle\) product state. Therefore
	\begin{equation}
		\langle (Z_jZ_{j+1})(Z_\ell Z_{\ell+1})\rangle_+=0,\qquad j\neq\ell.
		\label{eq:SM_bond_cross_terms}
	\end{equation}
	
	Using Eqs.~\eqref{eq:SM_bond_expectations} and \eqref{eq:SM_bond_cross_terms}, we obtain
	\begin{equation}
		\langle H_{ZZ}\rangle_+
		=
		\sum_{j=1}^{L-1}\langle B_j\rangle_+
		=
		0,
	\end{equation}
	and
	\begin{equation}
		\langle H_{ZZ}^2\rangle_+
		=
		\sum_{j=1}^{L-1}\langle B_j^2\rangle_+
		+
		\sum_{j\neq\ell}\langle B_jB_\ell\rangle_+
		=
		L-1.
	\end{equation}
	Hence
	\begin{equation}
		\operatorname{Var}_+(H_{ZZ})
		=
		\langle H_{ZZ}^2\rangle_+-\langle H_{ZZ}\rangle_+^2
		=
		L-1.
		\label{eq:SM_var_HZZ}
	\end{equation}
	
	It remains to combine the two pieces. Expanding the variance of
	\(H_{\rm TFIM}=JH_{ZZ}+hH_X\) gives
	\begin{equation}
		\operatorname{Var}_+(H_{\rm TFIM})
		=
		J^2\operatorname{Var}_+(H_{ZZ})
		+
		h^2\operatorname{Var}_+(H_X)
		+
		2Jh\,\operatorname{Cov}_+(H_{ZZ},H_X).
	\end{equation}
	We already know that \(\operatorname{Var}_+(H_X)=0\). The covariance also vanishes because \(H_X\) acts as the scalar \(L\) on \(|+\rangle^{\otimes L}\):
	\begin{equation}
		\langle H_{ZZ}H_X\rangle_+
		=
		L\langle H_{ZZ}\rangle_+,
		\qquad
		\langle H_{ZZ}\rangle_+\langle H_X\rangle_+
		=
		L\langle H_{ZZ}\rangle_+,
	\end{equation}
	so
	\begin{equation}
		\operatorname{Cov}_+(H_{ZZ},H_X)=0.
	\end{equation}
	Therefore only the Ising variance remains, and
	\begin{equation}
		\operatorname{Var}_+(H_{\rm TFIM})=J^2(L-1).
		\label{eq:SM_var_TFIM}
	\end{equation}
	
	Finally, using \(F_Q=4\,\operatorname{Var}\) for pure states, we arrive at
	\begin{equation}
		\boxed{	F_Q(|+\rangle^{\otimes L},H_{\rm TFIM})=4J^2(L-1) }.
		\label{eq:SM_initial_QFI}
	\end{equation}
	This is the same initial QFI quoted in the main text: the transverse field does not contribute because the initial state is already an eigenstate of \(H_X\), so all of the initial curvature comes from the commuting Ising bonds.
	
	\subsection{Magic curvature from the exact formula}
	
	The exact open-chain formula gives
	\begin{equation}
		M_2^{(L)}(t,h=0)
		=(L-1)\left[-\log_2\left(1-\frac{1}{4}\sin^2(4Jt)\right)\right].
		\label{eq:SM_exact_curvature_formula}
	\end{equation}
	For small \(t\),
	\begin{equation}
		\sin(4Jt)=4Jt+O(t^3),
		\label{eq:SM_sin_small_t}
	\end{equation}
	so
	\begin{equation}
		\frac{1}{4}\sin^2(4Jt)=4J^2t^2+O(t^4).
		\label{eq:SM_sin_squared_small_t}
	\end{equation}
	Using
	\begin{equation}
		-\log_2(1-x)=\frac{x}{\ln2}+O(x^2),
		\label{eq:SM_log_expansion}
	\end{equation}
	we obtain
	\begin{equation}
		M_2^{(L)}(t,h=0)=\frac{4J^2(L-1)}{\ln2}t^2+O(t^4).
		\label{eq:SM_magic_curvature_h0}
	\end{equation}
	The tangent theorem extends the same quadratic coefficient to \(H_{\rm TFIM}\), because the transverse field has zero initial variance and zero initial covariance in \(|+\rangle^{\otimes L}\):
	\begin{equation}
		\boxed{	M_2^{(L)}(t,h)=\frac{4J^2(L-1)}{\ln2}t^2+O(t^3)}.
		\label{eq:SM_magic_curvature_TFIM}
	\end{equation}
	Equivalently,
	\begin{equation}
		\boxed{	(\ln2)*\lim_{t\to0}\frac{M_2^{(L)}(t,h)}{t^2}=4J^2(L-1)=F_Q(|+\rangle^{\otimes L},H_{\rm TFIM}) }.
		\label{eq:SM_magic_QFI_agreement}
	\end{equation}

	\subsection{Connection to the independent QFI checks}
	
	The model-specific result in Eq.~\eqref{eq:SM_magic_QFI_agreement} is checked using the three independent QFI routes described in Sec.~\ref{subsec:SM_independent_QFI_benchmarks}: the direct variance, the squared-fidelity curvature, and the zero-temperature susceptibility. For the initial TFIM curvature, all four routes give the same value,
	\begin{equation}
		F_Q=4J^2(L-1).
		\label{eq:SM_QFI_all_match}
	\end{equation}
	This agreement verifies that the coefficient of the initial \(M_2\) growth is not extracted only from magic data, but also from standard QFI diagnostics.

	
	
	
	
	\section{Perturbative lifting of the quench revival}
	\label{sec:app_perturbative_quench}
	
	\subsection{Interaction-picture generator at \(t_\star=\pi/(4J)\)}
	
	At \(h=0\), the first nontrivial open-chain magic revival occurs at
	\begin{equation}
		t_\star=\frac{\pi}{4J},
		\qquad
		Jt_\star=\frac{\pi}{4}.
		\label{eq:SM_tstar}
	\end{equation}
	At this time, the unperturbed Ising evolution is Clifford, and therefore maps \(|+\rangle^{\otimes L}\) to a stabilizer state. To study how this revival is lifted by a small transverse field, we use
	\begin{equation}
		H_{\rm TFIM}=JH_{ZZ}+hH_X,
		\qquad
		H_X=\sum_{j=1}^{L}X_j.
		\label{eq:SM_TFIM_lifting}
	\end{equation}
	Set
	\begin{equation}
		\eta=\frac{h}{J},
		\qquad
		u=Js.
		\label{eq:SM_eta_u}
	\end{equation}
	
	We now rewrite the evolution in the interaction picture with respect to the unperturbed Ising part \(JH_{ZZ}\). In general, for a decomposition \(H=H_0+V\), one factors the evolution as \(U(t)=e^{-iH_0t}U_I(t)\), where \(U_I(t)\) evolves with the rotated perturbation \(V_I(t)=e^{iH_0t}Ve^{-iH_0t}\). Equivalently, \(U_I(t)\) is defined by the differential equation \(i\,\partial_t U_I(t)=V_I(t)U_I(t)\) with initial condition \(U_I(0)=I\), whose formal solution is the time-ordered exponential. Here \(H_0=JH_{ZZ}\) and \(V=hH_X\). Starting from
	\begin{equation}
		U(t,h)=e^{-it(JH_{ZZ}+hH_X)},
	\end{equation}
	this gives
	\begin{equation}
		U(t,h)
		=
		e^{-iJtH_{ZZ}}\,
		\mathcal T\exp\!\left[-ih\int_0^t ds\,H_X(s)\right],
	\end{equation}
	where
	\begin{equation}
		H_X(s)=e^{iJsH_{ZZ}}H_Xe^{-iJsH_{ZZ}}.
	\end{equation}
	Evaluating this at \(t=t_\star\), and using \(Jt_\star=\pi/4\), gives
	\begin{equation}
		U(t_\star,h)
		=
		e^{-i(\pi/4)H_{ZZ}}\,
		\mathcal T\exp\!\left[-ih\int_0^{t_\star} ds\,H_X(s)\right].
	\end{equation}
	Finally, changing variables from \(s\) to \(u=Js\), so that \(ds=du/J\) and the upper limit becomes \(u=Jt_\star=\pi/4\), yields
	\begin{equation}
		U(t_\star,h)
		=e^{-i(\pi/4)H_{ZZ}}\,
		\mathcal T\exp\left[-i\eta\int_0^{\pi/4}du\,H_X(u)\right],
		\label{eq:SM_interaction_picture_quench}
	\end{equation}
	where
	\begin{equation}
		H_X(u)=e^{iuH_{ZZ}}H_Xe^{-iuH_{ZZ}}.
		\label{eq:SM_HX_u}
	\end{equation}
	
	The first factor in Eq.~\eqref{eq:SM_interaction_picture_quench} is Clifford at the revival. By Clifford invariance, it does not affect the SRE. Thus, to leading order in the small parameter \(\eta=h/J\), the only operator that matters is the integrated interaction-picture generator
	\begin{equation}
		K_L=\int_0^{\pi/4}du\,H_X(u).
		\label{eq:SM_KL_definition}
	\end{equation}
	Indeed, expanding the time-ordered exponential to first order gives
	\begin{equation}
		\mathcal T\exp\left[-i\eta\int_0^{\pi/4}du\,H_X(u)\right]
		=
		I-i\eta K_L+O(\eta^2).
	\end{equation}
	After stripping off the Clifford prefactor, the state at the revival is therefore a small unitary deformation of a stabilizer state generated by \(K_L\). Equation~\eqref{eq:SM_tangent_formula} then applies with \(\theta=\eta\) and \(K=K_L\). Since for a pure state \(F_Q=4(\langle K_L^2\rangle_+-\langle K_L\rangle_+^2)\), the tangent formula gives
	\begin{equation}
		M_2^{(L)}(t_\star,h)
		=\frac{4}{\ln2}\eta^2 A_L+O(\eta^3),
		\qquad
		A_L=\langle K_L^2\rangle_+-\langle K_L\rangle_+^2 ,
		\label{eq:SM_quench_lifting_AL}
	\end{equation}
	where \(\langle\cdots\rangle_+\) denotes expectation in \(|+\rangle^{\otimes L}\).
	
	It remains to compute \(K_L\), so we next determine \(H_X(u)\). Since \(H_X=\sum_{j=1}^L X_j\), this reduces to conjugating each single-site operator \(X_j\) by the Ising evolution. Only the bonds touching site \(j\) matter; all other bonds commute with \(X_j\) and therefore drop out.
	
	For this purpose, we use the following elementary identity. Let \(A\) be a Pauli string with \(A^2=I\), and let \(B\) anticommute with \(A\), so that \(AB=-BA\). Since
	\begin{equation}
		e^{iuA}=\cos u\,I+i\sin u\,A, \qquad (A^2 = I)
	\end{equation}
	one finds
	\begin{equation}
		\begin{aligned}
			e^{iuA}Be^{-iuA}
			&=(\cos u\,I+i\sin u\,A)\,B\,(\cos u\,I-i\sin u\,A)\\
			&=B(\cos^2u-\sin^2u)+iAB(2\sin u\cos u)\\
			&=B\cos(2u)+iAB\sin(2u).
		\end{aligned}
		\label{eq:SM_anticommuting_conjugation}
	\end{equation}
	If \(A\) commutes with \(B\), the operator is unchanged under conjugation.
	
	As a first example, consider the left edge operator \(X_1\). Only the bond \(Z_1Z_2\) touches site \(1\), so
	\begin{equation}
		X_1(u)=e^{iuZ_1Z_2}X_1e^{-iuZ_1Z_2}.
	\end{equation}
	Using Eq.~\eqref{eq:SM_anticommuting_conjugation} with \(A=Z_1Z_2\) and \(B=X_1\), together with the single-site Pauli rule \(ZX=iY\) on site \(1\) and the fact that operators on different sites commute,
	\begin{equation}
		(Z_1Z_2)X_1=(Z_1X_1)Z_2=iY_1Z_2,
	\end{equation}
	gives
	\begin{equation}
		X_1(u) = e^{iuZ_1Z_2}X_1e^{-iuZ_1Z_2}
		=\cos(2u)X_1-\sin(2u)Y_1Z_2.
		\label{eq:SM_edge_conjugation_example}
	\end{equation}
	The same reasoning at the right edge gives
	\begin{equation}
		X_L(u)=\cos(2u)X_L-\sin(2u)Z_{L-1}Y_L.
		\label{eq:SM_right_edge_X_u}
	\end{equation}
	
	For a bulk site \(2\leq j\leq L-1\), there are exactly two adjacent bonds, \(Z_{j-1}Z_j\) and \(Z_jZ_{j+1}\). These two bond operators commute with each other, so we may conjugate successively:
	\begin{equation}
		X_j(u)
		=
		e^{iuZ_{j-1}Z_j}e^{iuZ_jZ_{j+1}}
		X_j
		e^{-iuZ_jZ_{j+1}}e^{-iuZ_{j-1}Z_j}.
	\end{equation}
	It is convenient to do this in two steps. First conjugate by the right bond:
	\begin{equation}
		e^{iuZ_jZ_{j+1}}X_je^{-iuZ_jZ_{j+1}}
		=
		\cos(2u)X_j-\sin(2u)Y_jZ_{j+1}.
	\end{equation}
	Now conjugate each term by the left bond \(Z_{j-1}Z_j\). The first term gives
	\begin{equation}
		e^{iuZ_{j-1}Z_j}X_je^{-iuZ_{j-1}Z_j}
		=
		\cos(2u)X_j-\sin(2u)Z_{j-1}Y_j.
	\end{equation}
	For the second term, note that \(Z_{j-1}Z_j\) anticommutes with \(Y_jZ_{j+1}\), and
	\begin{equation}
		(Z_{j-1}Z_j)(Y_jZ_{j+1})=-i\,Z_{j-1}X_jZ_{j+1},
	\end{equation}
	so
	\begin{equation}
		e^{iuZ_{j-1}Z_j}(Y_jZ_{j+1})e^{-iuZ_{j-1}Z_j}
		=
		\cos(2u)Y_jZ_{j+1}-\sin(2u)Z_{j-1}X_jZ_{j+1}.
	\end{equation}
	Substituting these two pieces and collecting terms gives
	\begin{equation}
		X_j(u)=\cos^2(2u)X_j-\cos(2u)\sin(2u)(Z_{j-1}Y_j+Y_jZ_{j+1})-\sin^2(2u)Z_{j-1}X_jZ_{j+1}.
		\label{eq:SM_bulk_X_u}
	\end{equation}
	
	Using Eqs.~\eqref{eq:SM_edge_conjugation_example}, \eqref{eq:SM_right_edge_X_u}, and \eqref{eq:SM_bulk_X_u}, we can now integrate term by term to obtain \(K_L\). The needed integrals are
	\begin{equation}
		\int_0^{\pi/4}du\,\cos(2u)=\frac{1}{2},
		\qquad
		\int_0^{\pi/4}du\,\sin(2u)=\frac{1}{2},
		\label{eq:SM_linear_integrals}
	\end{equation}
	and
	\begin{equation}
		\int_0^{\pi/4}du\,\cos^2(2u)=\frac{\pi}{8},
		\qquad
		\int_0^{\pi/4}du\,\sin^2(2u)=\frac{\pi}{8},
		\qquad
		\int_0^{\pi/4}du\,\cos(2u)\sin(2u)=\frac{1}{4}.
		\label{eq:SM_quadratic_integrals}
	\end{equation}
	Using Eq.~\eqref{eq:SM_KL_definition}, the two edge contributions to \(K_L\) are
	\begin{equation}
		\int_0^{\pi/4}du\,X_1(u)
		=
		\frac{1}{2}X_1-\frac{1}{2}Y_1Z_2,
		\qquad
		\int_0^{\pi/4}du\,X_L(u)
		=
		\frac{1}{2}X_L-\frac{1}{2}Z_{L-1}Y_L,
		\label{eq:SM_KL_edges}
	\end{equation}
	while, for each bulk site \(2\leq j\leq L-1\),
	\begin{equation}
		\int_0^{\pi/4}du\,X_j(u)
		=
		\frac{\pi}{8}X_j-\frac{1}{4}(Z_{j-1}Y_j+Y_jZ_{j+1})-\frac{\pi}{8}Z_{j-1}X_jZ_{j+1}.
		\label{eq:SM_KL_bulk}
	\end{equation}
	
	\subsection{Evaluation of \(A_L\)}
	
	The expectation rule in \(|+\rangle^{\otimes L}\) is simple:
	\begin{equation}
		\langle P\rangle_+\neq0
		\quad\text{only if every site of \(P\) is either \(I\) or \(X\)}.
		\label{eq:SM_plus_expectation_rule}
	\end{equation}
	In that case \(\langle P\rangle_+=1\). Equivalently, on a single site one has
	\begin{equation}
		\langle +|I|+\rangle=1,
		\qquad
		\langle +|X|+\rangle=1,
		\qquad
		\langle +|Y|+\rangle=0,
		\qquad
		\langle +|Z|+\rangle=0,
	\end{equation}
	and the expectation of a Pauli string factorizes over sites. Hence a string contributes only if no site carries a \(Y\) or a \(Z\).
	
	This rule immediately implies that the pure \(X\)-strings in \(K_L\) contribute to \(\langle K_L\rangle_+\), but they cancel out of the variance after subtracting \(\langle K_L\rangle_+^2\). It is therefore convenient to split
	\begin{equation}
		K_L=K_L^{(X)}+K_L^{({\rm nX})},
	\end{equation}
	where \(K_L^{(X)}\) contains the pure \(X\)-strings and \(K_L^{({\rm nX})}\) contains all other strings. Then
	\begin{equation}
		\langle K_L^{({\rm nX})}\rangle_+=0,
		\qquad
		\langle K_L^{(X)}K_L^{({\rm nX})}\rangle_+=0,
	\end{equation}
	so the variance reduces to
	\begin{equation}
		A_L=\langle (K_L^{({\rm nX})})^2\rangle_+.
	\end{equation}
	
	To see this explicitly, note first that every pure \(X\)-string \(P\) satisfies \(P|+\rangle^{\otimes L}=|+\rangle^{\otimes L}\). Therefore \(K_L^{(X)}\) acts diagonally on \(|+\rangle^{\otimes L}\),
	\begin{equation}
		K_L^{(X)}|+\rangle^{\otimes L}
		=
		\langle K_L^{(X)}\rangle_+\,|+\rangle^{\otimes L},
	\end{equation}
	and hence
	\begin{equation}
		\langle (K_L^{(X)})^2\rangle_+
		=
		\langle K_L^{(X)}\rangle_+^2.
	\end{equation}
	Thus the pure-\(X\) sector has zero variance. Next, every term in \(K_L^{({\rm nX})}\) contains at least one \(Y\) or \(Z\), so \(\langle K_L^{({\rm nX})}\rangle_+=0\). Moreover, multiplying a pure \(X\)-string by a non-\(X\) string cannot remove all \(Y\)'s and \(Z\)'s, because on one site
	\begin{equation}
		XY=iZ,\qquad XZ=-iY,
	\end{equation}
	so a factor \(Y\) or \(Z\) remains a \(Y\) or \(Z\) after multiplication by \(X\). Hence the mixed expectations also vanish, and only the non-\(X\) sector contributes to \(A_L\).
	
	Thus only the non-\(X\) strings in Eqs.~\eqref{eq:SM_KL_edges} and \eqref{eq:SM_KL_bulk} matter, together with those products of two such strings that simplify to a string made only of \(I\) and \(X\).
	
	For \(L\geq3\), the non-\(X\) strings are:
	\begin{itemize}
		\item the two edge strings \(-\frac12 Y_1Z_2\) and \(-\frac12 Z_{L-1}Y_L\);
		\item for each bulk site \(2\leq j\leq L-1\), the three strings
		\(-\frac14 Z_{j-1}Y_j\), \(-\frac14 Y_jZ_{j+1}\), and \(-\frac{\pi}{8}Z_{j-1}X_jZ_{j+1}\).
	\end{itemize}
	The direct-square contribution comes from squaring each of these strings individually. Since every Pauli string squares to the identity, each square contributes the square of its coefficient. Therefore
	\begin{equation}
		A_L^{\rm dir}
		=2\left(\frac{1}{2}\right)^2
		+(L-2)\left[
		2\left(\frac{1}{4}\right)^2
		+\left(\frac{\pi}{8}\right)^2
		\right].
		\label{eq:SM_AL_direct}
	\end{equation}
	
	We next look for nonzero cross terms. Most products of two distinct non-\(X\) strings still contain at least one \(Y\) or \(Z\), and therefore have zero expectation in \(|+\rangle^{\otimes L}\). The only surviving contractions are the adjacent ones in which a \(Y_jZ_{j+1}\) term from site \(j\) multiplies a \(Z_jY_{j+1}\) term from site \(j+1\):
	\begin{equation}
		(Y_jZ_{j+1})(Z_jY_{j+1})=X_jX_{j+1}.
		\label{eq:SM_adjacent_contraction}
	\end{equation}
	Here the multiplication is done site by site, using the fact that operators on different sites commute:
	\begin{equation}
		\begin{aligned}
			(Y_jZ_{j+1})(Z_jY_{j+1})
			&=(Y_jZ_j)(Z_{j+1}Y_{j+1})\\
			&=(iX_j)(-iX_{j+1})\\
			&=X_jX_{j+1}.
		\end{aligned}
	\end{equation}
	In the second line we used the single-site Pauli rules \(YZ=iX\) and \(ZY=-iX\). This resulting string contains only \(X\)'s, so its expectation is \(1\).
	
	There are two kinds of such contractions. At the left boundary, the string \(-\frac12 Y_1Z_2\) from the edge term contracts with the string \(-\frac14 Z_1Y_2\) from the \(j=2\) bulk term. Since both orderings appear in \(K_L^2\), the contribution is $2\left(\frac12\right)\left(\frac14\right)=\frac14$.

	The same happens at the right boundary, giving another \(\frac14\). In the bulk, for each \(j=2,\dots,L-2\), the string \(-\frac14 Y_jZ_{j+1}\) from site \(j\) contracts with the string \(-\frac14 Z_jY_{j+1}\) from site \(j+1\). Each such pair contributes $2\left(\frac14\right)\left(\frac14\right)=\frac18$.
	
	There are \(L-3\) such bulk pairs. Hence
	\begin{equation}
		A_L^{\rm cross}
		=\frac{1}{4}+\frac{L-3}{8}+\frac{1}{4}
		=\frac{L+1}{8}.
		\label{eq:SM_AL_cross}
	\end{equation}
	
	Combining the direct and cross pieces gives
	\begin{equation}
		A_L=\frac{L}{4}+\frac{3}{8}+\frac{(L-2)\pi^2}{64},
		\qquad L\geq3.
		\label{eq:SM_AL_Lge3}
	\end{equation}
	Equivalently,
	\begin{equation}
		A_L=\frac{16L+24+(L-2)\pi^2}{64},
		\qquad L\geq3.
		\label{eq:SM_AL_main_form}
	\end{equation}
	
	For \(L=2\), there are no bulk sites, so both sites are edge sites. In that case the only two non-\(X\) strings are \(Y_1Z_2\) and \(Z_1Y_2\), and they lie on the same bond:
	\begin{equation}
		(Y_1Z_2)(Z_1Y_2)=X_1X_2.
		\label{eq:SM_AL_L2_contraction}
	\end{equation}
	Again this follows site by site:
	\begin{equation}
		(Y_1Z_2)(Z_1Y_2)=(Y_1Z_1)(Z_2Y_2)=(iX_1)(-iX_2)=X_1X_2.
	\end{equation}
	The two direct squares contribute \(2\left(\frac12\right)^2=\frac12\) and the cross term contributes \(2\left(\frac12\right)\left(\frac12\right)=\frac12\). Thus
	\begin{equation}
		A_2=1.
		\label{eq:SM_AL_L2}
	\end{equation}
	Therefore
	\begin{equation}
		\boxed{	A_L=
			\begin{cases}
				1, & L=2,\\[3pt]
				\dfrac{16L+24+(L-2)\pi^2}{64}, & L\geq3.
		\end{cases} }
		\label{eq:SM_AL_final}
	\end{equation}
	Substituting this into Eq.~\eqref{eq:SM_quench_lifting_AL} gives
	\begin{equation}
		\boxed{	M_2^{(L)}(t_\star,h)
			=\frac{4A_L}{\ln2}\left(\frac{h}{J}\right)^2+O((h/J)^3)}.
		\label{eq:SM_quench_lifting_final}
	\end{equation}
	
	\subsection{Thermodynamic-limit coefficient}
	
	We now extract the large-\(L\) coefficient. Dividing Eq.~\eqref{eq:SM_quench_lifting_final} by \(L\) gives
	\begin{equation}
		\frac{M_2^{(L)}(t_\star,h)}{L}
		=
		\frac{4}{\ln2}\frac{A_L}{L}\left(\frac{h}{J}\right)^2
		+O((h/J)^3).
	\end{equation}
	Using Eq.~\eqref{eq:SM_AL_final}, one has for \(L\geq3\),
	\begin{equation}
		\frac{A_L}{L}
		=
		\frac{1}{4}+\frac{3}{8L}+\frac{(L-2)\pi^2}{64L}.
	\end{equation}
	Taking \(L\to\infty\) therefore yields
	\begin{equation}
		\lim_{L\to\infty}\frac{A_L}{L}
		=
		\frac{1}{4}+\frac{\pi^2}{64}.
	\end{equation}
	Hence the lifting of the magic density at the quench revival is
	\begin{equation}
		\boxed{	\lim_{L\to\infty}\frac{M_2^{(L)}(t_\star,h)}{L}
			=\frac{4}{\ln2}\left(\frac{1}{4}+\frac{\pi^2}{64}\right)\left(\frac{h}{J}\right)^2+O((h/J)^3) }.
		\label{eq:SM_quench_lifting_density}
	\end{equation}
	The coefficient is finite. Thus the exact many-body revival is lifted at order \((h/J)^2\), with a nonzero thermodynamic magic-density curvature.

	
	
	
	
	\section{Perturbative lifting of the two-kick Floquet revival}
	\label{sec:app_perturbative_floquet}
	
	\subsection{Effective detuning generator at \(\theta=\pi/8\)}
	
	We now consider the kicked-Ising circuit
	\begin{equation}
		U_F(g,\theta)=e^{-i\theta H_{ZZ}}e^{-igH_X}.
		\label{eq:SM_floquet_lifting_unitary}
	\end{equation}
	This is the same Floquet family discussed earlier, and here we zoom in on the first nontrivial two-kick revival. The reason for focusing on two kicks is that, at the special angle \(\theta=\pi/8\), the accumulated Ising angle after two kicks is \(2\theta=\pi/4\), which is the first nontrivial Clifford angle of the open-chain Ising evolution. Thus the two-kick state is exactly on a magic revival at the Clifford kick value \(g=\pi/2\), and the present goal is to determine how this revival is lifted when \(g\) is detuned slightly away from \(\pi/2\).
	
	We therefore write
	\begin{equation}
		g=\frac{\pi}{2}+\varepsilon.
		\label{eq:SM_floquet_epsilon}
	\end{equation}
	Let
	\begin{equation}
		V=e^{-i\theta H_{ZZ}},\qquad
		G=e^{-i(\pi/2)H_X}.
		\label{eq:SM_VG_definition}
	\end{equation}
	Since \(H_X\) commutes with itself, the detuned kick separates exactly into a small non-Clifford part and the Clifford kick,
	\begin{equation}
		e^{-igH_X}=e^{-i\varepsilon H_X}G.
		\label{eq:SM_detuned_kick_split}
	\end{equation}
	At \(g=\pi/2\), the kick \(G\) is Clifford. More explicitly, because \(H_X=\sum_{j=1}^L X_j\) is a sum of commuting single-site operators, one has
	\begin{equation}
		G=e^{-i(\pi/2)H_X}=(-i)^L\prod_{j=1}^L X_j.
	\end{equation}
	As discussed earlier in the Floquet section, this global \(X\equiv \prod_{j=1}^L X_j\) operator commutes with every Ising bond \(Z_jZ_{j+1}\), and it leaves \(|+\rangle^{\otimes L}\) invariant up to an overall phase. Therefore, for the purpose of computing the SRE, all factors of \(G\) may be stripped away: they are Clifford and act trivially on the reference stabilizer state up to phase.
	
	With this notation, the two-kick evolution is
	\begin{equation}
		U_F(g,\theta)^2
		=
		\left(Ve^{-i\varepsilon H_X}G\right)
		\left(Ve^{-i\varepsilon H_X}G\right).
	\end{equation}
	Using that \(G\) commutes with \(V\) and also with \(e^{-i\varepsilon H_X}\), we may move the two \(G\)'s to the far right:
	\begin{equation}
		U_F(g,\theta)^2
		=
		Ve^{-i\varepsilon H_X}Ve^{-i\varepsilon H_X}G^2.
	\end{equation}
	Next insert \(VV^{-1}=I\) between the first \(e^{-i\varepsilon H_X}\) and the second \(V\),
	\begin{equation}
		U_F(g,\theta)^2
		=
		V^2\left(V^{-1}e^{-i\varepsilon H_X}V\right)e^{-i\varepsilon H_X}G^2.
	\end{equation}
	Since \(G^2\) is only an overall phase, it is irrelevant for the state and for the SRE. This gives
	\begin{equation}
		(Ve^{-i\varepsilon H_X}G)^2
		\simeq V^2\left(V^{-1}e^{-i\varepsilon H_X}V\right)e^{-i\varepsilon H_X},
		\label{eq:SM_two_kick_stripped}
	\end{equation}
	up to Clifford factors and an overall phase.
	
	At \(\theta=\pi/8\), one has \(V^2=e^{-i(\pi/4)H_{ZZ}}=\prod_{j=1}^{L-1}e^{-i(\pi/4)Z_jZ_{j+1}}\). Each factor is Clifford, since a \(\pi/4\) rotation generated by a Pauli string maps Pauli strings to Pauli strings under conjugation by Eq.~\eqref{eq:SM_anticommuting_conjugation}; hence \(V^2\) is Clifford.\footnote{More explicitly, if a Pauli string \(A\) satisfies \(A^2=I\), then setting \(u=\pi/4\) in Eq.~\eqref{eq:SM_anticommuting_conjugation} gives \(e^{iuA}Be^{-iuA}=B\) when \(A\) and \(B\) commute, and \(e^{iuA}Be^{-iuA}=iAB\) when they anticommute. In either case a Pauli string is mapped to another Pauli string, which is precisely the Clifford property recalled in Eq.~\eqref{eq:SM_clifford_map}. Applied to \(A=Z_jZ_{j+1}\), this shows that each gate \(e^{-i(\pi/4)Z_jZ_{j+1}}\) is Clifford; this is also the bondwise content of the earlier Clifford-angle statement in Eq.~\eqref{eq:SM_clifford_angles}.} Therefore the only non-Clifford part that survives after stripping off the Clifford factors is the product of the two small unitaries
	\begin{equation}
		\left(V^{-1}e^{-i\varepsilon H_X}V\right)e^{-i\varepsilon H_X}.
	\end{equation}
	To first order in \(\varepsilon\), each factor is expanded as
	\begin{equation}
		V^{-1}e^{-i\varepsilon H_X}V
		=
		I-i\varepsilon\,V^{-1}H_XV+O(\varepsilon^2),
		\qquad
		e^{-i\varepsilon H_X}=I-i\varepsilon H_X+O(\varepsilon^2),
	\end{equation}
	and multiplying the two gives
	\begin{equation}
		\left(V^{-1}e^{-i\varepsilon H_X}V\right)e^{-i\varepsilon H_X}
		=
		I-i\varepsilon\left(H_X+V^{-1}H_XV\right)+O(\varepsilon^2).
	\end{equation}
	Hence the first non-Clifford tangent generator at the two-kick revival is
	\begin{equation}
		K_F=H_X+V^{-1}H_XV.
		\label{eq:SM_KF_definition}
	\end{equation}
	
	We may now invoke the tangent theorem, Eq.~\eqref{eq:SM_tangent_formula}. After the Clifford factors are removed, the two-kick state is a small unitary deformation of a stabilizer state generated by \(K_F\), with small parameter \(\varepsilon\). Therefore
	\begin{equation}
		M_{2,{\rm F}}^{(L)}(k=2)
		=\frac{4}{\ln2}B_L\varepsilon^2+O(\varepsilon^3),
		\qquad
		B_L=\langle K_F^2\rangle_+-\langle K_F\rangle_+^2,
		\label{eq:SM_floquet_lifting_BL}
	\end{equation}
	where \(\langle\cdots\rangle_+\) denotes expectation in \(|+\rangle^{\otimes L}\).
	
	It remains to compute \(K_F\). The operator \(V^{-1}H_XV\) is the same Ising conjugation problem already encountered in the quench subsection, and the needed identity is again Eq.~\eqref{eq:SM_anticommuting_conjugation}. Since
	\begin{equation}
		V^{-1}H_XV=e^{i\theta H_{ZZ}}H_Xe^{-i\theta H_{ZZ}},
		\label{eq:SM_V_inverse_HX_V}
	\end{equation}
	we may reuse the earlier edge and bulk formulas for the conjugated operators \(X_j\), now evaluated at \(u=\theta\). At \(\theta=\pi/8\), $\cos(2\theta)=\sin(2\theta)=\frac{1}{\sqrt2}$.
	
	For the two edge sites, the earlier conjugation formulas give
	\begin{equation}
		V^{-1}X_1V=\frac{1}{\sqrt2}X_1-\frac{1}{\sqrt2}Y_1Z_2,
		\qquad
		V^{-1}X_LV=\frac{1}{\sqrt2}X_L-\frac{1}{\sqrt2}Z_{L-1}Y_L.
	\end{equation}
	Adding the first \(H_X\) term in Eq.~\eqref{eq:SM_KF_definition}, namely the bare \(X_1\) and \(X_L\), gives the two edge contributions to \(K_F\):
	\begin{equation}
		X_1+V^{-1}X_1V
		=
		\left(1+\frac{1}{\sqrt2}\right)X_1-\frac{1}{\sqrt2}Y_1Z_2,
		\qquad
		X_L+V^{-1}X_LV
		=
		\left(1+\frac{1}{\sqrt2}\right)X_L-\frac{1}{\sqrt2}Z_{L-1}Y_L.
		\label{eq:SM_KF_edges}
	\end{equation}
	
	For a bulk site \(2\leq j\leq L-1\), the earlier bulk formula gives
	\begin{equation}
		V^{-1}X_jV
		=
		\cos^2(2\theta)X_j
		-\cos(2\theta)\sin(2\theta)\bigl(Z_{j-1}Y_j+Y_jZ_{j+1}\bigr)
		-\sin^2(2\theta)Z_{j-1}X_jZ_{j+1}.
	\end{equation}
	At \(\theta=\pi/8\), this becomes
	\begin{equation}
		V^{-1}X_jV
		=
		\frac{1}{2}X_j
		-\frac{1}{2}\bigl(Z_{j-1}Y_j+Y_jZ_{j+1}\bigr)
		-\frac{1}{2}Z_{j-1}X_jZ_{j+1}.
	\end{equation}
	Adding the first \(H_X\) term again yields, for each bulk site,
	\begin{equation}
		X_j+V^{-1}X_jV
		=
		\frac{3}{2}X_j-\frac{1}{2}(Z_{j-1}Y_j+Y_jZ_{j+1})-\frac{1}{2}Z_{j-1}X_jZ_{j+1}.
		\label{eq:SM_KF_bulk}
	\end{equation}
	Equations~\eqref{eq:SM_KF_edges} and \eqref{eq:SM_KF_bulk} completely determine \(K_F\).
	
	\subsection{Evaluation of \(B_L\)}
	
	The expectation rule is the same as in the quench subsection, Eq.~\eqref{eq:SM_plus_expectation_rule}: in \(|+\rangle^{\otimes L}\), a Pauli string has nonzero expectation only if every site carries \(I\) or \(X\), and in that case the expectation is \(1\). It is useful to repeat the single-site version explicitly,
	\begin{equation}
		\langle +|I|+\rangle=1,
		\qquad
		\langle +|X|+\rangle=1,
		\qquad
		\langle +|Y|+\rangle=0,
		\qquad
		\langle +|Z|+\rangle=0.
	\end{equation}
	Since the expectation of a Pauli string factorizes over sites, any string containing at least one \(Y\) or \(Z\) has zero expectation.
	
	As in the quench calculation, we split
	\begin{equation}
		K_F=K_F^{(X)}+K_F^{({\rm nX})},
	\end{equation}
	where \(K_F^{(X)}\) contains the pure \(X\)-strings and \(K_F^{({\rm nX})}\) contains all other strings. The pure \(X\)-strings do contribute to \(\langle K_F\rangle_+\), but they do not contribute to the variance after subtracting \(\langle K_F\rangle_+^2\). Indeed, every pure \(X\)-string acts as \(+1\) on \(|+\rangle^{\otimes L}\), so \(|+\rangle^{\otimes L}\) is an eigenstate of \(K_F^{(X)}\), which implies
	\begin{equation}
		\langle (K_F^{(X)})^2\rangle_+=\langle K_F^{(X)}\rangle_+^2.
	\end{equation}
	Also, every term in \(K_F^{({\rm nX})}\) contains at least one \(Y\) or \(Z\), so
	\begin{equation}
		\langle K_F^{({\rm nX})}\rangle_+=0.
	\end{equation}
	Finally, multiplying a pure \(X\)-string by a non-\(X\) string cannot remove all \(Y\)'s and \(Z\)'s: on one site, \(XY=iZ\) and \(XZ=-iY\), so a \(Y\) or \(Z\) remains a \(Y\) or \(Z\) after multiplication by \(X\). Therefore the mixed expectations also vanish. Hence
	\begin{equation}
		B_L=\langle (K_F^{({\rm nX})})^2\rangle_+.
	\end{equation}
	
	We can therefore discard the pure \(X\)-strings and keep only the non-\(X\) strings from Eqs.~\eqref{eq:SM_KF_edges} and \eqref{eq:SM_KF_bulk}. For \(L\geq3\), these are
	\begin{itemize}
		\item the two edge strings \(-\frac{1}{\sqrt2}Y_1Z_2\) and \(-\frac{1}{\sqrt2}Z_{L-1}Y_L\);
		\item for each bulk site \(2\leq j\leq L-1\), the three strings
		\(-\frac12 Z_{j-1}Y_j\), \(-\frac12 Y_jZ_{j+1}\), and \(-\frac12 Z_{j-1}X_jZ_{j+1}\).
	\end{itemize}
	
	The direct-square contribution comes from squaring each of these strings individually. Every Pauli string squares to the identity, so each square contributes the square of its coefficient. Thus
	\begin{equation}
		B_L^{\rm dir}
		=2\left(\frac{1}{\sqrt2}\right)^2
		+(L-2)\left[
		2\left(\frac{1}{2}\right)^2
		+\left(\frac{1}{2}\right)^2
		\right]
		=\frac{3L}{4}-\frac{1}{2}.
		\label{eq:SM_BL_direct}
	\end{equation}
	
	We next examine the cross terms in \((K_F^{({\rm nX})})^2\). Most products of two distinct non-\(X\) strings still contain at least one \(Y\) or \(Z\), and therefore have zero expectation in \(|+\rangle^{\otimes L}\). The only surviving contractions are again the adjacent pairs in which a \(Y_jZ_{j+1}\) term from site \(j\) multiplies a \(Z_jY_{j+1}\) term from site \(j+1\):
	\begin{equation}
		(Y_jZ_{j+1})(Z_jY_{j+1})=X_jX_{j+1}.
		\label{eq:SM_BL_adjacent_contraction}
	\end{equation}
	This is evaluated site by site. Operators on different sites commute, so
	\begin{equation}
		(Y_jZ_{j+1})(Z_jY_{j+1})
		=
		(Y_jZ_j)(Z_{j+1}Y_{j+1}).
	\end{equation}
	Using the single-site Pauli multiplication rules
	\begin{equation}
		YZ=iX,\qquad ZY=-iX,
	\end{equation}
	one obtains
	\begin{equation}
		(Y_jZ_j)(Z_{j+1}Y_{j+1})
		=
		(iX_j)(-iX_{j+1})
		=
		X_jX_{j+1}.
	\end{equation}
	The result contains only \(X\)'s, so its expectation is \(1\).
	
	There are two kinds of such contractions. At the left boundary, the string \(-\frac{1}{\sqrt2}Y_1Z_2\) from the edge term contracts with the string \(-\frac12 Z_1Y_2\) from the \(j=2\) bulk term. Since both orderings appear in \(K_F^2\), the contribution is $2\left(\frac{1}{\sqrt2}\right)\left(\frac{1}{2}\right)
	=
	\frac{1}{\sqrt2}$. The same happens at the right boundary, giving another \(1/\sqrt2\).
	
	In the bulk, for each \(j=2,\dots,L-2\), the string \(-\frac12 Y_jZ_{j+1}\) from site \(j\) contracts with the string \(-\frac12 Z_jY_{j+1}\) from site \(j+1\). Again both orderings appear in \(K_F^2\), so each such pair contributes $2\left(\frac12\right)\left(\frac12\right)=\frac12$. There are \(L-3\) such bulk pairs. Hence
	\begin{equation}
		B_L^{\rm cross}
		=\frac{1}{\sqrt2}+\frac{L-3}{2}+\frac{1}{\sqrt2}
		=\sqrt2+\frac{L-3}{2}.
		\label{eq:SM_BL_cross}
	\end{equation}
	
	Combining the direct and cross pieces gives
	\begin{equation}
		B_L=\frac{5L}{4}-2+\sqrt2,
		\qquad L\geq3.
		\label{eq:SM_BL_Lge3}
	\end{equation}
	
	For \(L=2\), there are no bulk sites, so both sites are edge sites. The only two non-\(X\) strings are \(Y_1Z_2\) and \(Z_1Y_2\), and they lie on the same bond:
	\begin{equation}
		(Y_1Z_2)(Z_1Y_2)=X_1X_2.
		\label{eq:SM_BL_L2_contraction}
	\end{equation}
	The direct squares contribute $2\left(\frac{1}{\sqrt2}\right)^2=1$,
	and the cross term contributes $2\left(\frac{1}{\sqrt2}\right)\left(\frac{1}{\sqrt2}\right)=1$.
	Thus
	\begin{equation}
		B_2=2.
		\label{eq:SM_BL_L2}
	\end{equation}
	Therefore
	\begin{equation}
		\boxed{	B_L=
			\begin{cases}
				2, & L=2,\\[3pt]
				\dfrac{5L-8+4\sqrt2}{4}, & L\geq3.
		\end{cases} }
		\label{eq:SM_BL_final}
	\end{equation}
	The two-kick Floquet revival is lifted as
	\begin{equation}
		\boxed{	M_{2,{\rm F}}^{(L)}(k=2)
			=\frac{4B_L}{\ln2}\varepsilon^2+O(\varepsilon^3)}.
		\label{eq:SM_floquet_lifting_final}
	\end{equation}
	
	\subsection{Thermodynamic-limit coefficient}
	
	We now divide by \(L\) and take the thermodynamic limit. Using Eq.~\eqref{eq:SM_BL_final}, one has for \(L\geq3\),
	\begin{equation}
		\frac{B_L}{L}
		=
		\frac{5}{4}-\frac{2}{L}+\frac{\sqrt2}{L}.
	\end{equation}
	Hence
	\begin{equation}
		\lim_{L\to\infty}\frac{B_L}{L}=\frac{5}{4}.
	\end{equation}
	Substituting this into Eq.~\eqref{eq:SM_floquet_lifting_final} gives
	\begin{equation}
		\boxed{	\lim_{L\to\infty}\frac{M_{2,{\rm F}}^{(L)}(k=2)}{L}
			=\frac{5}{\ln2}\varepsilon^2+O(\varepsilon^3) }.
		\label{eq:SM_floquet_lifting_density}
	\end{equation}
	Thus the two-kick magic revival is also lifted quadratically, with finite thermodynamic magic-density curvature.

	
	
	
	\section{Local instability of the magic-density minima}
	
	We now analyze the local shape of the exact magic-density minima identified in the preceding quench and Floquet sections. The point of this section is to show that the revivals discussed earlier are not merely isolated zeros of \(M_2\), but isolated \emph{quadratic} minima of the magic density under the natural local detunings of the quench and Floquet protocols. Since the relevant revival states are stabilizer states up to Clifford factors, the appropriate tool is the tangent theorem, Eq.~\eqref{eq:SM_tangent_formula}, which converts the leading local growth of \(M_2\) into the variance of the corresponding Clifford-stripped tangent generator.
	
	For the open-chain quench, the first nontrivial revival occurs at \(h=0\) and \(Jt=\pi/4\). At this value of the accumulated Ising angle, the Ising evolution is Clifford, so the magic vanishes exactly and the revival point lies on the stabilizer manifold. For the kicked-Ising Floquet problem, the first nontrivial two-kick revival occurs at \(g=\pi/2\), \(\theta=\pi/8\), and kick number \(k=2\). The reason for focusing on two kicks is that \(\theta=\pi/8\) is the first positive angle for which one kick already probes a nontrivial point away from the revival, while two kicks accumulate the Clifford angle \(2\theta=\pi/4\), producing the first nontrivial stroboscopic magic revival. In both problems, then, the local instability is controlled by the tangent geometry around a stabilizer point.
	
	\subsection{Quench expansion in \(\delta=Jt-\pi/4\) and \(\eta=h/J\)}
	
	Near the first quench revival, we introduce the local coordinates
	\begin{equation}
		\delta=Jt-\frac{\pi}{4},
		\qquad
		\eta=\frac{h}{J}.
		\label{eq:SM_delta_eta}
	\end{equation}
	These parameters measure the two physically distinct ways of moving away from the revival. The parameter \(\delta\) detunes the commuting Ising angle away from the Clifford value \(Jt=\pi/4\), while \(\eta\) turns on the noncommuting transverse field relative to the Ising scale.
	
	It is useful to isolate these two directions separately before combining them. First consider the pure Ising-angle detuning, so \(\eta=0\). Then
	\begin{equation}
		e^{-iJtH_{ZZ}}
		=
		e^{-i(\pi/4+\delta)H_{ZZ}}
		=
		e^{-i(\pi/4)H_{ZZ}}e^{-i\delta H_{ZZ}},
	\end{equation}
	because \(H_{ZZ}\) commutes with itself. At the revival point \(Jt=\pi/4\), the first factor is Clifford. Therefore, after stripping away this Clifford factor, the first-order non-Clifford deformation in the \(\delta\)-direction is generated simply by
	\begin{equation}
		H_{ZZ}.
	\end{equation}
	
	Next consider the transverse-field direction, so \(\delta=0\). This was already analyzed above in the perturbative quench section Sec.~\ref{sec:app_perturbative_quench}. There the revival lifting at fixed \(Jt=\pi/4\) was shown to be controlled by the interaction-picture generator \(K_L\), with
	\begin{equation}
		M_2^{(L)}(t_\star,h)
		=
		\frac{4}{\ln2}A_L\eta^2+O(\eta^3),
		\qquad
		A_L=\langle K_L^2\rangle_+-\langle K_L\rangle_+^2,
	\end{equation}
	as stated in Eq.~\eqref{eq:SM_quench_lifting_AL}. Thus the tangent generator in the \(\eta\)-direction is \(K_L\).
	
	When both detunings are present, the first-order stripped deformation is the superposition of the two first-order tangent directions. Hence the combined Clifford-stripped tangent generator is
	\begin{equation}
		K_{\rm Q}=\delta H_{ZZ}+\eta K_L.
		\label{eq:SM_quench_two_param_generator}
	\end{equation}
	This is the quench analogue of the familiar fact that, to first order, nearby deformations in two independent parameters add linearly at the level of the generator.
	
	We may therefore apply the tangent theorem, Eq.~\eqref{eq:SM_tangent_formula}, directly at the revival point. Since the stripped reference state is a stabilizer state, the leading quadratic growth of \(M_2\) is governed by the variance of \(K_{\rm Q}\):
	\begin{equation}
		M_2^{(L)}(\delta,\eta)
		=\frac{4}{\ln2}\operatorname{Var}_+(K_{\rm Q})
		+O_L((|\delta|+|\eta|)^3).
		\label{eq:SM_quench_local_variance}
	\end{equation}
	Here \(O_L((|\delta|+|\eta|)^3)\) means third or higher total order in the small detunings, with a coefficient that may depend on the finite system size \(L\).
	
	To expand the variance, one must expand both terms in its definition,
	\begin{equation}
		\operatorname{Var}_+(X)=\langle X^2\rangle_+-\langle X\rangle_+^2.
	\end{equation}
	For Hermitian operators \(A\) and \(B\),
	\begin{equation}
		\begin{aligned}
			\operatorname{Var}_+(A+B)
			&=\langle (A+B)^2\rangle_+-\langle A+B\rangle_+^2\\
			&=\langle A^2\rangle_++\langle B^2\rangle_+
			+\langle AB+BA\rangle_+\\
			&\quad-\langle A\rangle_+^2-\langle B\rangle_+^2
			-2\langle A\rangle_+\langle B\rangle_+\\
			&=\operatorname{Var}_+(A)+\operatorname{Var}_+(B)
			+2\,\operatorname{Cov}_+(A,B),
		\end{aligned}
	\end{equation}
	where
	\begin{equation}
		\operatorname{Cov}_+(A,B)
		\equiv
		\frac{1}{2}\langle AB+BA\rangle_+
		-\langle A\rangle_+\langle B\rangle_+.
	\end{equation}
	The symmetrized product appears because \(AB\) need not be Hermitian when \(A\) and \(B\) do not commute, whereas \(AB+BA\) is Hermitian and is precisely the mixed operator generated by expanding \((A+B)^2\). Applying this identity to
	\begin{equation}
		K_{\rm Q}=\delta H_{ZZ}+\eta K_L
	\end{equation}
	gives
	\begin{equation}
		\operatorname{Var}_+(K_{\rm Q})
		=
		\delta^2\operatorname{Var}_+(H_{ZZ})
		+\eta^2\operatorname{Var}_+(K_L)
		+2\delta\eta\,\operatorname{Cov}_+(H_{ZZ},K_L).
	\end{equation}
	We now evaluate the diagonal terms explicitly.
	
	Write
	\begin{equation}
		H_{ZZ}=\sum_{j=1}^{L-1}B_j,
		\qquad
		B_j=Z_jZ_{j+1}.
	\end{equation}
	Then
	\begin{equation}
		\langle H_{ZZ}\rangle_+
		=
		\sum_{j=1}^{L-1}\langle B_j\rangle_+.
	\end{equation}
	Each \(B_j\) contains \(Z\) operators, so \(\langle B_j\rangle_+=0\) in \(|+\rangle^{\otimes L}\). Hence
	\begin{equation}
		\langle H_{ZZ}\rangle_+=0.
	\end{equation}
	For the square,
	\begin{equation}
		H_{ZZ}^2
		=
		\sum_{j=1}^{L-1}B_j^2
		+\sum_{j\neq k}B_jB_k.
	\end{equation}
	Since every Pauli string squares to the identity,
	\begin{equation}
		B_j^2=I,
	\end{equation}
	so each diagonal term contributes \(1\) in expectation value. It remains to show that every cross term has zero expectation. If \(B_j\) and \(B_k\) are disjoint, then the expectation factorizes and vanishes because each factor contains \(Z\)'s. If they are adjacent, then, for example,
	\begin{equation}
		B_jB_{j+1}
		=
		Z_jZ_{j+1}Z_{j+1}Z_{j+2}
		=
		Z_jZ_{j+2},
	\end{equation}
	which still has zero expectation in \(|+\rangle^{\otimes L}\). Therefore all cross terms vanish, and
	\begin{equation}
		\langle H_{ZZ}^2\rangle_+=L-1.
	\end{equation}
	Thus
	\begin{equation}
		\operatorname{Var}_+(H_{ZZ})
		=
		\langle H_{ZZ}^2\rangle_+-\langle H_{ZZ}\rangle_+^2
		=
		L-1.
	\end{equation}
	
	The second diagonal term was already computed in the quench-lifting section~\ref{sec:app_perturbative_quench}:
	\begin{equation}
		\operatorname{Var}_+(K_L)=A_L,
	\end{equation}
	with the exact finite-\(L\) expression for $A_L$ given in Eq.~\eqref{eq:SM_AL_final}. The mixed covariance \(\operatorname{Cov}_+(H_{ZZ},K_L)\) will be shown below to vanish. Substituting these ingredients into the variance expansion gives
	\begin{equation}
		\boxed{	M_2^{(L)}(\delta,\eta)
			=\frac{4}{\ln2}\left[(L-1)\delta^2+A_L\eta^2\right]
			+O_L((|\delta|+|\eta|)^3) }.
		\label{eq:SM_quench_local_expansion}
	\end{equation}
	
	We now pass to the magic density. Dividing by \(L\) and taking \(L\to\infty\), we use Eq.~\eqref{eq:SM_AL_final}, which gives for \(L\geq3\),
	\begin{equation}
		A_L=\frac{16L+24+(L-2)\pi^2}{64}.
	\end{equation}
	Therefore
	\begin{equation}
		\frac{A_L}{L}
		=
		\frac{1}{4}+\frac{\pi^2}{64}+\frac{24-2\pi^2}{64L}
		\longrightarrow
		\frac{1}{4}+\frac{\pi^2}{64}.
	\end{equation}
	Since \((L-1)/L\to1\), the thermodynamic magic-density expansion near the quench revival is
	\begin{equation}
		\boxed{\mu_2^{\rm Q}(\delta,\eta)
			=\frac{4}{\ln2}\left[
			\delta^2+\left(\frac{1}{4}+\frac{\pi^2}{64}\right)\eta^2
			\right]+o(\delta^2+\eta^2)}.
		\label{eq:SM_quench_local_density}
	\end{equation}
	This shows that the quench revival is an isolated quadratic minimum of the magic density in the two natural local directions \(\delta\) and \(\eta\).
	
	\subsection{Floquet expansion in \(\delta_\theta=\theta-\pi/8\) and \(\varepsilon=g-\pi/2\)}
	
	We now repeat the same analysis for the kicked-Ising Floquet revival. Near the first nontrivial two-kick revival, we define
	\begin{equation}
		\delta_\theta=\theta-\frac{\pi}{8},
		\qquad
		\varepsilon=g-\frac{\pi}{2}.
		\label{eq:SM_delta_theta_epsilon}
	\end{equation}
	These are the natural local coordinates around the revival point. The parameter \(\delta_\theta\) detunes the Ising angle per kick away from the special value \(\theta=\pi/8\), while \(\varepsilon\) detunes the transverse kick away from the Clifford value \(g=\pi/2\).
	
	The reason the Ising-angle detuning enters with a factor of two is that the revival occurs after \emph{two} kicks. Over two kicks, the Ising part accumulates to
	\begin{equation}
		e^{-i\theta H_{ZZ}}e^{-i\theta H_{ZZ}}
		=
		e^{-i2\theta H_{ZZ}},
	\end{equation}
	since the two factors commute. Hence replacing \(\theta=\pi/8\) by \(\theta=\pi/8+\delta_\theta\) changes the total two-kick Ising angle from
	\begin{equation}
		2\theta=\frac{\pi}{4}
		\qquad\text{to}\qquad
		2\theta=\frac{\pi}{4}+2\delta_\theta.
	\end{equation}
	At accumulated angle \(\pi/4\), the two-kick Ising factor is Clifford. Therefore, after stripping away that Clifford part, the tangent generator in the \(\delta_\theta\)-direction is
	\begin{equation}
		2H_{ZZ}.
	\end{equation}
	
	The \(\varepsilon\)-direction was already analyzed in the preceding section~\ref{sec:app_perturbative_floquet} on perturbative lifting of the two-kick Floquet revival. There it was shown that, at fixed \(\theta=\pi/8\), detuning \(g=\pi/2+\varepsilon\) produces a leading non-Clifford deformation generated by the operator \(K_F\) defined in Eq.~\eqref{eq:SM_KF_definition}, and that
	\begin{equation}
		M_{2,{\rm F}}^{(L)}(k=2)
		=
		\frac{4}{\ln2}B_L\varepsilon^2+O(\varepsilon^3),
		\qquad
		B_L=\langle K_F^2\rangle_+-\langle K_F\rangle_+^2,
	\end{equation}
	as summarized in Eq.~\eqref{eq:SM_floquet_lifting_BL}. Thus the tangent generator in the \(\varepsilon\)-direction is \(K_F\).
	
	Combining the two directions, the full Clifford-stripped tangent generator near the two-kick Floquet revival is
	\begin{equation}
		G_{\rm F}=2\delta_\theta H_{ZZ}+\varepsilon K_F.
		\label{eq:SM_floquet_two_param_generator}
	\end{equation}
	Applying the tangent theorem, Eq.~\eqref{eq:SM_tangent_formula}, gives
	\begin{equation}
		M_{2,{\rm F}}^{(L)}(k=2)
		=\frac{4}{\ln2}\operatorname{Var}_+(G_{\rm F})
		+O_L((|\delta_\theta|+|\varepsilon|)^3).
		\label{eq:SM_floquet_local_variance}
	\end{equation}
	Expanding the variance yields
	\begin{equation}
		\operatorname{Var}_+(G_{\rm F})
		=
		4\delta_\theta^2\operatorname{Var}_+(H_{ZZ})
		+\varepsilon^2\operatorname{Var}_+(K_F)
		+4\delta_\theta\varepsilon\,\operatorname{Cov}_+(H_{ZZ},K_F).
		\label{eq:SM_floquet_variance_expanded}
	\end{equation}
	
	The first diagonal term is exactly the same bond variance computed above:
	\begin{equation}
		\operatorname{Var}_+(H_{ZZ})=L-1.
	\end{equation}
	The second diagonal term is
	\begin{equation}
		\operatorname{Var}_+(K_F)=B_L,
	\end{equation}
	with \(B_L\) given explicitly in Eq.~\eqref{eq:SM_BL_final}. The mixed covariance \(\operatorname{Cov}_+(H_{ZZ},K_F)\) will again be shown below to vanish. Therefore
	\begin{equation}
		\boxed{	M_{2,{\rm F}}^{(L)}(k=2)
			=\frac{1}{\ln2}\left[16(L-1)\delta_\theta^2+4B_L\varepsilon^2\right]
			+O_L((|\delta_\theta|+|\varepsilon|)^3) }. 
		\label{eq:SM_floquet_local_expansion}
	\end{equation}
	
	We again divide by \(L\) and take the thermodynamic limit. From Eq.~\eqref{eq:SM_BL_final}, for \(L\geq3\),
	\begin{equation}
		B_L=\frac{5L-8+4\sqrt2}{4},
	\end{equation}
	so
	\begin{equation}
		\frac{B_L}{L}
		=
		\frac{5}{4}+\frac{-8+4\sqrt2}{4L}
		\longrightarrow
		\frac{5}{4}.
	\end{equation}
	Hence
	\begin{equation}
		\boxed{	\mu_2^{\rm F}(\delta_\theta,\varepsilon)
			=\frac{1}{\ln2}\left(16\delta_\theta^2+5\varepsilon^2\right)
			+o(\delta_\theta^2+\varepsilon^2)}.
		\label{eq:SM_floquet_local_density}
	\end{equation}
	Thus the two-kick Floquet revival is also an isolated quadratic minimum of the thermodynamic magic density.
	
	\subsection{Absence of quadratic cross terms}
	
	It remains to justify the absence of the mixed quadratic terms used in the previous two subsections. For two Hermitian tangent generators \(A\) and \(B\), the relevant quadratic cross coefficient is the symmetrized covariance
	\begin{equation}
		\operatorname{Cov}_+(A,B)
		=\frac{1}{2}\langle AB+BA\rangle_+-\langle A\rangle_+\langle B\rangle_+.
		\label{eq:SM_covariance_definition}
	\end{equation}
	In the present problem, the only mixed terms that can appear are
	\begin{equation}
		\operatorname{Cov}_+(H_{ZZ},K_L),
		\qquad
		\operatorname{Cov}_+(H_{ZZ},K_F).
	\end{equation}
	
	The first simplification is immediate:
	\begin{equation}
		\langle H_{ZZ}\rangle_+=0,
	\end{equation}
	because \(H_{ZZ}\) is a sum of bond strings \(Z_jZ_{j+1}\), each of which has zero expectation in \(|+\rangle^{\otimes L}\). Therefore, for either \(K=K_L\) or \(K=K_F\),
	\begin{equation}
		\operatorname{Cov}_+(H_{ZZ},K)
		=
		\frac{1}{2}\langle H_{ZZ}K+KH_{ZZ}\rangle_+.
	\end{equation}
	
	Now write
	\begin{equation}
		H_{ZZ}=\sum_{j=1}^{L-1}B_j,
		\qquad
		B_j=Z_jZ_{j+1}.
	\end{equation}
	By linearity, it is enough to show that for every bond term \(B_j\) and every Pauli-string term \(P\) appearing in \(K_L\) or \(K_F\),
	\begin{equation}
		\langle B_jP+PB_j\rangle_+=0.
	\end{equation}
	The structure of the proof is the same in the quench and Floquet cases, because the only ingredients are the explicit Pauli-string forms of \(K_L\) and \(K_F\), already derived earlier in Eqs.~\eqref{eq:SM_KL_edges}, \eqref{eq:SM_KL_bulk}, \eqref{eq:SM_KF_edges}, and \eqref{eq:SM_KF_bulk}, together with the expectation rule of Eq.~\eqref{eq:SM_plus_expectation_rule}.
	
	There are two cases.
	
	\paragraph*{1. Anticommuting case.}
	If \(B_j\) anticommutes with \(P\), then
	\begin{equation}
		B_jP+PB_j=0
	\end{equation}
	identically. Hence this pair contributes nothing to the covariance. This occurs whenever the bond \(B_j\) overlaps with \(P\) on an odd number of sites carrying \(X\) or \(Y\). Typical examples are \(P=X_j\), \(P=Y_jZ_{j+1}\), \(P=Z_{j-1}Y_j\), and \(P=Z_{j-1}X_jZ_{j+1}\) with the bond chosen to touch the site where the nontrivial anticommutation occurs.
	
	\paragraph*{2. Commuting case.}
	If \(B_j\) commutes with \(P\), then
	\begin{equation}
		B_jP+PB_j=2B_jP.
	\end{equation}
	So the question becomes whether \(B_jP\) can have nonzero expectation in \(|+\rangle^{\otimes L}\). By Eq.~\eqref{eq:SM_plus_expectation_rule}, such an expectation is nonzero only if the resulting Pauli string contains only \(I\) and \(X\) on every site.
	
	For the strings that actually appear in \(K_L\) and \(K_F\), this never happens in the commuting case:
	\begin{itemize}
		\item If \(P\) is a pure \(X\)-string, then any commuting bond \(B_j\) must be disjoint from the support of those \(X\)'s. The product \(B_jP\) therefore still contains explicit \(Z_jZ_{j+1}\) factors and has zero expectation.
		\item If \(P\) is one of the non-\(X\) strings appearing in \(K_L\) or \(K_F\), then \(P\) already contains at least one \(Y\) or \(Z\). In the commuting case the product \(B_jP\) cannot remove all such factors. Therefore \(B_jP\) still contains at least one \(Y\) or \(Z\), and its expectation also vanishes by Eq.~\eqref{eq:SM_plus_expectation_rule}.
	\end{itemize}
	
	Thus every bond-string pair gives zero contribution to the symmetrized expectation, either because the anticommutator vanishes identically or because the resulting Pauli string has zero expectation in \(|+\rangle^{\otimes L}\). Summing over all bonds and all Pauli-string terms in \(K_L\) or \(K_F\), we conclude that
	\begin{equation}
		\operatorname{Cov}_+(H_{ZZ},K_L)=0,
		\qquad
		\operatorname{Cov}_+(H_{ZZ},K_F)=0.
		\label{eq:SM_cross_covariances_zero}
	\end{equation}
	
	Substituting Eq.~\eqref{eq:SM_cross_covariances_zero} into the variance expansions above yields the diagonal quadratic forms in Eqs.~\eqref{eq:SM_quench_local_expansion} and \eqref{eq:SM_floquet_local_expansion}, and hence the thermodynamic-density expansions in Eqs.~\eqref{eq:SM_quench_local_density} and \eqref{eq:SM_floquet_local_density}. Since all quadratic coefficients are strictly positive, the quench and Floquet revival points are isolated quadratic minima of the thermodynamic magic density.

	\section{Many-body nature of the SRE}
	
	\subsection{Free-fermion solvability of the transverse-field Ising chain}
	
	The transverse-field Ising Hamiltonian used in the main text is
	\begin{equation}
		H_{\rm TFIM}=J\sum_{j=1}^{L-1}Z_jZ_{j+1}+h\sum_{j=1}^{L}X_j.
		\label{eq:SM_TFIM_free_fermion}
	\end{equation}
	To bring this into the standard Jordan--Wigner form, it is convenient to relabel the local spin axes as
	\begin{equation}
		\tau_j^x=Z_j,\qquad
		\tau_j^z=X_j,\qquad
		\tau_j^y=-Y_j.
		\label{eq:SM_tau_rotation}
	\end{equation}
	This is simply a rotation of spin coordinates, chosen so that the Ising coupling becomes an \(x\!-\!x\) term and the transverse field becomes a \(z\)-field, matching the conventional presentation of the quantum Ising chain in the Jordan--Wigner literature \cite{Mbeng_Russomanno_Santoro_2024}. In these rotated axes,
	\begin{equation}
		H_{\rm TFIM}=J\sum_{j=1}^{L-1}\tau_j^x\tau_{j+1}^x+h\sum_{j=1}^{L}\tau_j^z.
		\label{eq:SM_TFIM_tau}
	\end{equation}
	
	For an open chain, we now introduce the Jordan--Wigner Majorana operators \cite{Mbeng_Russomanno_Santoro_2024}
	\begin{equation}
		\gamma_{2j-1}=\left(\prod_{\ell<j}\tau_\ell^z\right)\tau_j^x,\qquad
		\gamma_{2j}=-\left(\prod_{\ell<j}\tau_\ell^z\right)\tau_j^y.
		\label{eq:SM_majoranas}
	\end{equation}
	They satisfy
	\begin{equation}
		\gamma_a^\dagger=\gamma_a,\qquad
		\{\gamma_a,\gamma_b\}=2\delta_{ab}.
		\label{eq:SM_majorana_algebra}
	\end{equation}
	The inverse identities needed here are
	\begin{equation}
		\tau_j^z=i\gamma_{2j-1}\gamma_{2j},\qquad
		\tau_j^x\tau_{j+1}^x=i\gamma_{2j}\gamma_{2j+1}.
		\label{eq:SM_majorana_inverse}
	\end{equation}
	Substituting these into Eq.~\eqref{eq:SM_TFIM_tau} gives
	\begin{equation}
		H_{\rm TFIM}
		=iJ\sum_{j=1}^{L-1}\gamma_{2j}\gamma_{2j+1}
		+ih\sum_{j=1}^{L}\gamma_{2j-1}\gamma_{2j}.
		\label{eq:SM_TFIM_majorana}
	\end{equation}
	This is quadratic in Majorana operators. Equivalently,
	\begin{equation}
		H_{\rm TFIM}=\frac{i}{4}\sum_{a,b=1}^{2L}A_{ab}\gamma_a\gamma_b,
		\label{eq:SM_quadratic_majorana_form}
	\end{equation}
	where \(A\) is a real antisymmetric matrix. The Heisenberg equations are therefore linear:
	\begin{equation}
		\frac{d\gamma_a(t)}{dt}=\sum_bA_{ab}\gamma_b(t),
		\qquad
		\gamma(t)=e^{At}\gamma(0).
		\label{eq:SM_majorana_linear_evolution}
	\end{equation}
	Hence the Majorana covariance matrix
	\begin{equation}
		\Gamma_{ab}(t)=\frac{i}{2}\langle[\gamma_a(t),\gamma_b(t)]\rangle
		\label{eq:SM_majorana_covariance}
	\end{equation}
	evolves as
	\begin{equation}
		\Gamma(t)=O(t)\Gamma(0)O(t)^T,\qquad O(t)=e^{At}.
		\label{eq:SM_covariance_evolution}
	\end{equation}
	This gives an exact finite-size free-fermion solution of the state dynamics for all \(h\).
	
	\subsection{Why \(M_2\) remains a many-body Pauli moment}
	
	Free-fermion solvability of the state does not make the stabilizer R\'enyi entropy a single-particle observable. The quantity computed in the main text is
	\begin{equation}
		M_2(\psi)=-\log_2\left[2^{-L}\sum_{P\in\mathcal P_L}\langle P\rangle_\psi^4\right].
		\label{eq:SM_M2_many_body_again}
	\end{equation}
	The sum runs over all \(4^L\) spin Pauli strings. Under the Jordan--Wigner map, a spin Pauli string generally becomes a Majorana monomial,
	\begin{equation}
		P=\zeta(P)\gamma_{i_1}\gamma_{i_2}\cdots\gamma_{i_m},
		\label{eq:SM_P_majorana_monomial}
	\end{equation}
	where \(\zeta(P)\) is a phase and the length \(m\) can grow with system size.
	
	For a parity-definite Gaussian state, every odd Majorana monomial has zero expectation. For an even monomial, Wick's theorem gives the expectation value as the Pfaffian of the covariance submatrix built from the Majorana indices appearing in that monomial. Writing
	\begin{equation}
		I(P)=\{i_1,i_2,\dots,i_m\}
	\end{equation}
	for the ordered set of Majorana indices appearing in Eq.~\eqref{eq:SM_P_majorana_monomial}, one has
	\begin{equation}
		\langle P\rangle_\psi=\zeta(P)\operatorname{Pf}\Gamma_{I(P)}.
		\label{eq:SM_Pfaffian_expectation}
	\end{equation}
	Here \(\Gamma_{I(P)}\) denotes the \(m\times m\) antisymmetric submatrix of \(\Gamma\) obtained by restricting to the rows and columns labeled by the indices in \(I(P)\).
	
	Thus, with odd monomials contributing zero,
	\begin{equation}
		\boxed{	M_2(\psi)
			=-\log_2\left[
			2^{-L}\sum_{P\in\mathcal P_L}
			\left|\operatorname{Pf}\Gamma_{I(P)}\right|^4
			\right] }.
		\label{eq:SM_M2_Pfaffian_sum}
	\end{equation}
	The free-fermion solution makes each correlator accessible, but the SRE still requires a global fourth moment over the full spin-Pauli distribution. This is why the exact product formula in the commuting open-chain problem is not merely a consequence of free-fermion solvability. It relies on the stronger Clifford normal form, which turns the full Pauli distribution into a product distribution.

	
	\section{Numerical methods and checks}
	
	\subsection{Exact Pauli enumeration}
	
	For small systems we evaluated \(M_\alpha\) directly from the definition. We enumerated all phase-free Pauli strings
	\begin{equation}
		P_a=\sigma^{a_1}\otimes\cdots\otimes\sigma^{a_L},
		\qquad a_j\in\{0,1,2,3\},
		\label{eq:SM_exact_enum_pauli}
	\end{equation}
	and formed
	\begin{equation}
		\Xi_{P_a}(\psi)=2^{-L}|\langle\psi|P_a|\psi\rangle|^2 .
		\label{eq:SM_exact_enum_probability}
	\end{equation}
	The stabilizer R\'enyi entropy was then computed as
	\begin{equation}
		M_\alpha(\psi)
		=\frac{1}{1-\alpha}\log_2\sum_{P_a\in\mathcal P_L}\Xi_{P_a}(\psi)^\alpha-L,
		\label{eq:SM_exact_enum_Malpha}
	\end{equation}
	with \(\alpha=1\) taken by continuity. For the numerical checks in this work we used \(\alpha=2\), for which
	\begin{equation}
		M_2(\psi)
		=-\log_2\left[
		2^{-L}\sum_{P_a\in\mathcal P_L}
		\langle\psi|P_a|\psi\rangle^4
		\right].
		\label{eq:SM_exact_enum_M2}
	\end{equation}
	This brute-force sum scales as \(4^L\), so it was used only for small systems. In practice we used \(L=6,8,10\) for direct checks of the Pauli-basis implementation. In that small-\(L\) validation, the replica Pauli-MPS evaluation was run with Pauli-MPS bond dimension \(\chi_{\rm P}=400\), replica contraction bond dimension \(\chi_2=400\), and Pauli-MPS truncation cuoff \(\epsilon_{\rm P}=10^{-12}\), so that the comparison isolated algorithmic correctness rather than truncation error.
	
	\subsection{Pauli-basis MPS evaluation of \(M_2\)}
	
	For larger systems we evaluated \(M_2\) with matrix-product-state methods for magic in the Pauli basis~\cite{haug2023mpsmagic,lami2023perfectsampling,tarabunga2024paulibasis}. Our implementation follows the replica Pauli-MPS construction of Ref.~\cite{tarabunga2024paulibasis}. We write
	\begin{equation}
		\rho=|\psi\rangle\langle\psi|
		=2^{-L}\sum_{a_1,\ldots,a_L=0}^{3}
		r_{a_1\cdots a_L}\,
		\sigma^{a_1}\otimes\cdots\otimes\sigma^{a_L},
		\label{eq:SM_pauli_expansion_coefficients}
	\end{equation}
	where
	\begin{equation}
		r_{a_1\cdots a_L}
		=\langle\psi|\sigma^{a_1}\otimes\cdots\otimes\sigma^{a_L}|\psi\rangle .
		\label{eq:SM_pauli_coefficients}
	\end{equation}
	Then
	\begin{equation}
		M_2(\psi)
		=-\log_2\left[
		2^{-L}\sum_{a_1,\ldots,a_L=0}^{3}
		r_{a_1\cdots a_L}^4
		\right].
		\label{eq:SM_pauli_mps_M2}
	\end{equation}
	The tensor \(r_{a_1\cdots a_L}\) was stored as a Pauli-basis MPS,
	\begin{equation}
		r_{a_1\cdots a_L}
		=\sum_{\beta_1,\ldots,\beta_{L-1}}
		A_1^{a_1}{}_{\beta_1}
		A_2^{a_2}{}_{\beta_1\beta_2}
		\cdots
		A_L^{a_L}{}_{\beta_{L-1}} .
		\label{eq:SM_pauli_MPS}
	\end{equation}
	The Pauli-basis MPS is built from the physical state MPS and then compressed. For \(M_2\), the fourth moment in Eq.~\eqref{eq:SM_pauli_mps_M2} is obtained by replica contraction, so the full Pauli fourth moment is evaluated without explicitly storing all \(4^L\) Pauli strings.
	
	The state evolution was performed with TDVP for open chains using the ITensor~\cite{Fishman2022Aug} convention \(Z=2S^z\) and \(X=2S^x\). Unless otherwise stated, we used
	\begin{equation}
		J=1,\qquad
		\Delta t=\frac{\pi}{600J},\qquad
		\chi_\psi=[200\,\, \text{to }450],\qquad
		\epsilon_\psi=10^{-12},\qquad
		n_{\rm sweep}=6 .
		\label{eq:SM_TDVP_parameters}
	\end{equation}
	Here \(\chi_\psi\) is the maximum bond dimension of the state MPS set to 200,300,400,450 for system sizes 32,48,64,96 respectively, \(\epsilon_\psi\) is the TDVP truncation cutoff, and \(n_{\rm sweep}\) is the number of TDVP sweeps per time step. The Pauli-basis calculation used
	\begin{equation}
		\chi_{\rm P}=[200, 300, 400, 450],\qquad
		\chi_2=80,\qquad
		\epsilon_{\rm P}=10^{-10},
		\label{eq:SM_pauli_MPS_parameters}
	\end{equation}
	where $\chi_P$ is the bond dimension of the compressed Pauli-MPS (truncated from the raw bond dimension $\chi_\psi^2$, using the same system size dependent values as $\chi_\psi$ above), $\chi_2$
	is the max bond dimension set after the replica contraction step for $M_2$, and \(\epsilon_{\rm P}\) is the Pauli-MPS truncation cutoff. The local physical dimension of the Pauli MPS is \(4\).

	
	\subsection{Finite-size and benchmark checks}
	
	At \(h=0\), the finite-size dependence is known exactly:
	\begin{equation}
		\frac{M_\alpha^{(L)}(\vartheta)}{L}
		=\left(1-\frac{1}{L}\right)m_\alpha(\vartheta).
		\label{eq:SM_exact_finite_size_scaling}
	\end{equation}
	This identity was used as the main check of the open-chain data and gives the finite-size scaling behind the main-text inset. For the main-text size scans we used
	\begin{equation}
		L=32,\ 48,\ 64,\ 96 .
		\label{eq:SM_MPS_system_sizes}
	\end{equation}
	The exactly solvable open-chain point \(h=0\) was checked against Eq.~\eqref{eq:SM_open_chain_m2}. We also monitored the Pauli-vector norm before the replica contraction; for a normalized pure state this norm should be one. This provides a direct check that the Pauli-basis truncation has not changed the state at the level relevant for \(M_2\).
	
	For the open-chain quench at \(h=0\), we additionally compared the numerical \(M_2(t)\) curve against the exact closed form. For the entanglement benchmarks, the von Neumann entropy was computed from the Schmidt values across the chosen bond, and at \(h=0\) we checked the exact cut-independent formula on two interior cuts.
	
	\subsection{Checks of the initial QFI benchmark and the tangent theorem}
	
	The initial curvature and the tangent theorem were benchmarked independently in four ways. First, we used the exact value \(F_Q=4J^2(L-1)\). Second, we evaluated the variance directly as an MPO expectation value of \(H_{ZZ}\). Third, we extracted the tangent coefficient from the numerically computed small-\(t\) data for \(M_2(t)\). Fourth, we evaluated the squared-fidelity curvature from the overlap definition using
	\begin{equation}
		\varepsilon_{\rm fid}=10^{-2},\ 10^{-3},\ 10^{-4},\ 10^{-5}.
	\end{equation}
	We also checked the same QFI using the zero-temperature dynamical susceptibility. For this we evolved the correlation function on a real-time grid with
	\begin{equation}
		\Delta t_\chi=0.05,\qquad t_{\max,\chi}=80,
	\end{equation}
	used damping values
	\begin{equation}
		\eta=0.2,\ 0.1,\ 0.05,\ 0.02,\ 0.01,
	\end{equation}
	and performed the frequency integral on a grid \(\omega\in[10^{-3},\,4L+8]\) with 4000 points.
	
	\subsection{Checks of the quench revival lifting coefficient}
	
	The perturbative lifting coefficient for the static quench revival was checked independently from the analytic derivation by approaching the revival point numerically. For the static quench revival,
	\begin{equation}
		\frac{\ln2}{4}\frac{M_2^{(L)}(t_\star,h)}{(h/J)^2}
		\longrightarrow A_L,
		\qquad
		h/J\to0,
		\label{eq:SM_state_vector_AL_check}
	\end{equation}
	with \(A_L\) given in Eq.~\eqref{eq:SM_AL_final}. The numerical scan used \(t_\star=\pi/(4J)\) and
	\begin{equation}
		h/J=0,\ 0.005,\ 0.010,\ldots,\ 0.250 .
		\label{eq:SM_h_scan_values}
	\end{equation}
	Time traces illustrating the same lifting used
	\begin{equation}
		h/J=0,\ 0.05,\ 0.10,\ 0.15,\ 0.20,\ 0.25,\ 0.30,
		\qquad
		0\leq t\leq 1.3\,t_\star .
		\label{eq:SM_h_trace_values}
	\end{equation}
	
\subsection{Checks of the Floquet revival lifting coefficient}

For the numerical scan shown in Fig.~2(d) of the main text, we fixed \(L=64\), \(\theta=\pi/8\), and \(k=2\), and parametrized the transverse kick as \(g=\pi/2+\varepsilon\). The detuning values used were
\begin{equation}
	\begin{aligned}
		\varepsilon={}&
		0.005,\ 0.0075,\ 0.010,\ 0.015,\ 0.0175,\ 0.020,\ 0.025,\\
		&0.0275,\ 0.030,\ 0.035,\ 0.0375,\ 0.040,\ 0.045,\ 0.0475,\\
		&0.050,\ 0.055,\ 0.0575,\ 0.060,\ 0.065,\ 0.070,\ 0.075,\\
		&0.080,\ 0.085,\ 0.090,\ 0.095,\ 0.100,\ 0.150,\ 0.200 .
	\end{aligned}
	\label{eq:SM_floquet_detuning_scan_values}
\end{equation}
The resulting two-kick magic was compared with the leading quadratic
prediction in Eq.~\eqref{eq:SM_floquet_lifting_BL}.

	\bibliography{refs}